%% file: InterleavedMatching.tex
\documentclass{JHEP3}
\usepackage{axodraw}
\usepackage{amsmath}
\usepackage[utf8]{inputenc}

\input ./mydefs

\renewcommand{\aux}{\ensuremath{z}}

\preprint{arXiv:1109.4829 [hep-ph] (March 7, 2012)}

\skip\footins = 1\bigskipamount plus 2pt minus 4pt                              

\title{Matching Tree-Level Matrix Elements with Interleaved Showers}

\author{Leif Lönnblad and Stefan Prestel\\
  Dept.~of Astronomy and Theoretical Physics, Lund University, Sweden, and\\
  CERN Theory Division, Geneva, Switzerland\\
  E-mail: \email{Leif.Lonnblad@thep.lu.se}
    and \email{Stefan.Prestel@thep.lu.se}}
  
  \abstract{We present an implementation of the so-called \ckkwl
    merging scheme for combining multi-jet tree-level matrix elements
    with parton showers. The implementation uses the
    transverse-momentum-ordered shower with interleaved multiple
    interactions as implemented in \pytppp. We validate our procedure
    using $\ee$-annihilation into jets and vector boson production
    in hadronic collisions, with special attention to details in the
    algorithm which are formally sub-leading in character, but may
    have visible effects in some observables.

    We find substantial merging scale dependencies induced by the
    enforced rapidity ordering in the default \pytppp shower. If this
    rapidity ordering is removed the merging scale dependence is
    almost negligible. We then also find that the shower does a
    surprisingly good job of describing the hardness of multi-jet
    events, as long as the hardest couple of jets are given by the
    matrix elements.

    The effects of using interleaved multiple interactions as compared
    to more simplistic ways of adding underlying-event effects in
    vector boson production are shown to be negligible except in a few
    sensitive observables.

    To illustrate the generality of our implementation, we also give
    some example results from di-boson production and pure QCD jet
    production in hadronic collisions.

  }

\begin{document}
 
\sloppy
 
\section{Introduction}
\label{sec:intro}

Production rates for multi-jet events at the LHC are very large,
and the understanding of such events is important, not least as
most discovery channels for new physics involve jets. The main
irreducible, and often huge, background for such signals comes from 
QCD processes. To distill a signal one therefore
needs to make complicated cuts to decrease the QCD background,
sometimes by several orders of magnitude. For this it is very
important that we have a good understanding, not only of the average
behaviour of multi-jet processes, but also the fluctuations and very
rare events coming from standard QCD.

The state-of-the-art for simulating multi-jet final states with Monte
Carlo event generators is to use \ckkw-based algorithms to
combine exact tree-level matrix elements (ME) with parton showers (PS) in a
consistent way. Here, the matrix elements describe accurately the
production of several hard, well-separated partons, while the parton
shower encodes how these are evolved into partonic jets by accurately
modelling the soft and collinear partonic emissions, in a way such
that standard hadronisation models can be applied to produce realistic
exclusive hadronic multi-jet final states.

However, \ckkw merging algorithms mainly focus on the jets produced
in the primary interaction, and little attention is normally given to
jets which may arise from rare, but hard fluctuations in the
underlying events. If at all, the underlying-event contribution is
typically added to the merged sample assuming that the additional
scatterings are completely independent of the primary interaction.
This may be a good approximation in most cases, but it is clear that
there are correlations between the primary interaction and the
underlying event, which we think are important to investigate carefully.

The multiple interaction model in \pytppp is arguably the most
advanced model for the underlying event today. It contains several
sources of correlations between the primary interaction and the
underlying event. In particular, the model for multiple scatterings
is tightly tied to the parton shower in that additional scatterings
are \emph{interleaved} with the parton evolution.

In this paper we implement the \ckkwl algorithm for merging parton
showers with tree-level matrix elements in \pytppp, and in doing
so we consider possible effects of the fact that the \pytppp
shower is interleaved with multiple interactions. Although the effects
turn out to be small, we note that there may be more sources of
correlations which are currently not taken into account by
\pytppp, and our scheme is a way to automatically take into
account any such correlations also in the merging with tree-level
matrix elements.

It should be noted that this is not the first implementation of
matrix-element merging with the \pythia shower. Interfaces exists for
the FORTRAN version of \pythia to the \alpgen~\cite{Mangano:2002ea}
program by employing the MLM matching prescription~\cite{MLM}, and to
\madgraph/ \madevent~\cite{Maltoni:2002qb} using so-called
Pseudo-Shower merging~\cite{Mrenna:2003if}.

The outline of this article is as follows. First in sections
\ref{sec:ckkw-l} and \ref{sec:interleaved} we briefly recapitulate the
main features of the \ckkwl merging scheme and the interleaved showers
respectively, before we describe the details of our \pytppp
implementation in section~\ref{sec:implementation}. Then we present
results in section~\ref{sec:results}, starting with some control plots
to validate the implementation before we study the effects of multiple
interactions and other formally sub-leading features on the production
of vector bosons with additional jets at the LHC. We end with showing
some comparisons with data, and some preliminary results also from
di-boson and pure QCD jet production. Finally we present our
conclusions in section~\ref{sec:outlook}.

\section{The \ckkwl merging scheme}
\label{sec:ckkw-l}

Here we will present the main features of the \ckkwl merging
procedure. For a more detailed discussion of \ckkwl and other similar
merging algorithms we refer to \cite{Lavesson:2007uu,Alwall:2007fs}
and the original publications \cite{Catani:2001cc,Lonnblad:2001iq}. 

The starting point for \ckkwl is that we have a tree-level
matrix-element generator capable of generating the Born-level process
of interest, as well as the same process with up to $N$ additional
partons. The matrix elements used are regularised with a jet cutoff
which we refer to as the merging scale, $\tms$. To the states
generated in this way we want to add a parton shower to \emph{dress
  up} the hard partons with emissions below the
merging scale in a way such that the soft and collinear emissions are
properly modelled.

As the matrix elements are inclusive, in that they give the cross
section for states with \emph{at least} $n$ additional partons
resolved above the merging scale, it is obvious that we cannot simply
add the event samples generated with different parton
multiplicities. Instead we want to make the samples exclusive by
reweighting them with Sudakov form factors taken to be the no-emission
probabilities the parton shower would have used to produce the same
partonic states.

To calculate the form factor we first have to reconstruct a parton-shower
history for the states with $n$ additional partons, \state{n}, given
by the matrix element generator. This means that we have to answer the
question, \emph{how would my parton shower have generated this state?}
The answer to this question is not necessarily unique. The parton
shower may produce a given final parton state in several ways,
just as a given state may be represented by many different Feynman
diagrams. In \ckkwl, these different path are considered by reconstructing
all possible parton shower histories, and picking one of them according to 
probabilities calculated from the relevant splitting functions.

Doing this, we arrive at a history, in which a sequence of parton
shower emissions are specified by the ordering scale of each emission
$\ord_i$ and other splitting variables such as the energy fractions,
and azimuthal angles, denoted by $\aux_i$. We also obtain a sequence
of intermediate parton states, $\state{i}$.  The requirement on the
parton shower is therefore that it must have complete on-shell
intermediate parton states between each splitting. Until fairly
recently this was only true for the \ariadne
program\cite{Lonnblad:1992tz}, which also was the first to use the
\ckkwl merging\cite{Lonnblad:2001iq}.

Let us denote Sudakov form factors by
\begin{equation}
  \label{eq:sud}
  \sud{i}(\ord_i,\ord_{i+1})=
  \exp\left[-\int_{\rho_{i+1}}^{\rho_i}d\rho\int d\aux
    \as(\rho) \splitP_i(\rho,\aux)\right]~.
\end{equation}
This is the probability that there are no parton shower emissions
from the state $\state{i}$ between the scales $\rho_i$, and
$\rho_{i+1}$. The reweighting with Sudakov form factors now
proceeds by starting the parton shower at a given intermediate state
$\state{i}$, setting $\ord_i$ as the maximum scale, and generating one emission
$(\rho,\aux)$. The probability that this emission is above
$\rho_{i+1}$ is exactly $1-\sud{i}(\ord_i,\ord_{i+1})$, so throwing
away the event if the emission is above $\rho_{i+1}$ is equivalent to 
reweighting with the Sudakov form factor.

A special treatment is called for in the Sudakov between the last
emission scale, $\rho_n$, and the merging scale, in the case the
cutoff in the matrix elements is not defined in terms of the parton
shower ordering variable. In the case of $n<N$, the event is rejected
if the trial emission from the state $\state{n}$ is above the matrix
element cut-off, irrespective of how it is defined. In the case of
$n=N$, however, no Sudakov-reweighting is done.

The $n$-parton state is typically generated using matrix elements with
a fixed $\as(\mu)$, so that we also reweight the event with
\begin{equation}
  \label{eq:alphasreweight}
  \prod_{i=1}^n\frac{\as(\rho_i)}{\as(\mu)}
\end{equation}
to obtain the same running of \as\ as in the shower.

Finally, note that for initial-state parton-shower
splittings, the no-emission probability \Pnoem is not the same as the
Sudakov form factor needed to reweight the matrix-element generated
state. Instead we have \cite{Krauss:2002up,Lavesson:2005xu},
\begin{equation}
  \label{eq:sudnoem}
  \sud{i}(\ord_i,\ord_{i+1})=
  \frac{f(x,\ord_{i})}{f(x,\ord_{i+1})}\times
  \noem{i}(\ord_i,\ord_{i+1}),
\end{equation}
and the corresponding ratios of parton density functions are included
as an additional weight.

We have thus constructed exclusive final states with an arbitrary
number of partons resolved above the parton shower cutoff scale
$\ord_c$. The distribution of these states are resummed to all orders
in \as, according the precision of the parton shower. However, the
$n\le N$ emissions which are considered hardest in the parton-shower
sense, and are above the merging scale as well, will have their
splitting functions corrected to reproduce the correct tree-level
matrix element.

It should be noted that if the merging scale is defined in the same
way as the parton shower evolution scale, the \ckkwl is equivalent to
standard \ckkw, as long as the latter is used with a shower which is
properly vetoed and truncated\cite{Nason:2004rx}. In
Appendix~\ref{sec:comm-logar-accur} we elaborate on how the
logarithmic accuracy of the shower is preserved in \ckkwl and compare
with the case of standard \ckkw using truncated showers.

\section{Interleaved showers}
\label{sec:interleaved}

As mentioned in the previous section, the requirement on a parton
shower to be used in the \ckkwl procedure is that it gives complete
on-shell partonic states between each emission. In this respect, the
transverse-momentum ordered shower in \pytppp\cite{Sjostrand:2007gs}
is perfectly suited. However, it is not completely straight forward to
implement \ckkwl with \pytppp, as the parton shower in the case of
hadron collisions is \emph{interleaved} with multiple interactions.

The philosophy behind the interleaved shower is that processes with a
high scale in some sense happen \emph{before} processes at lower
scales. As the emissions in a parton shower are not completely
independent in that every emission will give rise to recoils and will
carry away some energy and momentum, it is important that the
emissions are performed in the right order. It is, for example, not
reasonable that an emission of a gluon with small transverse momentum
removes so much energy as to make an emission with a higher transverse
momentum impossible. The argument is based on formation times --- a
final state parton with large transverse momentum is to some extend
formed long before one with small transverse momentum.

If we consider standard QCD jet production in proton collisions, a
parton shower is typically initiated by a hard $2\to2$ matrix element
at some transverse momentum. The parton shower then evolves these hard
jets by emitting final-state radiation from the outgoing partons and
initial-state radiation from the incoming partons. This is done
iteratively, ordering the emissions in transverse momentum.

There is also a chance for a second (semi-)hard
interaction between the colliding protons. Also, one of the outgoing
partons from the hard interaction can rescatter with one of the 
spectator partons in one of the colliding protons, and in addition,
outgoing partons are allowed to rescatter among themselves. In \pytppp such 
scatterings are included in the shower procedure such that an additional 
scattering at a scale $\ord\sub{MI}$ will happen before \eg\ an initial-state 
splitting at a scale, $\ord\sub{i}<\ord\sub{\tiny{MI}}$.

This means that the no-emission probabilities are modified
in \pytppp, and now consist of several pieces,
\begin{equation}
  \label{eq:inteleavedsud}
  \noem{i}(\ord_i,\ord)=
  \noem{i}\sup{PS}(\ord_i,\ord)\noem{i}\sup{MI}(\ord_i,\ord)
  \noem{i}\sup{RS}(\ord_i,\ord),
\end{equation}
where the superscript refers to the standard parton shower (PS),
multiple interactions (MI) and rescattering (RS). If we have resolved a state 
\state{i} at a scale $\ord_{i}$, the probability for a change of 
type $a$ at scale $\ord$ is 
\begin{equation}
  \label{eq:interleavedprob}
  {\cal P}^a(\ord)=\splitP^a(\ord)\times
  \sud{i}\sup{PS}(\ord_i,\ord)\sud{i}\sup{MI}(\ord_i,\ord)
  \sud{i}\sup{RS}(\ord_i,\ord),
\end{equation}
where $\splitP^a$ is the inclusive probability.

\pytppp uses an interleaved treatment of spacelike
(initial-state radiation --- ISR) and timelike showers (final-state
radiation --- FSR), so that the no-emission probability
$\noem{i}\sup{PS}$ is further subdivided as
\begin{equation}
  \label{eq:inteleavedsudps}
  \noem{i}\sup{PS}(\ord_i,\ord)=\noem{i}\sup{ISR}(\ord_i,\ord)
  \noem{i}\sup{FSR}(\ord_i,\ord).
\end{equation}

The ordering scale, \ord, is defined in different ways for different
processes, but they all correspond to a relative transverse momentum
of emitted partons. For ISR the scale is
\begin{equation}
\label{eq:ordISR}
\ord^{\textnormal{ISR}} = (1-z)~Q^2,
\end{equation}
where $-Q^2$ is the virtuality of the incoming original parton and $z$
is its momentum fraction, and for FSR we have
\begin{equation}
\label{eq:ordFSR}
\ord^{\textnormal{FSR}} = z(1-z)~Q^2 ~,
\end{equation}
where $Q^2$ is the invariant mass of the radiating parton, and $z$ the
energy fraction (in the dipole rest frame) of the emitted parton. For
MI and RS the scale is simply given by the squared transverse momentum
of the emitted partons.

The full interleaving of all shower components makes \pytppp ideal for
our prescription of matrix element merging, since the full no-emission
probability can, as will be explained below, easily be generated in
only one step.

\section{Implementation in \pytppp}
\label{sec:implementation}

Due to the requirement of fully on-shell intermediate states,
\ckkwl merging has so far only been implemented in the \ariadne
shower. Here, we present a new implementation within \pytppp, which is
conceptually equivalent to the former, but differs in details relating
to the differences in the parton showers.

\subsection{Constructing the parton shower history}
\label{sec:history}

A key concept of the merging algorithm is the assignment of a shower
history --- a sequence of shower states and evolution scales --- to each
$n$-particle configuration supplied by the matrix element
generator. In the \ckkwl approach, this is done by constructing every
possible path from a core Born-level process to the current
$n$-particle state.

Here we encounter the first difference between \ariadne and
\pytppp. The first gluon emission is particularly simple in \tee in
\ariadne, where the evolution variable and the splitting kernel for
the first splitting are symmetrical between both outgoing legs, thus
resulting in only one possible path: One dipole splitting into two
dipoles. In \pytppp, the approach is slightly different. Also here a
dipole-like approach is used, but the emission is explicitly divided
up into two contributions stemming from each of the dipole ends, where
the radiation close to one end of the dipole is considered more likely
to come from this dipole end itself. Different splitting probabilities
for either dipole end will thus result in two different ways in which
\pytppp could have arrived at the $+1$-parton state. In general there
are more possible paths in \pytppp than in \ariadne. We therefore
try to investigate in some detail the effects of different ways of
choosing a path.

What we basically want to do is to reconstruct which Feynman diagram
gives the largest contribution to the state produced by the matrix
element generator. The preferred option would be to ask the
matrix element generator itself, but this information is not always
easily accessible. Even if such details were available, it is not always 
enough, as a given Feynman diagram may also correspond to different parton 
shower histories.
As is discussed in Appendix \ref{app:prob-and-kin}, we approach
this issue by constructing all possible path of collinear splittings, and 
pick a path according to the product of splitting probabilities. For the
hardest emission, the splitting probability is supplemented so that the matrix
element transition probability is assured. More precisely, we choose a path
according to the probability
\begin{eqnarray}
  \label{eq:path-prob}
  w_p &=& \frac{ w_{1p}(z_{1p}) \prod_{i=2}^n
          \frac{\splitP_{ip}(z_{ip})}{\ord_{ip}}}%
  {\sum_{r} w_{1r}(z_{1r})
 \prod_{i=2}^n\frac{\splitP_{ir}(z_{ir})}{\ord_{ir}}}
\qquad\textnormal{where}
\end{eqnarray}
\begin{description}
\item[$\splitP_{ip}$]: Splitting kernel for splitting number $i$ in path number $p$,
\item[$\ord_{ip}$]: Evolution scale of splitting number $i$ in path number $p$,
\item[$z_{ip}$]: Energy fraction carried by the parton emitted in splitting 
number $i$ in path number $p$,
\item[$w_{1p}$]: Improved splitting probability for hardest splitting,
including weights of ME corrections in the shower. 
\end{description}
The precise forms of these terms are derived in Appendix 
\ref{app:prob-and-kin}, where we also elaborate on how the  
intermediate states $\state{i}$ in path $p$ are constructed.

It must be noted that in the limit of strong ordering, which is the
relevant limit when looking at the formal logarithmic accuracy of the
procedure, picking the most likely path is trivial. Hence, the
way a path is selected will only give sub-leading effects on any
observable. We will nevertheless investigate how large these effects
are by implementing two different schemes. One is similar to the
original \ariadne-implementation, and is based on
\eqref{eq:path-prob}. The other is inspired by the \ckkw-implementation
in \herpp \cite{Tully:2009}, where the path which has the smallest sum of
transverse momenta in the splittings is chosen exclusively. Clearly, in the
strongly ordered limit, both of these will find the ``right'' path,
but as we will see in section \ref{sec:results}, there are visible
differences.

For higher jet multiplicities, minor complications of the
path concept arise. First, we know that shower emissions are always
ordered in some scale variable $\ord$ (virtuality, angle, transverse
momentum). This is not always true for consecutive clusterings of jets
from a matrix element. We choose to interpret a sequence of such
unordered splittings as a single step in the algorithm such that all
steps will be ordered. We must then decide which scale to use for this
combined emission step. Assume that we have a sequence of
reconstructed scales given by $\ord_1>\ord_3>\ord_2>\ord_4$. The
combined emission then corresponds to $\ord_2$ and $\ord_3$ and we can
generate the total no-emission probability as
\begin{equation}
  \label{eq:unorderedmax}
  \noem{}(\rho_0,\rho_4)=
  \noem{0}(\rho_0,\rho_1)\noem{1}(\rho_1,\rho_3)\noem{3}(\rho_3,\rho_4)
\end{equation}
or
\begin{equation}
  \label{eq:unorderedmin}
  \noem{}(\rho_0,\rho_4)=
  \noem{0}(\rho_0,\rho_1)\noem{1}(\rho_1,\rho_2)\noem{3}(\rho_2,\rho_4)~.
\end{equation}
In the former case, the no-emission probability between the
scales $\rho_3$ and $\rho_2$ is calculated using the 1-parton state,
while in the latter, the 3-parton state is used. We will investigate 
the difference between using the higher ($\ord_3$) or lower scale 
($\ord_2$) as minimal scale for rejecting trial emissions off the 
1-parton state in section \ref{sec:validation}.

\FIGURE{
\centering
  \includegraphics[width=0.5\textwidth]{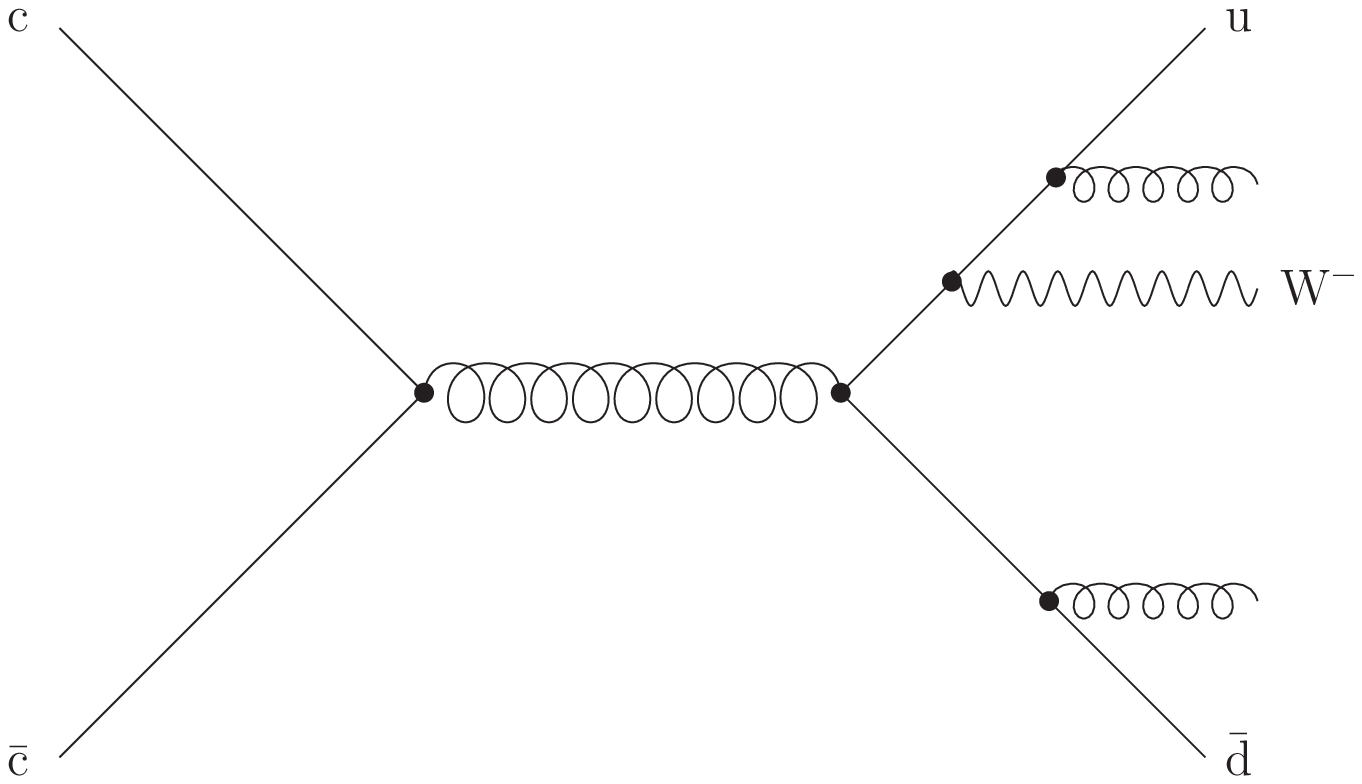}
\caption{\label{fig:incomplete}An example of a matrix element 
contribution without a complete shower history. In this case, only 
the two gluon emission can be reclustered, $\qc\cbar\to\qu\dbar\Wm$ 
is regarded a separate hard process.}
}

Some rare matrix element configurations, \eg\ massive electroweak
corrections to an underlying QCD process, as shown in 
Figure~\ref{fig:incomplete}, could never have been produced in the shower
algorithm. For such processes, clustering will be attempted as far as
possible. The last, irreducible, state will be treated as a new hard
process, and be assigned a shower starting scale in the same way
\pytppp normally would have assigned a scale when presented with such
processes. When handling externally generated processes, \pytppp would
by default start the evolution at the factorisation scale defined in
the matrix element evaluation. However, different user choices are
allowed. In section \ref{sec:validation}, we will also investigate the
effects of other scale choices.

\FIGURE{
\centering
  \includegraphics[width=0.7\textwidth]{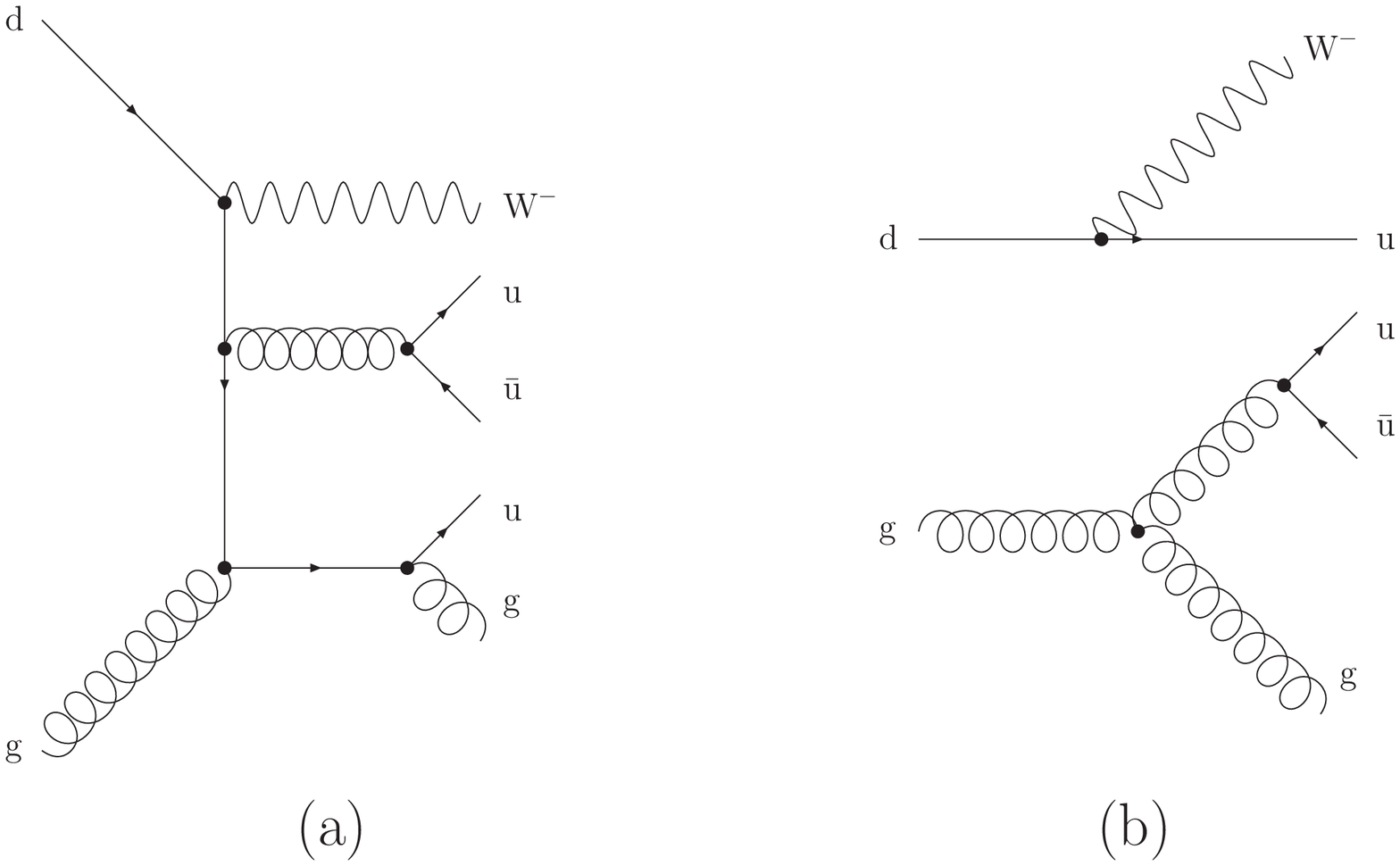}
  \caption{\label{fig:impossible}An example of two different ways the
    iterative clustering may interpret of a particular colour
    configuration in a $\qd\g\to\W^-\qu\ubar\qu\g$ process. Since (b) has
    disconnected external particles, no valid shower history can be
    found.}
}

On a more technical note, we disallow clusterings that will result in
a unreasonable Born-level process. An example would involve starting
from the configuration shown in Figure~\ref{fig:impossible} (a). From
only recombining colour and flavour, alternative (b) would be
identical, and, albeit being disconnected, allowed. The interpretation
of the configuration as either (a) or (b) is tied to which \qu\ubar\
pair is clustered to a gluon. Since we will always be able to find
sensible paths like (a), impossible paths (b) leading to disconnected
diagrams will be discarded.

In other merging prescriptions, these problems are addressed with
other strategies. In \herpp \cite{Tully:2009}, the authors found that
results where insensitive to the treatment of unordered or incomplete
paths and chose to retain incomplete contributions. \sherpa
\cite{Gleisberg:2008ta} follows a different approach in that no
incomplete histories are constructed, since if necessary, electroweak
bosons will be clustered as well. This would interpret the diagram in
Figure~\ref{fig:incomplete} as an electro--weak matrix element
correction to di-jet production.

\subsection{Interleaved multiple interactions}
\label{sec:interl-mult-inter}

At the LHC, events with only one parton--parton scattering per collision are 
highly improbable, and a lot of effort has gone in to the modelling of 
multiple scatterings in \pytppp. When merging the \pytppp shower 
with matrix elements, it is therefore desirable to keep the modelling of
multiple scatterings as intact as possible.

In \pytppp, multiple interactions and radiation compete for the
available phase space. To make sure that some part of phase space is
exclusively filled by matrix element configurations, another part by
shower radiation and multiple interactions, we minimally modify the
\ckkwl algorithm. The generation of no-emission probabilities has to
be slightly refined to keep the effect of multiple interactions on
the no-emission probability, while assuring the validity of our
algorithm. 

The formal proof that the merging scale dependence cancels to the 
accuracy of the shower rests on the assumption that the factorisation 
scheme defined by the shower evolution equation is uniform over all 
of phase space. In \ckkwl, this is realised by allowing trial 
emissions in the matrix element domain, \ie\ off reclustered states, 
without phase space restrictions, and vetoing events if the first 
emission off a ME configuration produced another ME configuration, 
\ie\ a parton above $\tms$. In this way, events with non-zero weight 
have been treated identically in the ME and PS regions. This 
prescription has to be generalised to include additional sources of 
emissions, \eg\ multiple interactions.

The requirement that the shower evolution is identical in ME and PS 
domains forces us to treat multiple interactions on equal footing 
with radiative emissions, once secondary scatterings are included in 
the evolution of partons by allowing for competition over phase 
space. When performing trial showers on a reclustered state, we thus 
treat multiple interactions identical to ``ordinary" emissions. The 
treatment of the first emission off ME configurations defines the 
border between ME and PS regions. We choose to slightly refine this 
definition by requiring that the matrix element region contains only
radiative emissions above a cut $\tms$. This means that once a 
different type of emission has been produced, we are in the parton 
shower domain, and we should continue the shower without any 
additional phase space restriction. More concretely, when checking 
the first shower evolution response from a ME configuration, we keep 
the state if an emission below $\tms$ or a secondary scattering has 
been generated. Hence, the lower bound on the 
matrix-element-corrected region is changed to 
$\tms'=\max(\tms, \ord_{\textnormal{\tiny{MI}}})$. The reason for 
this treatment is that we want to keep hard multiple interactions 
generated by the shower, rather than unjustifiably restricting 
them to be below $\tms$.

Let us describe our procedure with a specific example for merging up to three
additional jets. Consider a $\W+3$ gluon event, with scales
$\ord_1 \geq \ord_2 \geq \ord_3$. When only allowing QCD radiation
and multiple scatterings, this state could be produced by
\begin{enumerate}
\item Three gluon emissions off $\W$ production;
\item One gluon emission off $\W$ production, and one secondary $gg
  \to gg$ or $\q\qbar\to gg$ scattering;
\end{enumerate}
Clearly, the first possibility can and should be corrected with matrix
elements according to the standard \ckkwl procedure. In the second
case, the hardest scale can be attributed to either MI ($\ord_1 =
\ord_{\textnormal{\tiny{MI}}} =\ord_2 > \ord_3$) or an emission. In
the former, we think of the state as inside the PS domain. This means
that the shower would have produced the secondary interaction first,
``freeing" the subsequent emissions from phase space
restrictions. Thus, we have to generate this state from the 0-jet matrix 
element, and, to avoid double counting, veto it in trial showers off
reconstructed configurations. If the hardest scale was associated with
an emission ($\ord_1 > \ord_2 = \ord_{\textnormal{\tiny{MI}}}
=\ord_3$), we can distinguish two cases. If the hardest emission is in
the PS domain already, there is no reason to restrict the event
generation further by disallowing MI above a certain scale. In effect,
the configuration is taken from the evolution of the 0-jet ME sample,
while removing it from the 1-jet sample by vetoing configurations with
$\ord_{\textnormal{\tiny{MI}}} > \ord_{1,\textnormal{reclus}}$ in the
trial showers. Finally, the emission with $\ord_1$ can be in the
matrix element phase space. Adding one secondary interaction will
produce a state of two correlated $2 \to 2$ processes. Since no matrix
elements can include this state, it is unambiguously inside the PS
region, even without applying additional constraints related to a
merging scale. This reasoning leads us to define the cross-over of ME
and PS domains by a phase space cut for emissions, or the existence of
more than one $2 \to 2$ process. Coming back to the example, we will
generate this state from the 1-jet matrix element by adding a secondary
scattering. In order to avoid double counting, in trial showers off 
reconstructed states, we veto the event if the trial emission resulted in 
$\ord_{1,\textnormal{reclus}} > \ord_{\textnormal{\tiny{MI}}} >
\ord_{2,\textnormal{reclus}}$.

This example illustrates the algorithm and sheds light on how
particular configurations are generated. The bottom line is that every
event where the $n$ hardest (according to the parton shower ordering)
partons can produced in one of the matrix elements samples, it will be
taken from this sample. Hence, we are still true to the philosophy of
\ckkwl merging. Note that in this publication, we will only consider
merging matrix elements with additional QCD-induced jets. Therefore we
will \eg\ treat photon radiation in the shower in the same way as to
multiple interactions.

To validate our algorithm, we chose to implement an alternative
treatment of multiple interactions, which is similar to the
prescription applied in \sherpa \cite{Gleisberg:2008ta}. For this, we
exclude multiple interactions when performing trial showers on
reclustered states, keeping only the shower emissions in the
Sudakov form factors. Then, when showering the matrix element
configurations, we allow additional interactions below the scale
$\rho_1$ of the reclustered $2\to 2$ process. For the $+0$ jet
contribution, we choose $\rho_0=\tms$ as maximal scale. Differences
between both treatments are investigated in section
\ref{sec:validation}.

\subsection{The algorithm step-by-step}
\label{sec:algorithm-step-step}

After choosing a parton shower history for the matrix element state,
the weight the parton shower would have generated while evolving to
this state has to be calculated. This includes the running of $\as$ in
the shower, the no-emission probabilities generated by choosing
particular splittings and the way parton distribution functions guide
the space-like evolution. In the \ckkwl scheme, a seamless inclusion
of ME configurations into the parton shower is then achieved by
reweighting the state with the parton shower weight
\begin{eqnarray}
w_{\textnormal{\tiny{CKKWL}}} &=& 
 \tfrac{x_0^+f_0^+(x_0^+,\ord_0)}{x_n^+f_n^+(x_n^+,\mu_F^2)}
 \tfrac{x_0^-f_0^-(x_0^-,\ord_0)}{x_n^-f_n^-(x_n^-,\mu_F^2)}
 \times \left(\prod_{i=1}^{n}
  \tfrac{x_i^+f_{i}^+(x_i^+,\ord_i)}{x_{i-1}^+f_{i-1}^+(x_{i-1}^+,\ord_{i})}
 ~\tfrac{x_i^-f_{i}^-(x_i^-,\ord_i)}{x_{i-1}^-f_{i-1}^-(x_{i-1}^-,\ord_{i})}
  \right)\nonumber\\
&&\times\left(\prod_{i=1}^{n}\tfrac{\as(\ord_i)}{\asme}\right)
  \times\left(\prod_{i=1}^{n}\noem{i-1}(\ord_{i-1},\ord_i)\right)
  \times\noem{n}(\ord_n,\tms)\label{eq:ckkwl-wgt-old}\\
&=&  \frac{x_{n}^+f_{n}^+(x_{n}^+,\ord_n)}{x_{n}^+f_{n}^+(x_{n}^+,\mu_F^2)}
     \frac{x_{n}^-f_{n}^-(x_{n}^-,\ord_n)}{x_{n}^-f_{n}^-(x_{n}^-,\mu_F^2)}
     \nonumber\\
 &&\times
   \prod_{i=1}^{n} \Bigg[\frac{\as(\ord_i)}{\asme}
     \frac{x_{i-1}^+f_{i-1}^+(x_{i-1}^+,\ord_{i-1})}
          {x_{i-1}^+f_{i-1}^+(x_{i-1}^+,\ord_i)}
     \frac{x_{i-1}^-f_{i-1}^-(x_{i-1}^-,\ord_{i-1})}
          {x_{i-1}^-f_{i-1}^-(x_{i-1}^-,\ord_i)}\nonumber\\
&&\qquad \noem{i-1}(\ord_{i-1},\ord_i)\Bigg]\noem{n}(\ord_n,\tms)
  ~,\label{eq:ckkwl-wgt}
\end{eqnarray}
where $\ord_i$ are the reconstructed scales of the splittings. The
first PDF ratio in \eqref{eq:ckkwl-wgt-old} means that the total cross
section is given by the lowest order Born-level matrix element, which
is what the non-merged \pytppp shower uses. The PDF ratio in brackets comes
from of the fact that shower splitting probabilities are products of
splitting kernels and PDF factors. The running of $\as$ is correctly
included by the second bracket. Finally, the event is made exclusive
by multiplying no-emission probabilities. In our implementation, we
chose to reorder the PDF ratios according to \eqref{eq:ckkwl-wgt},
so that only PDFs of fixed flavour and x-values are divided, thus
making the weight piecewise numerically more stable. The algorithm to
calculate and apply this weight can be summarised as follows:
\begin{enumerate}
\item[I.] Produce \textit{Les Houches} event files (LHEF)
  \cite{Alwall:2006yp} with a matrix element generator for
  $n~=~0,1\ldots N$ extra jets with a regularisation cut-off, $\tms$,
  typically using a fixed factorisation scale, $\mu_F$, and a fixed
  $\asme$.
\item[II.] Pick a jet multiplicity, $n$, and a state $S_n$ according
  to the cross sections given by the matrix element generator.
  \begin{enumerate}
  \item[1.] Find all shower histories for the state $S_n$, pick a sequence 
    according to the product of splitting probabilities. Only pick un-ordered 
    sequences if no ordered sequence was found. Only pick incomplete paths if no
    complete path was constructed.
  \item[2.] Perform reweighting according to \eqref{eq:ckkwl-wgt}: For each
    $0\leqslant i-1<n$,
    \begin{enumerate}
    \item[i.] Start the shower off the state $S_{i-1}$ at
      $\ord_{i-1}$, generate a trial state $R_i$ with scale
      $\ord_{R_i}$. If $\ord_{R_i} > \ord_{i}$, veto the event and
      start again from II.
    \item[ii.] Calculate the weight factor
      \begin{equation}
        w_{i-1} = \frac{\as(\ord_{i})}{\asme}
        \frac{x_{i-1}^+f_{i-1}^+(x_{i-1}^+,\ord_{i-1} )}
              {x_{i-1}^+f_{i-1}^+(x_{i-1}^+,\ord_{i} )}
        \frac{x_{i-1}^-f_{i-1}^-(x_{i-1}^-,\ord_{i-1} )}
             {x_{i-1}^-f_{i-1}^-(x_{i-1}^-,\ord_{i} )}
      \end{equation}
    \end{enumerate}
  \item[3.] Start the shower from $S_{n}$ at $\ord_{n}$, giving a state 
    $R_{n+1}$ with the scale $\ord_{R_{n+1}}$.
    \begin{enumerate}
    \item[i.] If $n<N$, and $R_{n+1}$ was produced from $S_{n}$ by QCD
      radiation, and $k_\perp(R_{n+1}) > \tms$, reject the event and
      start again from II. Otherwise, accept the event and the
      emission and continue the shower. If a multiple interaction was
      generated, keep it and continue the shower without restrictions.
    \item[ii.] If $n=N$, continue the shower without vetoing.
    \end{enumerate}
  \end{enumerate}
\item[III.] If the event was not rejected, multiply the event weight by
  \begin{equation}
    \frac{x_{n}^+f_{n}^+(x_{n}^+,\ord_{n} )} {x_{n}^+f_{n}^+(x_{n}^+,\mu_F^2 )}
    \times
    \frac{x_{n}^-f_{n}^-(x_{n}^-,\ord_{n} )} {x_{n}^-f_{n}^-(x_{n}^-,\mu_F^2 )}
    \times\prod_{i=1}^n w_{i-1}
  \end{equation}
\item[IV.] Start again from II.
\end{enumerate}

Our merging approach is, with dynamically generated Sudakov factors,
tailored to always reproduce what \pytppp would most probably have
done to arrive at the current configuration. Starting scales are of
course no exception. Thus, we will start (trial) showering of
electroweak $2\to 2$ processes at the kinematical limit $\sqrt{s}$,
both for radiation and for multiple interactions, which is the default
procedure in \pytppp. In this way, the question for a starting scale
of multiple interactions when merging additional emissions is
irrelevant.

For jet production in the pure QCD case, by default we set the
transverse momentum of the outgoing partons in the $2\to 2$ process as
starting scale in the shower and multiple interactions. This should be
adequate as long as the merging scale is not too small. For very small
merging scales we have the option of including a Sudakov form factor
giving the probability that no additional scatterings are produced
between the maximum scale, $\sqrt{s}$, and the transverse momentum of
the $2\to 2$ process. This would make the primary process exclusive,
in the sense that we make sure that there are no harder scatterings in
the event.

Note also that in pure QCD, the Born-level $2\to2$ process is in
itself divergent and we must introduce a cutoff regularisation. This
cutoff need not be the same as the merging scale. In fact we will here
choose a much lower scale to avoid having a large fraction of the
reclustered multi-jet ME-states ending up below the cut and resulting
in un-ordered paths. In addition, the procedure must be changed
slightly since also the scale of the reclustered $2\to2$ state is
included in the classification of un-ordered histories.

In all cases, we implemented the scale settings such that user choices
(\eg\ forcing ``power showers'') are always transferred to the trial
showers off $2\to 2$ processes. For higher-order tree-level matrix
elements, we use the reconstructed splitting scale of the state as
starting point.

When comparing alternative MI treatments, special care is required 
when setting the starting scale. For the \sherpa-inspired 
prescription, we will set the scale $\rho_1$ of the reclustered 
$2\to 2$ process as the MI starting scale for states $\state{n>0}$, and allow
multiple scatterings below $\ord_0 = \tms$ for the $+0$ jet matrix element 
contributions.

\section{Results}
\label{sec:results}

We have implemented the necessary code for \ckkwl merging in \pytppp,
where it has been publicly available as of version 8.157.

In the following, we will first show some validation plots on parton
level for jet production in $\ee$ collisions and weak boson production
at hadron colliders. We then move to more realistic observables for
these processes, and compare to data. Thereafter, di-boson production and pure 
QCD jet production are examined.

As input matrix element kinematics, we choose Les Houches Event Files
generated with \madgraph/\madevent and the following
settings\footnote{Note that the values of \as\ and the
  factorisation scales used here are somewhat irrelevant, as they will
  nevertheless be divided out in \eqref{eq:ckkwl-wgt}.}:
\begin{itemize}
\item Fixed renormalisation scale $\mu_R = \mz$.
\item CTEQ6L1 parton distributions used for hadron collisions.
\item $\as(\mz) = 0.118$ for lepton collisions and to $\as(\mz) =
  0.129783$ for hadron collisions.
\item Fixed factorisation scale $\mu_F$ set to $\mw$ for $\W+$jets,  $\mz$ for 
$\Z+$jets, $\mw+\mz$ for $\W\Z+$jets and $\mz$ for pure QCD di-jets.
\item Durham/$k_\perp$-cut
\begin{equation}
\label{eq:tmsdefinition}
k_\perp^2 =
\begin{cases}
\min\left\{ 2 \cdot \min( E_i^2, E_j^2) ( 1 - \cos\theta_{ij})\right\} & 
  \textnormal{for $\ee\to$jets}\\
\min\left\{\min( p_{T,i}^2, p_{T,j}^2),\min( p_{T,i}^2, p_{T,j}^2) 
  \frac{(\Delta \eta_{ij})^2 + (\Delta\phi_{ij})^2}{D^2}\right\} & 
  \textnormal{for $\p\p (\p\pbar) \rightarrow$ (V+) jets.}
\end{cases}\nonumber
\end{equation}
  with $D = 0.4$, to regularise the QCD divergences and act as merging scale 
  $\tms$.
\item Require $p_{T,\ell} > 20$ GeV in $\Z+$ jets to avoid low momentum in 
  $\gamma$ propagators.
\item Require $p_{T,j} > 10$ GeV in QCD di-jet events.
\end{itemize}
For brevity, we will refer to results of merging of up to $N$ additional jets 
as ME$N$PS. Contributions for a fixed number $n \leq N$ of jets from the 
matrix element will be indicated by a superscript $n$, as in ME$^nN$PS. Also, 
we will write {\tt PYTHIA8} when talking about the default \pytppp behaviour. 
For all distributions, we use routines of the {\texttt{fastjet}} package 
\cite{Cacciari:2005hq} to define and analyse jets. If not otherwise indicated, 
we present plots at the parton level, \ie\ after shower and multiple 
interaction evolution, since merging effects are more visible without smearing
due to hadronisation.

\subsection{Validation}
\label{sec:validation}

We begin by considering the simplest case, with only one extra
parton added to the Born-level state. This is a very useful benchmark
for any matching or merging algorithm, as emphasised in
\cite{Lavesson:2007uu}, because many parton shower programs, such \ariadne
and \pytppp, implement directly the tree-level matching by modifying
the splitting functions for the first emission. Hence, when comparing
a merged parton shower with the matched one, it is very easy to see if
the merging algorithm, for example, has any non-trivial dependence on
the merging scale.

\subsubsection*{Merging scale dependence in $\ee\to jjj$}

The \pytppp parton cascade by default includes reweighting of the
first splitting of the hard process with the correct matrix element
expression, thus giving an excellent handle to check our
implementation of $\ee\to$ jets. To compare our result with \pytppp,
we however have to make a minor change to the shower. When supplied
with a $\ee\to\q\qbar$ state, \pytppp will use the three body matrix
element as splitting kernel for the first splitting of $\q$ \emph{and}
the first splitting of $\qbar$. This is done since the
$\ee\to\q\qbar\g$ matrix element provides a better estimate of the
dipole splitting kernel than the DGLAP kernel. However, when starting
from $\ee\to\q\qbar\g$ input, \pytppp will use DGLAP kernels in the
evolution of the quarks. Thus the showers response to LHEF input of
$\ee\to\q\qbar$ and $\ee\to\q\qbar\g$ will slightly differ when
constructing additional jets. Since we want to merge also higher jet
multiplicities with the \pytppp cascade, it is natural to exclude the
improvement in the $\ee\to\q\qbar$ case, and switch off the usage of
matrix element correction weights for more than three final partons. In the most
recent versions of \pytppp, such a switch is available for user input.

\FIGURE{
  \includegraphics[width=0.75\textwidth]
    {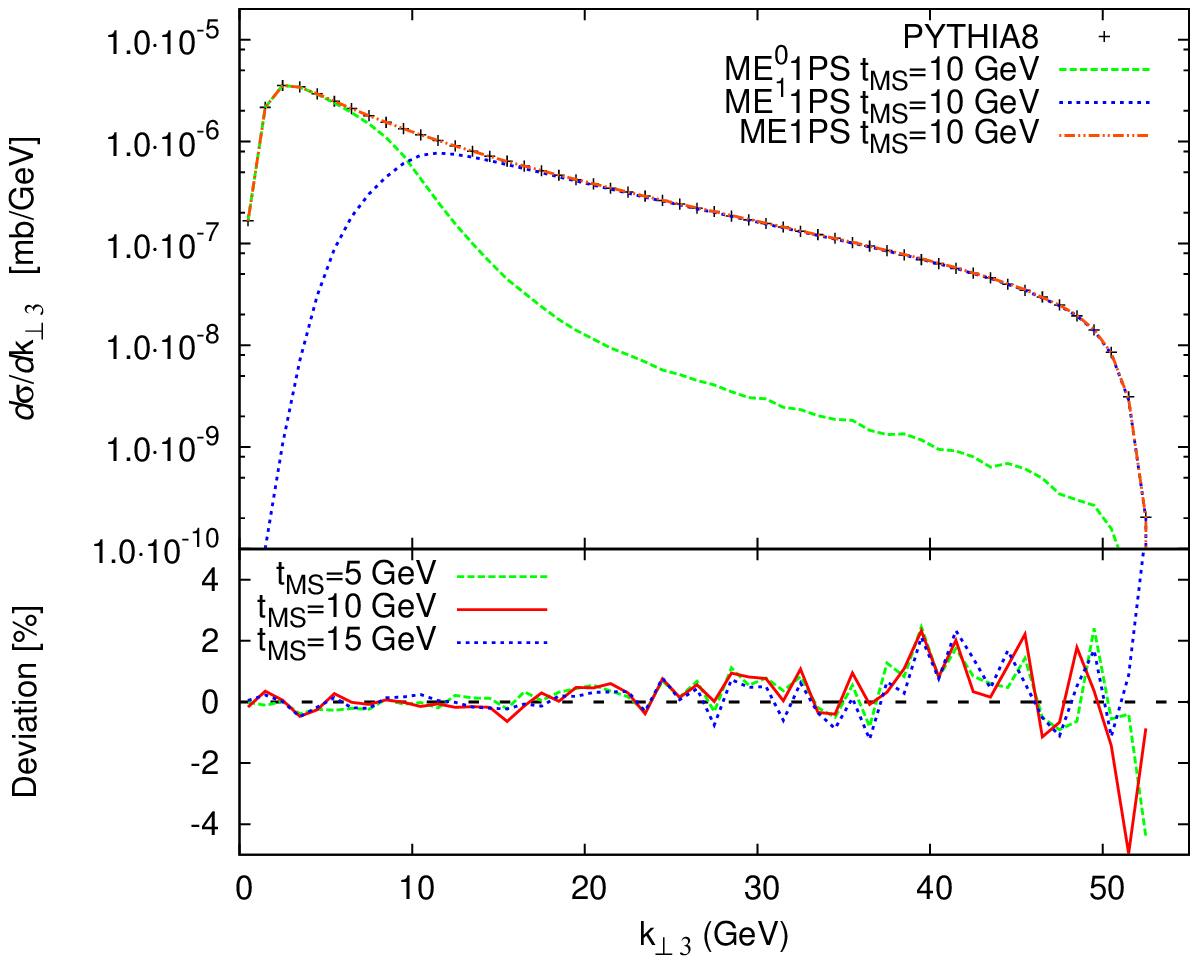}
\caption{\label{fig:ee:kTthird}$k_\perp$ separation of the third jet in $\ee$ 
collisions at $\ECM = 91.25$ GeV. Jets were defined with the Durham algorithm. 
Hadronisation was switched off. The bottom in-set shows the deviation of the 
merged samples for three different merging scales $\tms$ with respect to 
default, matrix-element-corrected \pytppp.}
}

Doing this, we can compare ME1PS with \pytppp. The variable used as a
separation cut $\tms$ between matrix element and parton shower domains
is most sensitive to the implementation of the merging procedure. In
Figure~\ref{fig:ee:kTthird}, we show the value of $k_\perp$ for which
three jets would be clustered to two jets. As desired, we find
excellent agreement, and, when examining different values of the
separation cut $\tms$, vanishing merging scale dependence.

\FIGURE{
\centerline{
  \includegraphics[width=0.75\textwidth]
    {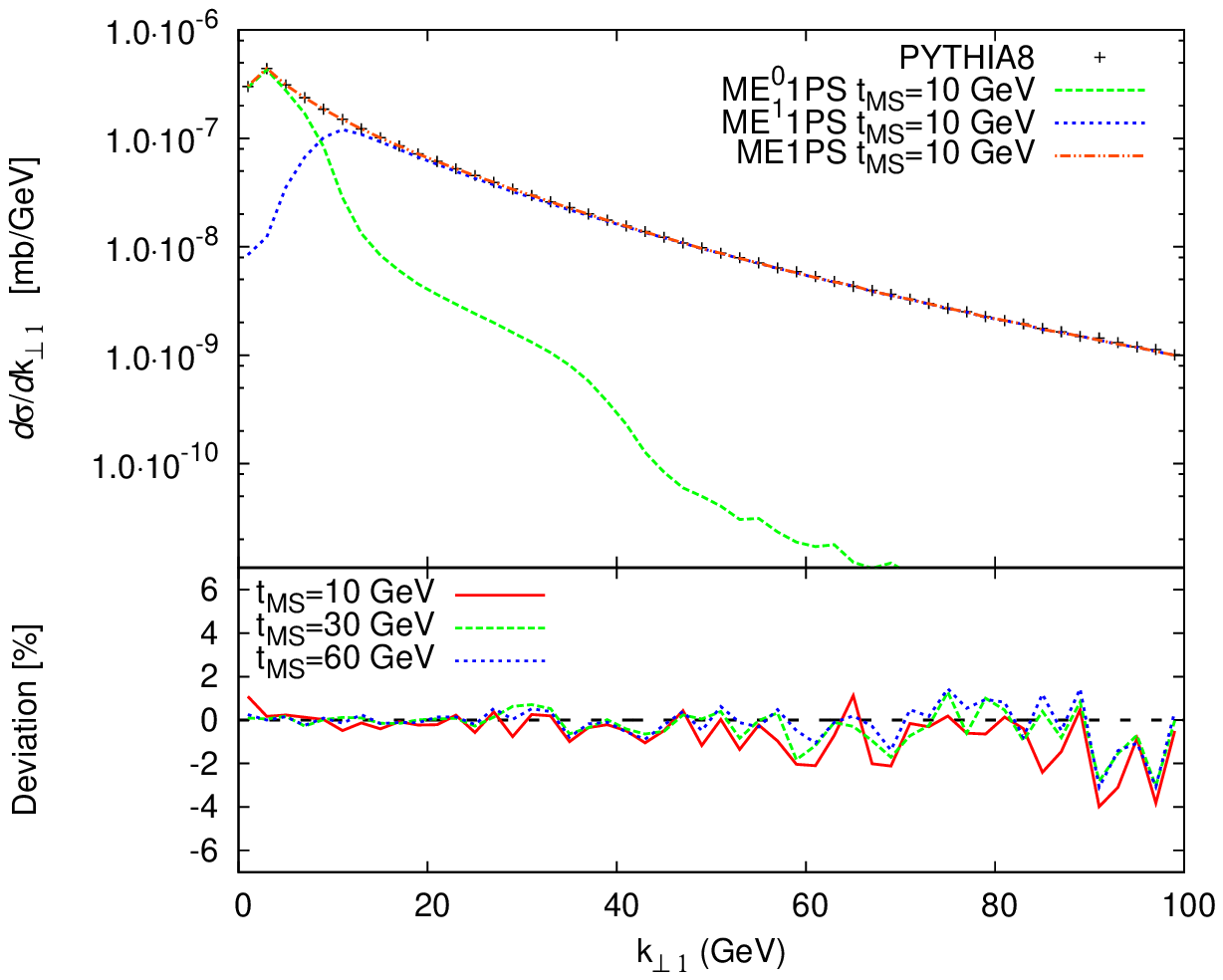}
}
  \caption{\label{fig:w:kTfirst}Transverse momentum of the hardest jet in 
$\W+1$ jet events at $\ECM = 7000$ GeV in $\p\p$ collisions. Jets were defined 
with the $k_\perp$-algorithm with $D = 0.4$. Multiple interactions and 
hadronisation have been switched off. The bottom in-set shows the deviation of 
the merged sample with respect to default matrix-element corrected \pytppp, for
three different merging scales.}
}

\subsubsection*{Merging scale dependence in $\p\p\to$ V + 1 jet}

Similarly, the implementation of V+1 jet merging can be validated against 
default \pytppp. In accordance with the discussion above, we switch off 
additional matrix element reweighting factors in default \pytppp after the first
 initial state emission. Further, it is important to note that in \pytppp, 
infrared divergences in space-like splittings are regularised by shifting the 
denominator of the integration measure in the evolution equation by a small 
$\ord_{reg}$. This shift is inspired by the interleaved evolution of space-like
splittings and multiple interactions, where colour screening will dampen the 
number of interactions. Not strictly perturbative effects like these will be 
present in the default \pytppp distributions, even at 
$p_{\perp} \approx \mathcal{O}(10\textnormal{ GeV})$. That the merging is well 
under control is shown in Figure~\ref{fig:w:kTfirst}, 
where we set $\ord_{reg}=0$ for the first splitting in default \pytppp to 
remove the deliberate mismatch in integration measures. We then find complete 
agreement in the $k_\perp$ distributions.

\FIGURE{
\centerline{
  \includegraphics[width=0.5\textwidth]
    {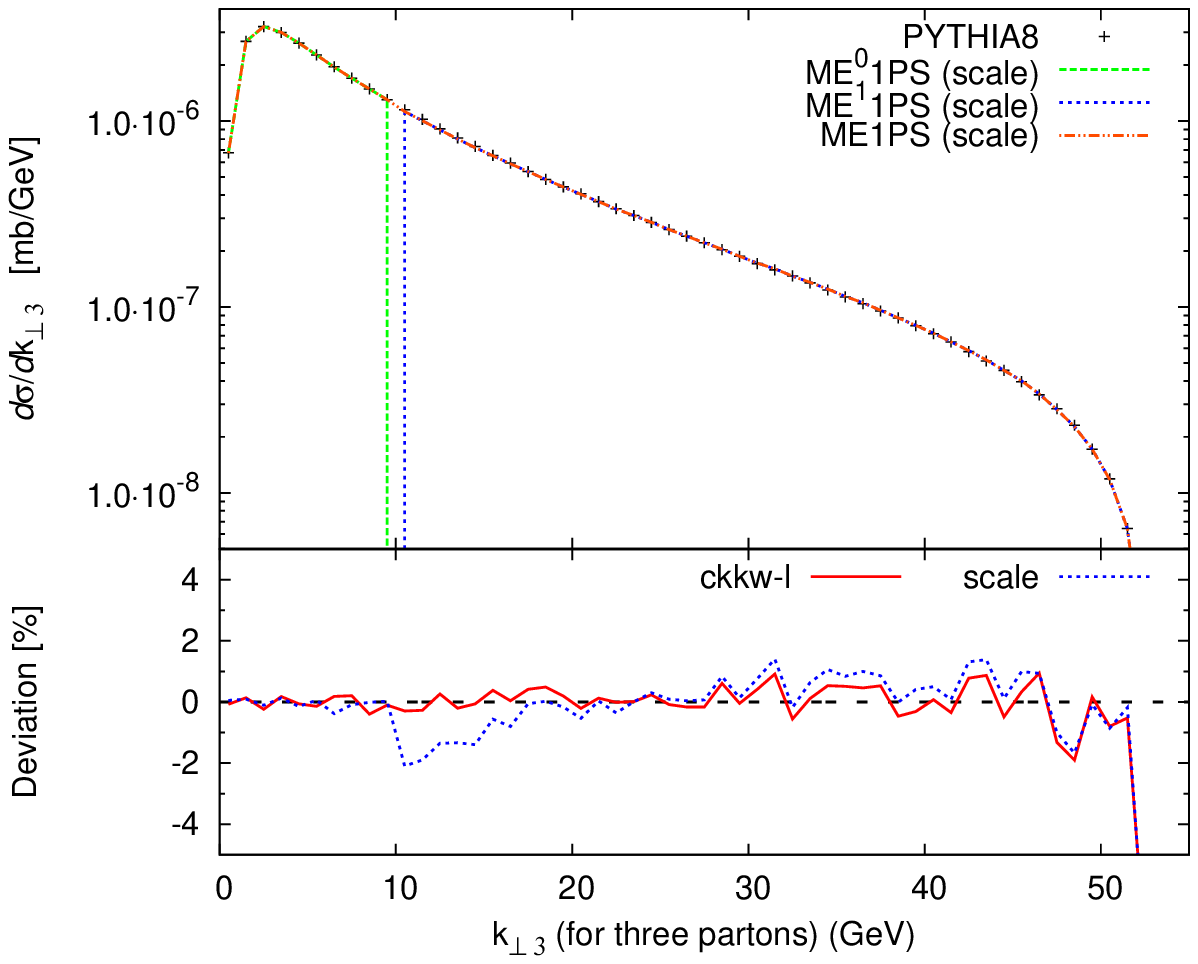}
  \includegraphics[width=0.5\textwidth]
    {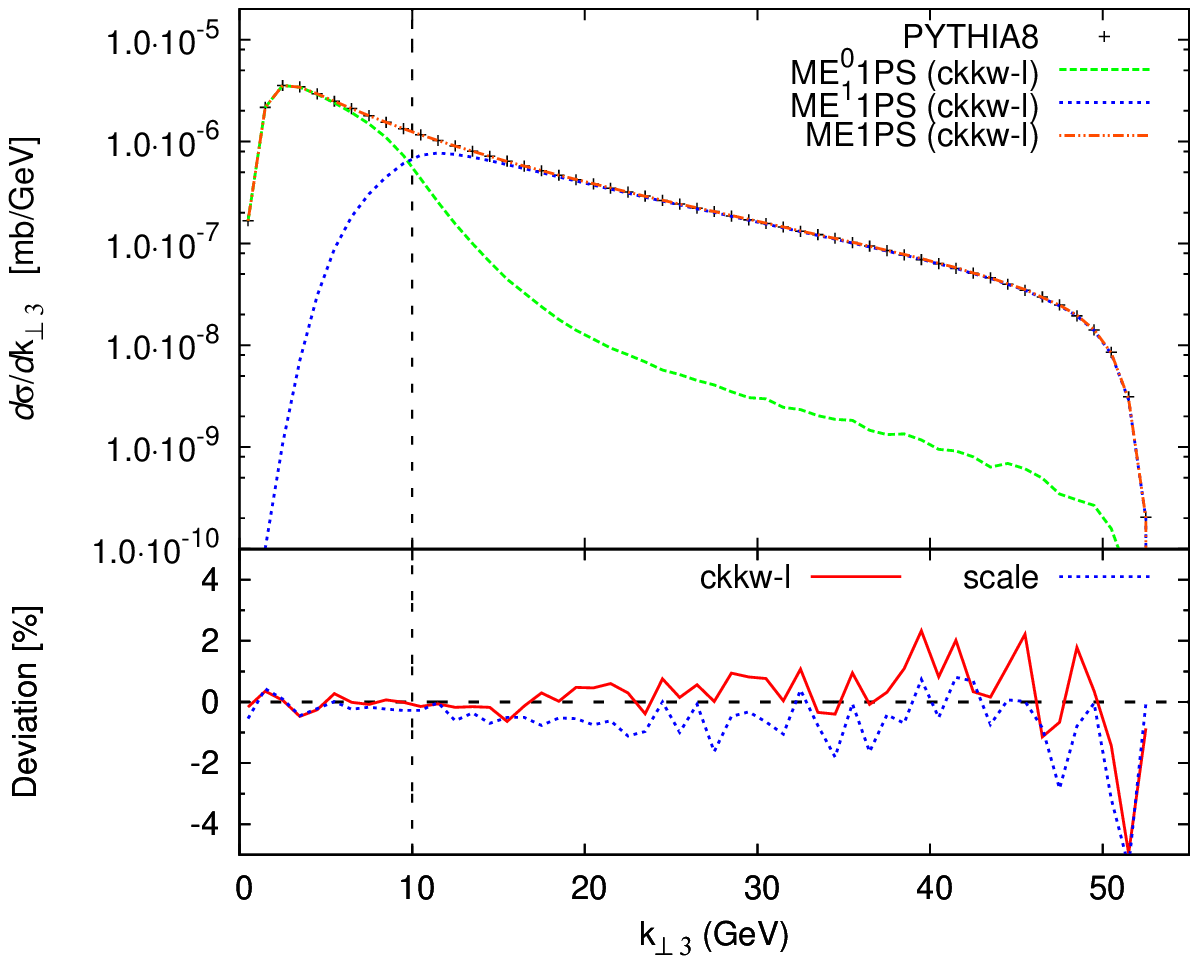}
}
\caption{\label{fig:ee:choose-by-scale-lep}A comparison of different
  prescriptions of choosing the history for $\ee\to$ 3 jets. Results
  for choosing in a probabilistic way, with splitting probabilities
  defined in eqs.\ \ref{eq:path-prob-1} and \ref{eq:path-prob-2}, are 
  labelled ``ckkw-l",
  while adopting a winner-takes-it-all strategy of picking the history
  with lowest scale carries the label ``scale". The plots were
  produced with a merging scale $\tms = \min\{k_{\perp i}\} =
  10\textnormal{ GeV}$. Hadronisation was switched off. The left panel shows the
  $k_{\perp}$-separation (in the Durham algorithm) between the third
  and second hardest parton in the first (reconstructed) emission. The
  distributions for ME$^0$1PS and ME$^1$1PS for a scale-dependent
  choice are shown in the upper part, whereas the bottom in-set gives
  the deviation of both prescriptions from default \pytppp. In the
  right panel we show the $k_{\perp}$-separation the third and second
  hardest \emph{jets} defined in the exclusive Durham algorithm for
  the probabilistic approach, with the bottom in-set again giving the
  deviation of both prescriptions from default \pytppp.}
}

\subsubsection*{Influence of the prescription on how to choose a shower history}

That different prescriptions to choose amongst reconstructed histories
differ only by sub-leading terms is exemplified in
Figure~\ref{fig:ee:choose-by-scale-lep}. We see a small merging scale
dependence when always choosing the history with the smallest sum of
transverse momenta. The smallness of the effect stems from the fact
the probabilistic choice --- on average giving the ``correct" shower
history --- is dominated by a $\frac{1}{\ord}$ factor, so that picking
a history by lowest scale $\ord$ or probabilistically almost equally
well answer the question ``\emph{how would my parton shower have
  generated this state}".

\FIGURE{
\centerline{
  \includegraphics[width=0.5\textwidth]
    {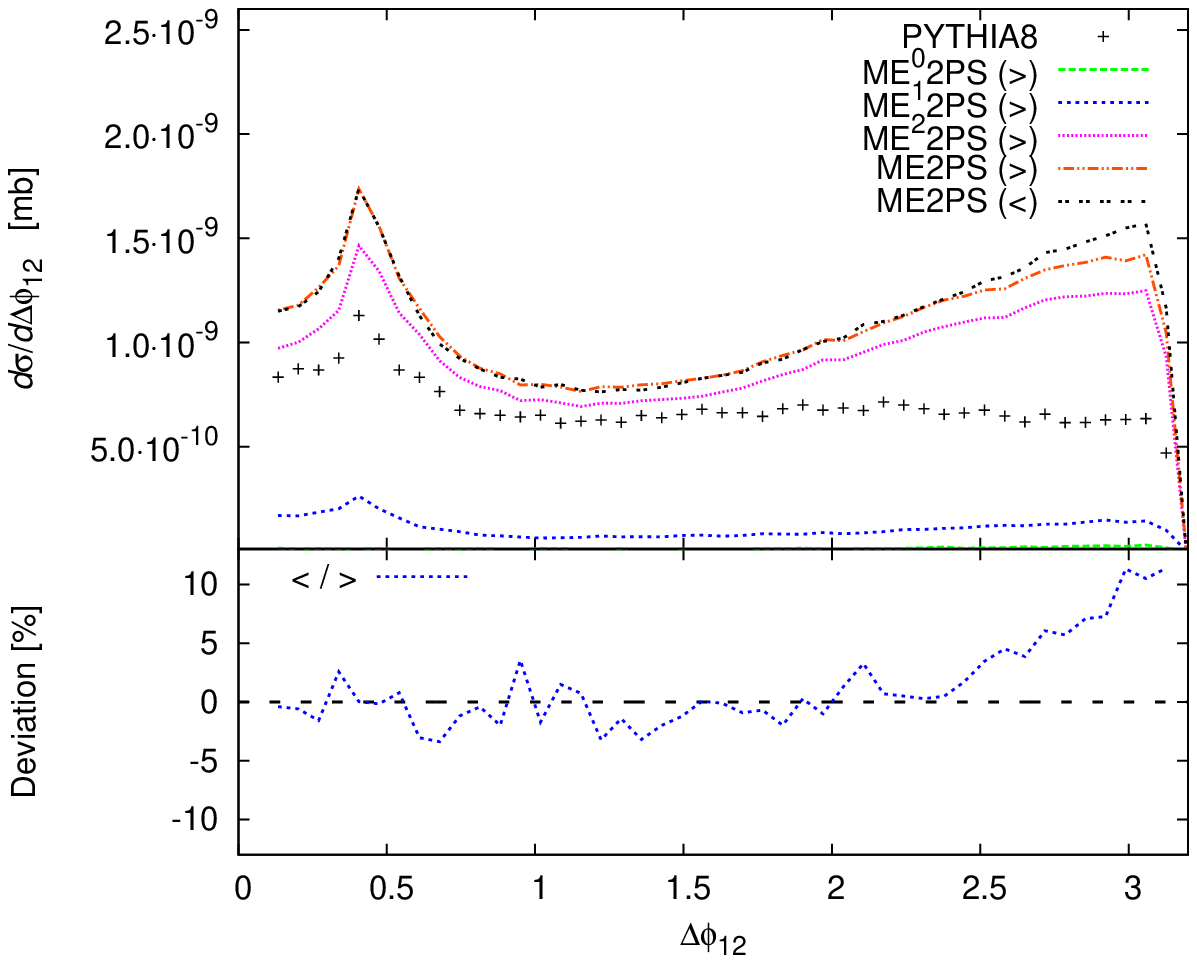}
  \includegraphics[width=0.5\textwidth]
    {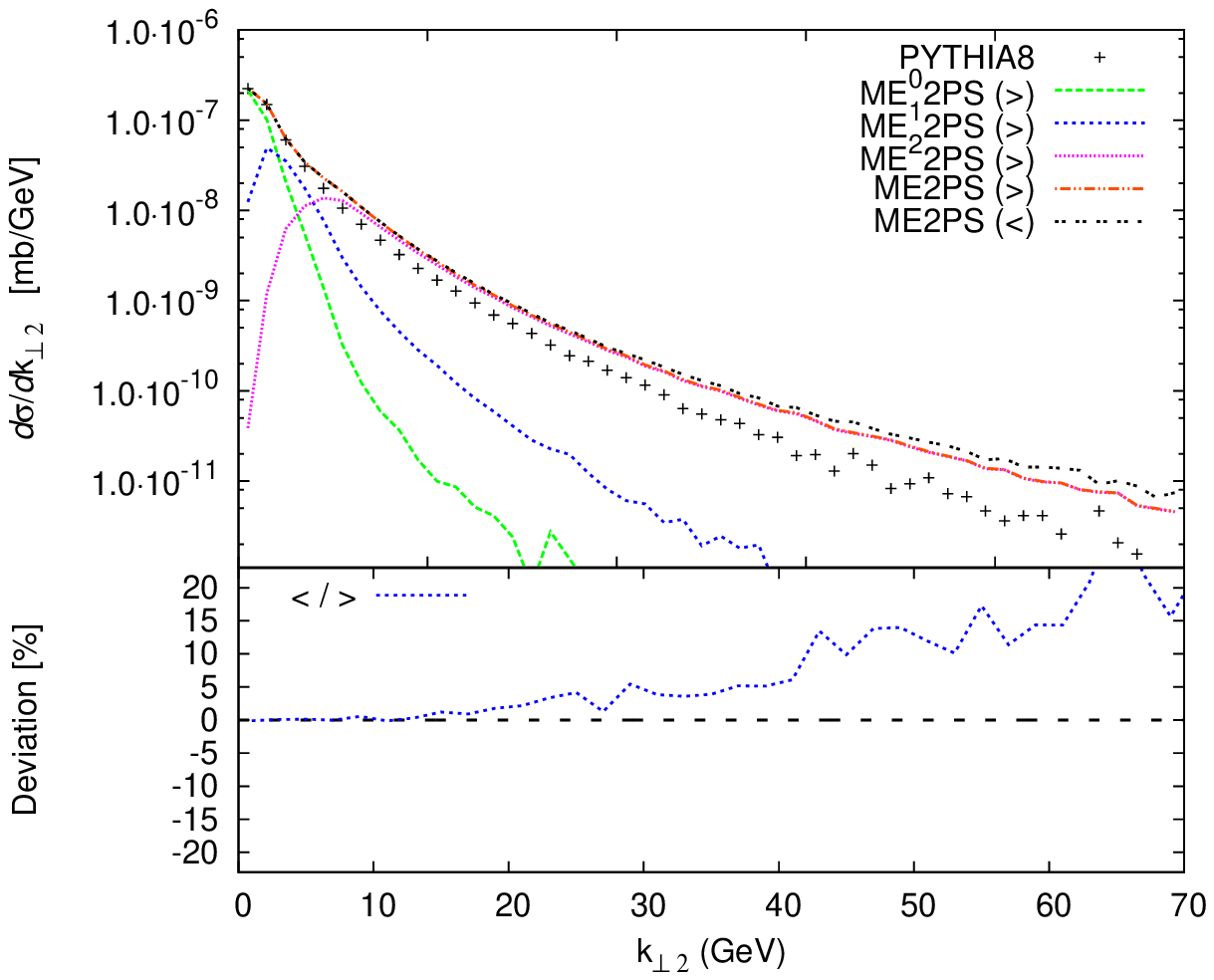}
}
\caption{\label{fig:wp:unord-scale-choice-tev}Two different ways of choosing a 
combined scale for unordered emissions, in $\W+2$ jet events at $\ECM = 1960$ 
GeV in $\p\pbar$ collisions. The merging scale is $\tms = 10\textnormal{ GeV}$. 
The curves are labelled with ``$>$" when assigning the higher scale 
$\ord_{\textnormal{combined}} = \max(\ord_i, \ord_{i+1})$, and with ``$<$" when 
assigning the lower scale 
$\ord_{\textnormal{combined}} = \min(\ord_i, \ord_{i+1})$, as the combined scale
 of two unordered emissions. The bottom in-sets show the deviation of the lower 
scale sample with respect to the higher scale sample. Jets were defined in the
 $k_\perp$-algorithm with $D = 0.4$, while multiple scatterings and 
hadronisation were turned off. The left panel shows the azimuthal difference 
$\Delta\phi_{12}$ between the hardest and second hardest jet. The right panel
shows the $k_\perp$ separation $k_{\perp 2}$ of the second hardest jet.}
}

\subsubsection*{Variation when changing the starting scales for un-ordered
histories}

In the following, we refrain from setting the infrared regularisation
parameter $\ord_{reg}$ to zero. When facing histories with unordered
emission sequences, different ways to assign an emission scale to the
combined splitting are conceivable, as discussed in section
\ref{sec:history}. To investigate this we turn to two-jet merging, the
lowest non-trivial jet multiplicity at which non-ordered histories may
occur. Figure~\ref{fig:wp:unord-scale-choice-tev} highlights that when
choosing the lower scale as a common scale, the transverse momentum of
the second jet has a harder tail compared to setting the higher of
both scales as the scale of the combined emission. Also, back-to-back
jets are more prominent. This is an effect of the reweighting with a
running coupling constant, which produces a more pronounced
enhancement of the cross section when choosing smaller scales. For all
further results, we will use the larger scale when evaluating
$\as(\ord)$.

\FIGURE{
\centerline{
  \includegraphics[width=0.75\textwidth]
    {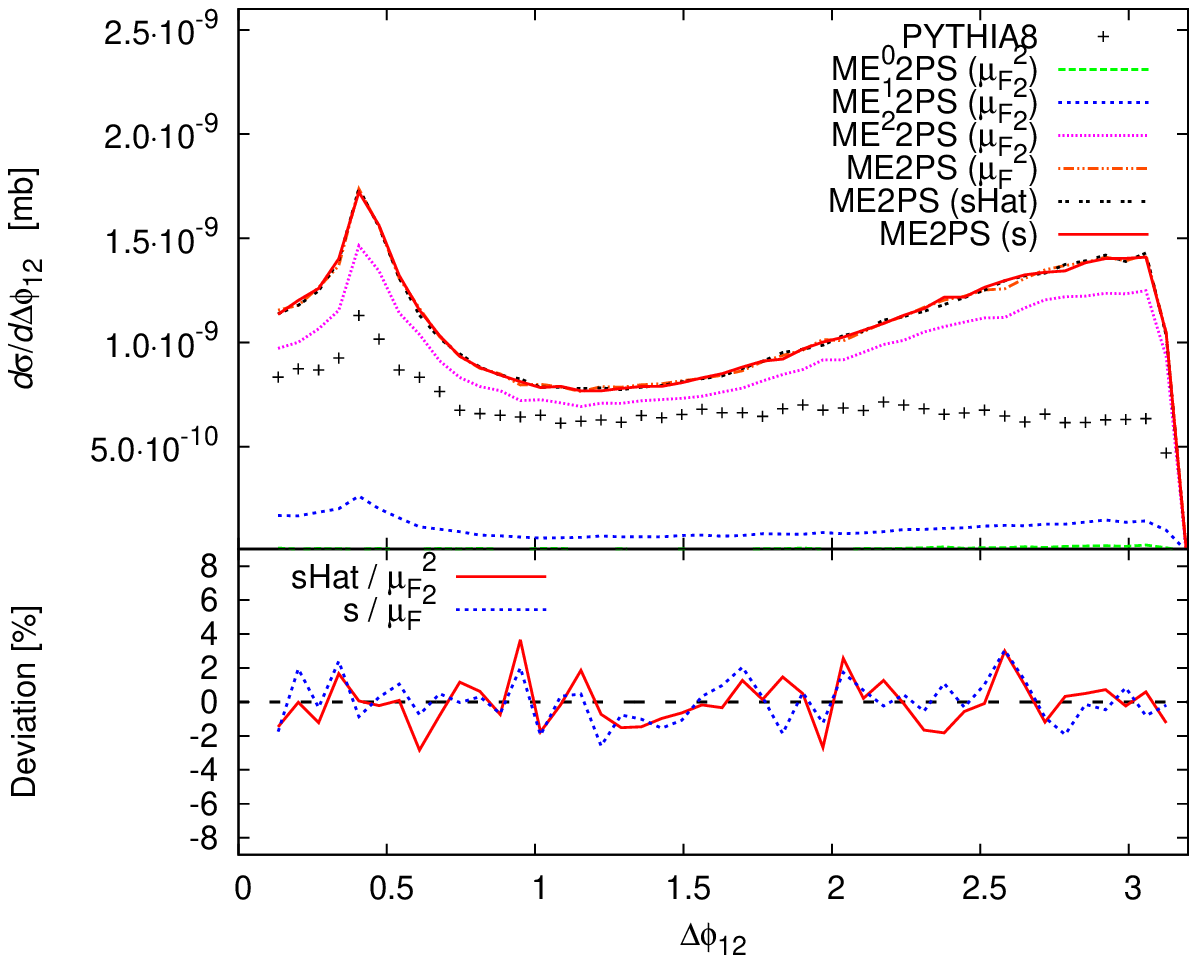}
}
\caption{\label{fig:wp:incomp-scale-choice-tev}Azimuthal difference 
$\Delta\phi_{12}$ between the hardest and second hardest jet, for three 
different ways of choosing the starting scale $\ord_0$ for incomplete histories,
 in $\W+2$ jet events at $\ECM = 1960$ GeV in $\p\pbar$ collisions. The merging 
scale is $\tms = \min\{k_{\perp i}\} =  10\textnormal{ GeV}$. The curves are 
labelled ``$\mu_F^2$" for $\ord_0 = \mu_F^2$, ``sHat" if $\ord_0 = \hat s$ and 
``s" if $\ord_0 = s$. Jets were defined by the $k_\perp$-algorithm with 
$D = 0.4$. Multiple scatterings and hadronisation were switched off. The 
bottom in-sets show the deviation of the $\ord_0 = \hat s$ and $\ord_0 = s$ 
samples with respect to the $\mu_F^2$ sample.}
}

\subsubsection*{Variation due the choice of starting scales for incomplete
histories}

Figure~\ref{fig:wp:incomp-scale-choice-tev} shows the consequence of
adopting different shower starting scales for incomplete
histories. Particularly the consistency of distributions for
$\rho_0=\mu_F^2 = (80.4 \textnormal{ GeV})^2$ and $\rho_0= s = (1960
\textnormal{ GeV})^2$ allows to conclude that the dependence on the
starting scale for incomplete emissions is negligible, which reflects
the fact that the corresponding states are very rare.

\FIGURE{
\centerline{
  \includegraphics[width=0.5\textwidth]
    {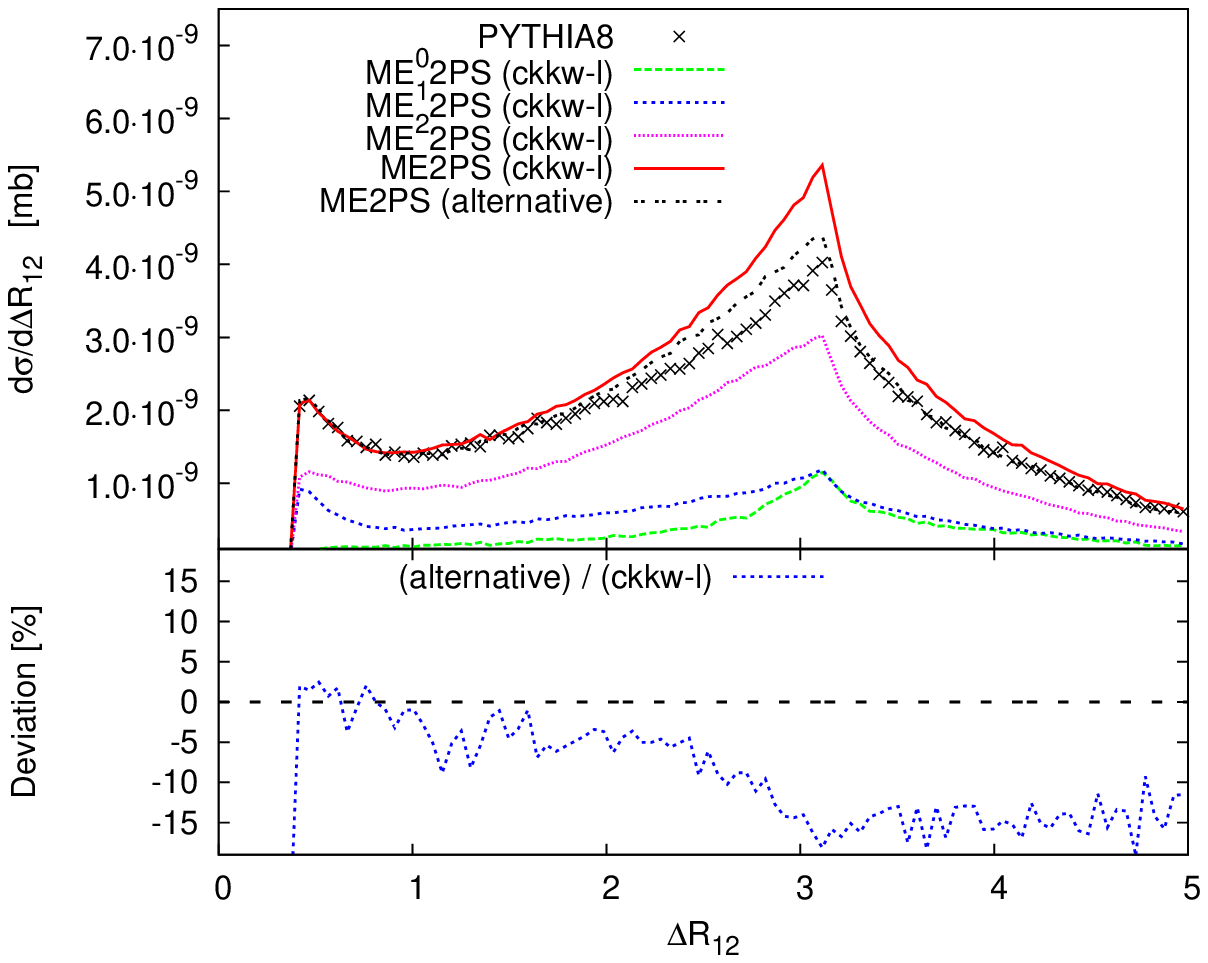}
  \includegraphics[width=0.5\textwidth]
    {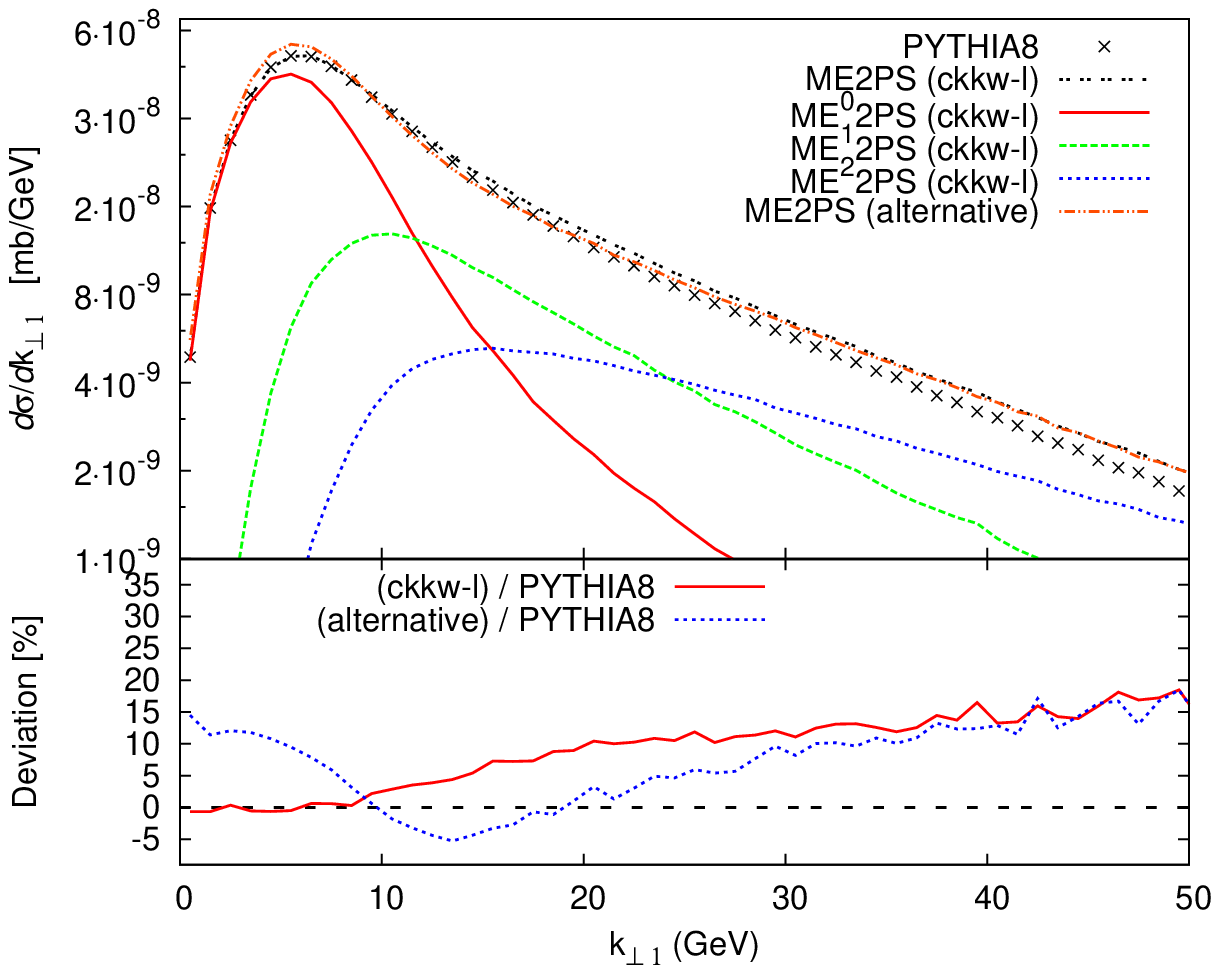}
}
\caption{\label{fig:dy:mi-treatment-lhc}Two examples for differences in the 
treatment of secondary interactions, for $\Z+2$ jet events at $\ECM = 7000$ GeV
in $\p\p$ collisions. The merging scale is 
$\tms = \min\{k_{\perp i}\} =  10\textnormal{ GeV}$. Jets were defined with the
$k_\perp$-algorithm with $D = 0.4$. The left panel shows the $R$ separation 
$\Delta R_{12}$ between the hardest and second hardest jet, with the bottom 
in-set giving the difference of the \sherpa-inspired sample with respect to the 
\ckkwl treatment. The right panel shows the jet $k_\perp$ after hadronisation, 
when clustering to exactly one jet. Ratios of the two MI treatments to default 
\pytppp are found in the bottom right in-set.}
}

\subsubsection*{Differences between treatments of multiple interactions}

Different treatments of multiple interactions are presented in
Figure~\ref{fig:dy:mi-treatment-lhc}, which illustrates that at the
LHC, variations of up to $10\%$ may occur between the default \ckkwl
recipe and the \sherpa-inspired alternative. Due to the phase space
restriction $\rho_0=\tms$ for additional scatterings in $\Z+0$ jet
matrix element samples, the alternative treatment produces fewer
multiple interactions. Thus, the $k_{\perp 1}$ spectrum for
intermediate scales $15 \textnormal{ GeV}<k_{\perp 1}< 30 \textnormal{
  GeV}$ is softer than the \ckkwl result. At scales $ k_{\perp 1} > 50
\textnormal{ GeV}$, the two prescriptions become
indistinguishable. The behaviour at low scales is also anticipated,
since the alternative sample does not include suppression due to MI
no-emission probabilities. Since these are present in default \pytppp,
the alternative recipe exhibits a higher maximum, whereas the default
prescription reproduces the showers low scale features closely.

Our goal when developing a generalisation of the \ckkwl method including
interleaved showers was to be as similar for low scales to the event generator
as possible, meaning that the modelling of \pytppp in regions where multiple 
interactions are important should be left unchanged. As pointed out in 
\ref{sec:interl-mult-inter}, this can be formally be achieved in \pytppp by
employing the ``\ckkwl\!\!" prescription. The discussion of the last paragraph 
also showed that in the implementation of the method, low scale features of 
\pytppp are retained. Hence, we choose the ``\ckkwl\!\!" prescription of adding 
the influence of multiple scatterings as the default. As can be inferred from 
Figure~\ref{fig:z-ue} below, this method succeeds in not changing the 
underlying event description of \pytppp.

Because in weak boson measurements at low scales, the
shape and position of maxima is unchanged in the \ckkwl\ approach, we also
minimise the need for changes of some tuning parameters, \eg\ primordial
$p_{\perp}$. This is not obviously true for the alternative method, in which
some changes in primordial $p_{\perp}$ might be necessary. Meanwhile, once 
hadronisation is added and experimental cuts are applied, $\Z+$ jets 
observables at the Tevatron show only little dependence on the strategy how 
multiple interactions are included in merged samples.

\TABLE{
\renewcommand{\tabcolsep}{1.7mm}
\begin{tabular}{|l|c|c|c|c|c|c|}
\hline
\small Process & \small $\tms$ & \small $2\to2$ & \small ME1PS & \small ME2PS & \small ME3PS & \small ME4PS\\\hline
\small $\ee\to$ jets & \small 5  GeV &                      & \small $32.92(2)$ nb & \small $32.50(2)$ nb & --- & --- \\
                     & \small 10 GeV & \small $32.91(3)$ nb & \small $32.93(2)$ nb & \small $32.81(2)$ nb & \small 32.79(2) nb & \small 32.87(3) nb \\
                     & \small 15 GeV &                      & \small $32.90(3)$ nb & \small $32.88(3)$ nb & \small 32.87(3) nb & \small 32.87(3) nb \\\hline
\small $\p\pbar\to\Z^0$+jets & \small 10 GeV &            & \small $194.9(5)$ pb & \small $199.7(5)$ pb & \small $200.3(5)$ pb & --- \\
                   & \small 15 GeV & \small $194.0(1)$ pb & \small $194.5(6)$ pb & \small $196.8(6)$ pb & \small $197.2(6)$ pb & --- \\
                   & \small 30 GeV &                      & \small $194.0(6)$ pb & \small $194.7(6)$ pb & \small $194.6(6)$ pb & --- \\
                   & \small 45 GeV &                      & \small $193.9(6)$ pb & \small $194.3(6)$ pb & \small $194.3(6)$ pb & --- \\\hline
\small $\p\pbar\to\W^+$+jets & \small 10 GeV &           & \small $1038(3)$ pb & \small $1066(3)$ pb & \small $1074(3)$ pb & \small $1076(3)$ pb \\
                   & \small 15 GeV & \small $1034(1)$ pb & \small $1034(3)$ pb & \small $1048(3)$ pb & \small $1051(3)$ pb & \small $1053(3)$ pb \\
                   & \small 30 GeV &                     & \small $1034(3)$ pb & \small $1039(3)$ pb & \small $1038(3)$ pb & \small $1039(3)$ pb \\
                   & \small 45 GeV &                     & \small $1034(3)$ pb & \small $1036(3)$ pb & \small $1036(3)$ pb & \small $1036(3)$ pb \\\hline
\end{tabular}
\caption{\label{tab:unitarity}Impact of changing the merging scale $\tms$ and 
maximum number of jets on the process cross sections, for three different
processes. $\ee\to$ jets is evaluated LEP energy ($\ECM = 91.25$), and 
cross sections for
$\p\pbar\to\Z$+jets and $\p\pbar\to\W^+$+jets are calculated at Tevatron 
Run II energy ($\ECM = 1960$). Results were produced with \pytppp Tune 4C. 
Multiple interactions and hadronisation were switched off.
}
}

\TABLE{
\renewcommand{\tabcolsep}{1.7mm}
\begin{tabular}{|l|c|c|c|c|c|c|}
\hline
\small Process & \small $\tms$ & \small $2\to2$ & \small ME1PS & \small ME2PS & \small ME3PS & \small ME4PS\\\hline
\small $\p\pbar\to\W^+$+jets & \small 10 GeV &           & \small $1037(3)$ pb & \small $1048(3)$ pb & \small $1047(3)$ pb & \small $1045(3)$ pb \\
                   & \small 15 GeV & \small $1034(1)$ pb & \small $1034(3)$ pb & \small $1043(3)$ pb & \small $1044(3)$ pb & \small $1043(3)$ pb \\
                   & \small 30 GeV &                     & \small $1034(3)$ pb & \small $1038(3)$ pb & \small $1038(3)$ pb & \small $1038(3)$ pb \\
                   & \small 45 GeV &                     & \small $1034(3)$ pb & \small $1036(3)$ pb & \small $1036(3)$ pb & \small $1036(3)$ pb \\\hline
\end{tabular}
\caption{\label{tab:unitarity-no-rap}Impact of changing the merging scale 
$\tms$ and 
maximum number of jets on the $\W +$ jets cross sections in $\p\pbar$ collisions
at $\ECM = 1960$. Multiple interactions and hadronisation were switched off. 
Results were produced using Tune 4C, with enforced rapidity ordering switched 
off.
}
}

\subsubsection*{Unitarity violations}

We finish our validation by discussing a theoretical issue.
Parton shower resummation alone does not change the cross section of the 
hard process, since the probability of having no emission, together with the
sum of probabilities to evolve into states with an arbitrary number of
emissions adds to unity --- a property dubbed
unitarity. This however is only true if the transition probabilities used in 
generating additional emissions are identical to the terms exponentiated in
Sudakov form factors. As pointed out in \cite{Hamilton:2010wh,Hoche:2010kg}, 
unitarity is violated by tree level merging due to the fact that the transition
probabilities above and below $\tms$ are different, while Sudakov factors are 
always generated with shower splitting kernels, \ie\ the transition 
probabilities below $\tms$. The magnitude of the resulting
unitarity violations for different merging scales is assessed for $\W+$ jets in 
Table~\ref{tab:unitarity}. We have also verified that the main points of the 
following discussion apply to all example processes used in this report. 

\FIGURE{
\centerline{
  \includegraphics[width=0.5\textwidth]
    {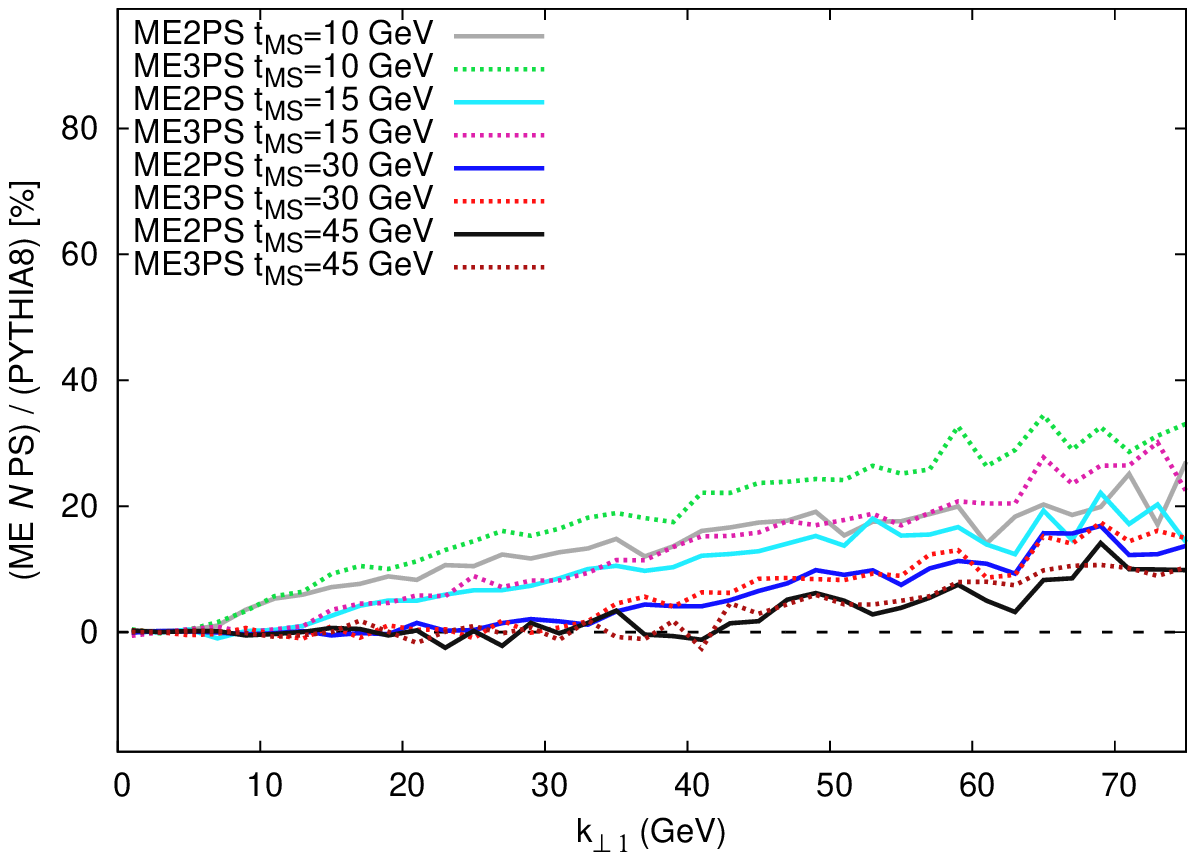}
  \includegraphics[width=0.5\textwidth]
    {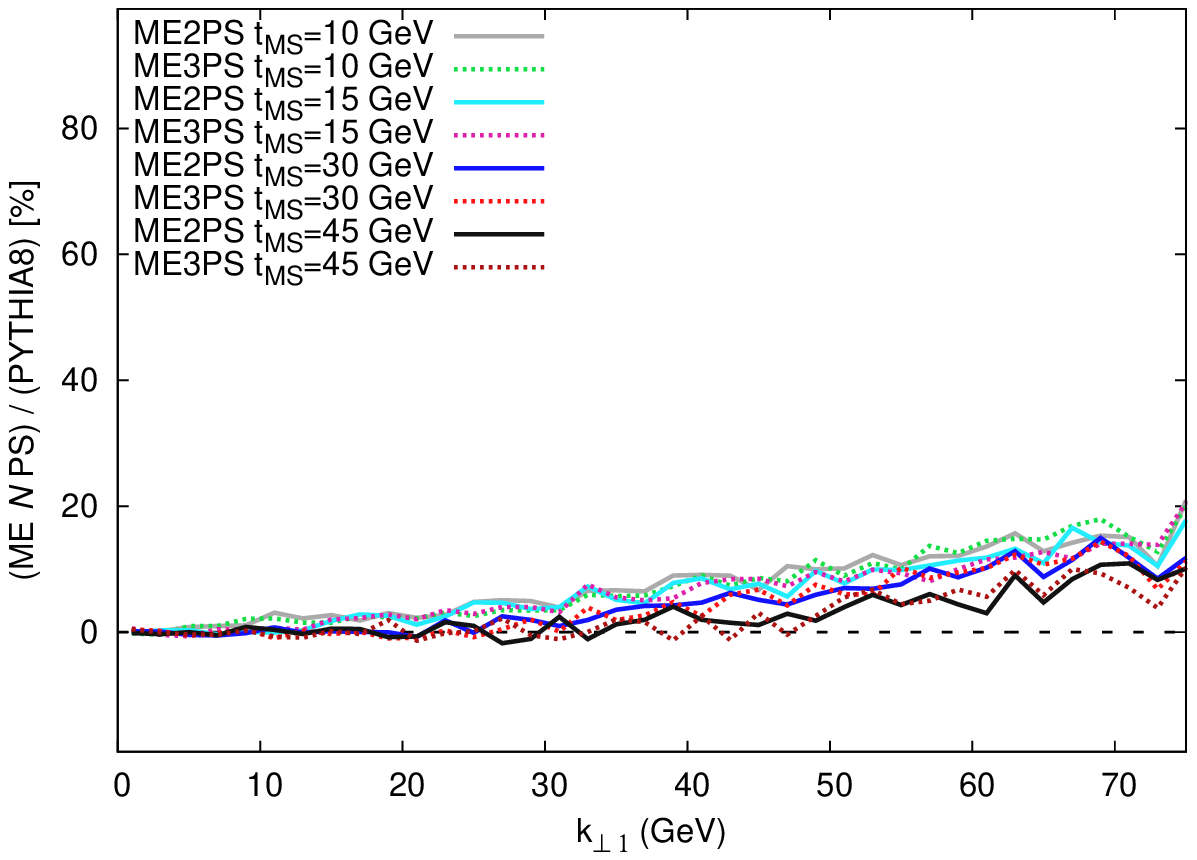}
}
\caption{\label{fig:wp:pt-jets-diff-njet-unitarity}Comparison of the $k_\perp$ 
of the first jet in $\W+$ jets events at 
$\ECM = 1960$ GeV in $\p\pbar$ collisions, between default, first-order 
corrected \pytppp, and \ckkwl, for different number of merged jets and 
different merging scales. Jets were defined with the $k_\perp$-algorithm with 
$D = 0.4$, clustering to exactly one jet. Multiple interactions and 
hadronisation were switched off. Left panel: Results when using Tune 4C, which 
by default includes ordering emissions in rapidity as well as $\ord$. 
Right panel: Results when using Tune 4C, with enforced rapidity ordering 
switched off.}
}

First, we note that including one additional jet does
not lead to unitarity violations for vector boson 
production, since \pytppp is already matrix element corrected, so that the full
tree-level splitting probability is exponentiated. When including more than one
jet, we observe smaller deviations from the hard process cross section as we 
increase the merging scale. This is expected since for larger $\tms$, the 
Sudakov form factors generated by trial showering quickly approach unity. 
Because of the higher merging scale, phase space regions with low scale 
emissions (where Sudakov factors differ from unity) are generated by the parton 
shower. Thus, identical splitting probabilities are used to generate the 
emissions and Sudakov form factors, and unitarity is preserved to reasonable 
accuracy. One immediate consequence is that we should not choose $\tms$ too 
low, since otherwise, sizable violations can occur.

Unitarity violations give a measure of how well the shower splitting
probability, integrated over the PS phase space (ordered in the
evolution variable), captures the matrix element features and the
allowed phase space.  Different choices of PS evolution variables can
lead to different regions of the full phase space --- which includes
unordered emissions --- being neglected in the parton shower
approximation. These regions of unordered emission sequences are
formally beyond the accuracy of the
shower. Figure~\ref{fig:wp:pt-jets-diff-njet-unitarity} shows the
differences in transverse momentum distributions between merged
distributions and \pytppp, for two different ways of ordering
emissions.  Deviations from unitarity are more significant if the
shower evolution is ordered both in $\ord$ and rapidity. This is due
to neglecting larger regions of the full phase space in the parton
shower. We have verified that when only keeping ME configurations for
which a history ordered in $\ord$ and rapidity can be constructed,
unitarity violations are greatly reduced. Nonetheless, we have to
conclude that ordering the cascade in these two variables makes the
parton shower approximation worse than ordering in $\ord$ alone. Only
ordering in $\ord$, we find in Table~\ref{tab:unitarity-no-rap} that
the inclusive $2\to2$ cross section is not changed drastically when
including additional jets. Also, the $k_{\perp 1}$ spectrum becomes only
slightly harder in this case, as is seen in the right panel of
Figure~\ref{fig:wp:pt-jets-diff-njet-unitarity}.

It should be noted that the rapidity ordering was introduced in
\pytppp to suppress the high transverse momentum emissions from
dipoles between incoming and outgoing partons. However, this is
now achieved through another damping mechanism described in
\cite{Corke:2010yf}, which means that the rapidity ordering is no
longer needed to achieve a reasonable description of data.

We checked that deviations from unitarity can be
even further reduced when excluding unordered emissions. However, in
\ckkwl, we want to include states which are out of the reach of the
shower, and thus, as discussed in section \ref{sec:history}, by
default keep ME configurations for which only unordered histories can
be found. For enthusiasts, switches for rejecting configurations with
unordered histories are available in the public code.  Provided
considerable unitarity violations remain after excluding differences
between the full allowed and the PS phase space, this could suggest
large higher-order corrections, since by choosing a low merging scale
we effectively include major parts of the real emission phase space of
an NLO calculation \cite{Hoeche:2009rj}. It should be noted that
unitarity is a parton shower concept and need not be fulfilled in
other contexts, see \eg\ \cite{Andersen:2011hs}.

\subsection{\ee\ four-jet observables}
\label{sec:ee-four-jet}

\FIGURE{
\centerline{
  \includegraphics[width=0.5\textwidth]
    {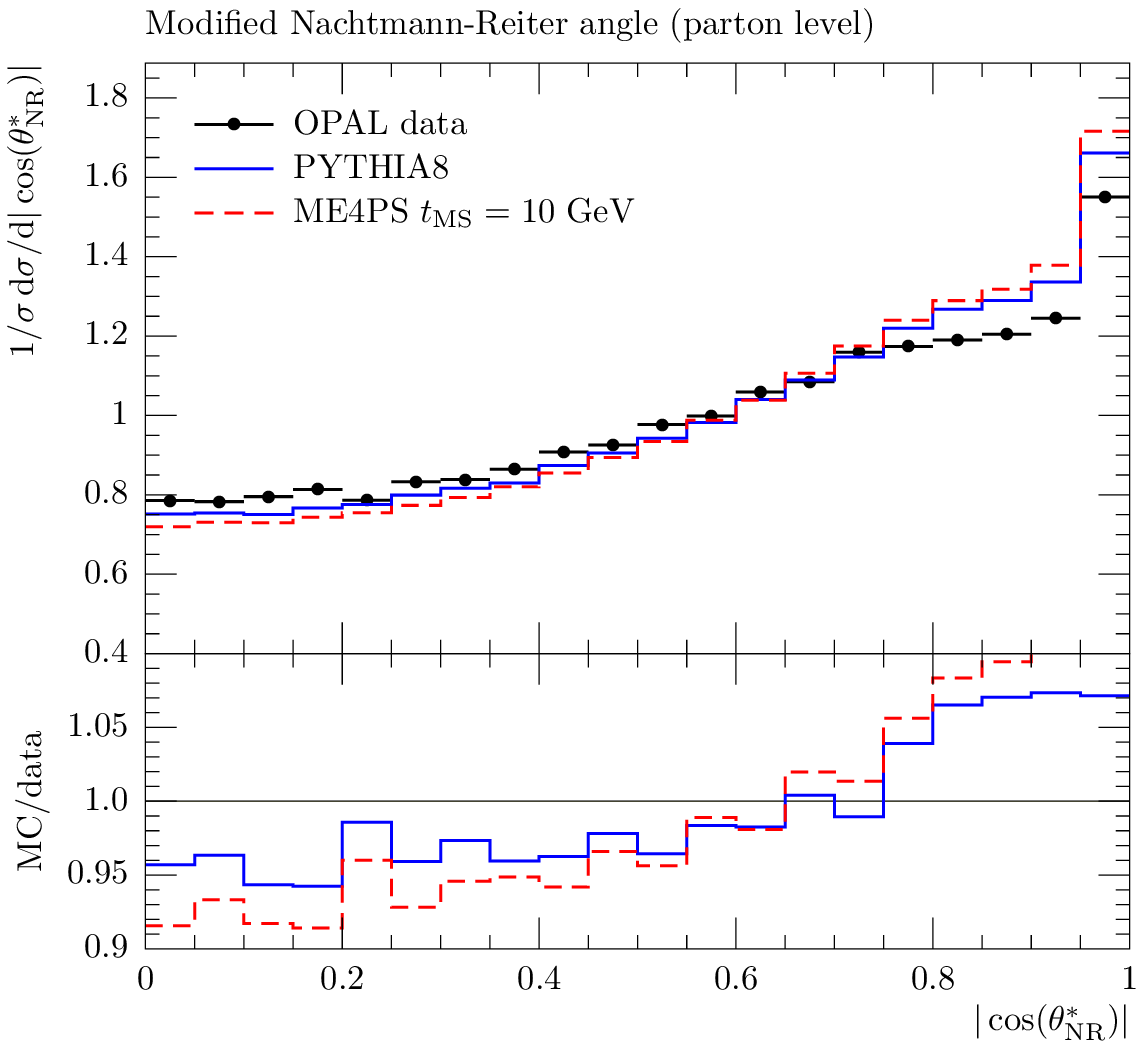}
  \includegraphics[width=0.5\textwidth]
    {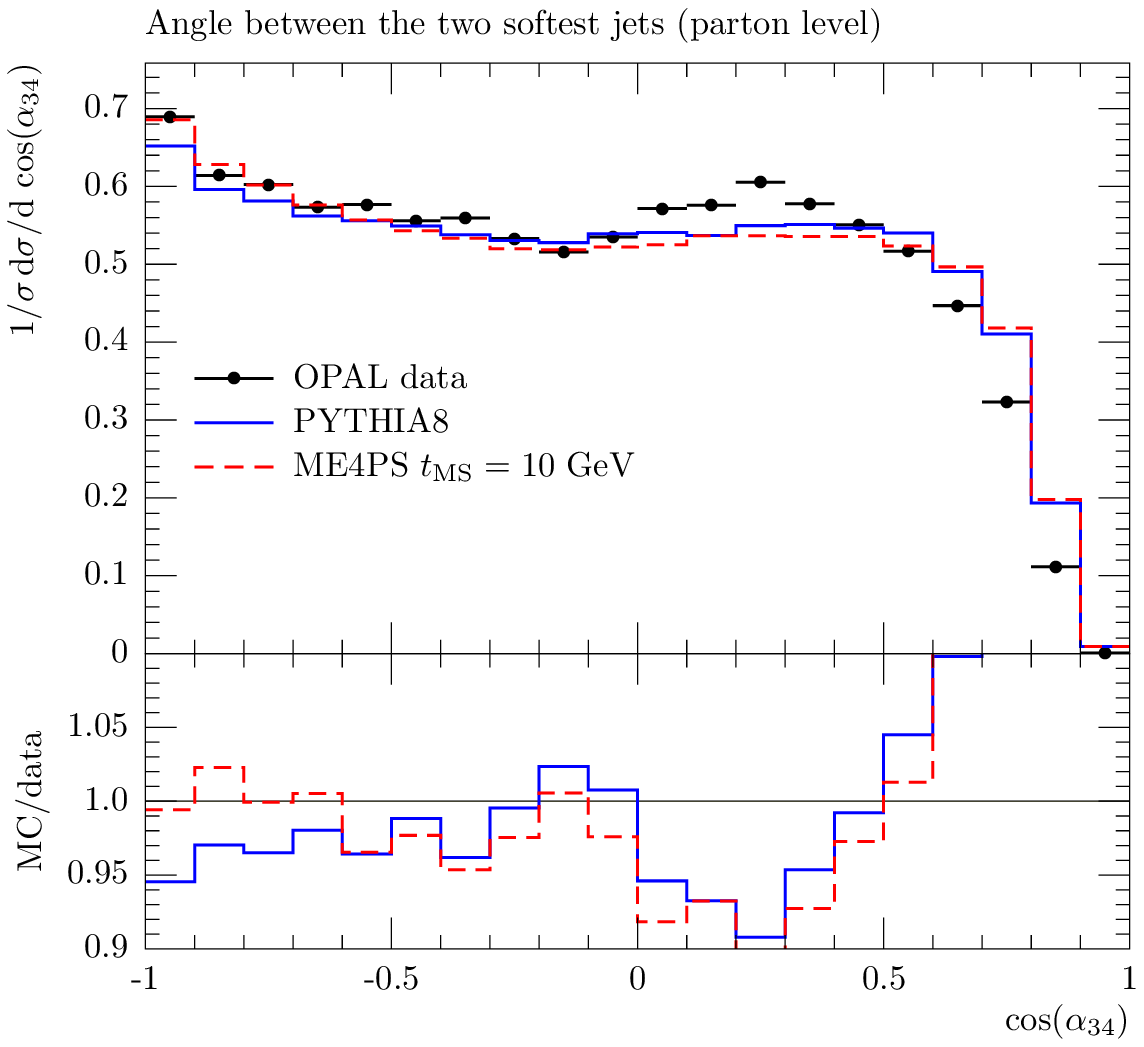}
}
\caption{\label{fig:ee:4j-angles}Four jet angular correlations in $\ee$ 
collisions at $\ECM = 91.25$ GeV, as measured by OPAL \cite{Abbiendi:2001qn}. 
Up to four additional jets were included in the merged samples. Effects of 
hadronisation are included. Left panel: Modified Nachtmann-Reiter angle 
$\left|\cos \theta_{NR}\right|$. Right panel: Angle between the two lowest 
energy jets $\cos \alpha_{34}$. The plots were produced
  with \rivet \cite{Buckley:2010ar}.}
}

Merging procedures aim for a better description of well separated 
jets in the parton shower. Historically, angular correlations in 
$\ee\to 4$ jet production have been used to investigate the 3-gluon vertex. 
The description of these observables should be improved when 
including additional jets. More specifically, we look at the 
the (modified) Nachtmann-Reiter angle
\begin{equation}
  \left|\cos \theta_{NR}\right| =
  \left|\frac{\left(\vec p_1 - \vec p_2 \right)\cdot
    \left(\vec p_3 - \vec p_3 \right)}
 {\left| \vec p_1 - \vec p_2 \right|\left| \vec p_3 - \vec p_4 \right|}\right|~,
\end{equation}
and the angle between the two lowest energy jets
\begin{equation}
  \cos \alpha_{34} =
  \frac{\vec p_3\cdot\vec p_4}{\left| \vec p_3\right|\left|\vec p_4 \right|},
\end{equation}
where $\vec p_i$ are the energy ordered three-vectors of the jets. As
shown in Figure~\ref{fig:ee:4j-angles}, the default \pytppp
description of these observables is fairly good to start with,
reflecting the fact that some azimuthal correlations are included in
the shower, and it is only slightly changed when merging additional
jets. We notice that $\left|\cos \theta_{NR}\right|$ becomes slightly
worse when including additional jets. The different shape of the
generator curves can be explained by the fact that the data was
corrected to the parton level, whereas the MC samples where generated
with full hadronisation.  In $\left|\cos \theta_{NR}\right|$, the
hadronisation corrections \cite{Abbiendi:2001qn} would change the MC
shapes towards a better agreement.  $\cos \alpha_{34}$ is captured
slightly better for $\cos \alpha_{34} \approx -1$, when including
additional jets. The trend to overshoot at $\cos \alpha_{34} \geq 0.5$
can again be explained by the fact that we have generated the
distributions at the hadron level, whereas the data was corrected to
the parton level. We have checked by excluding hadronisation that
these statements are true, and that the irregularities are
reduced. However, the general trends in both $\left|\cos
  \theta_{NR}\right|$ and $\cos \alpha_{34}$ remain, albeit less
pronounced. This might be explained with the fact that the
hadronisation corrections applied to the data are estimated with a
model different from the one used by \pytppp. Since the cross-over
from partonic to hadronic states is a highly model-dependent
statements, artifacts of the model used to estimate corrections could
be present in the data. Even so, we think
Figure~\ref{fig:ee:4j-angles} illustrates that when including
higher-order tree-level matrix elements in the description of $\ee\to$
jets, changes as compared to the default shower are fairly modest,
which indicates that \pytppp already nicely describes observables at
LEP. When checking further LEP observables, we find that \ckkwl does as good
or moderately better than default \pytppp. This means that when
developing a new tune including additional matrix elements, the
hadronisation parameters, which are predominantly constrained at LEP,
may not have to be touched.

\subsection{Vector boson production}
\label{sec:vect-boson-prod}

\FIGURE{
\centerline{
  \includegraphics[width=0.5\textwidth]
    {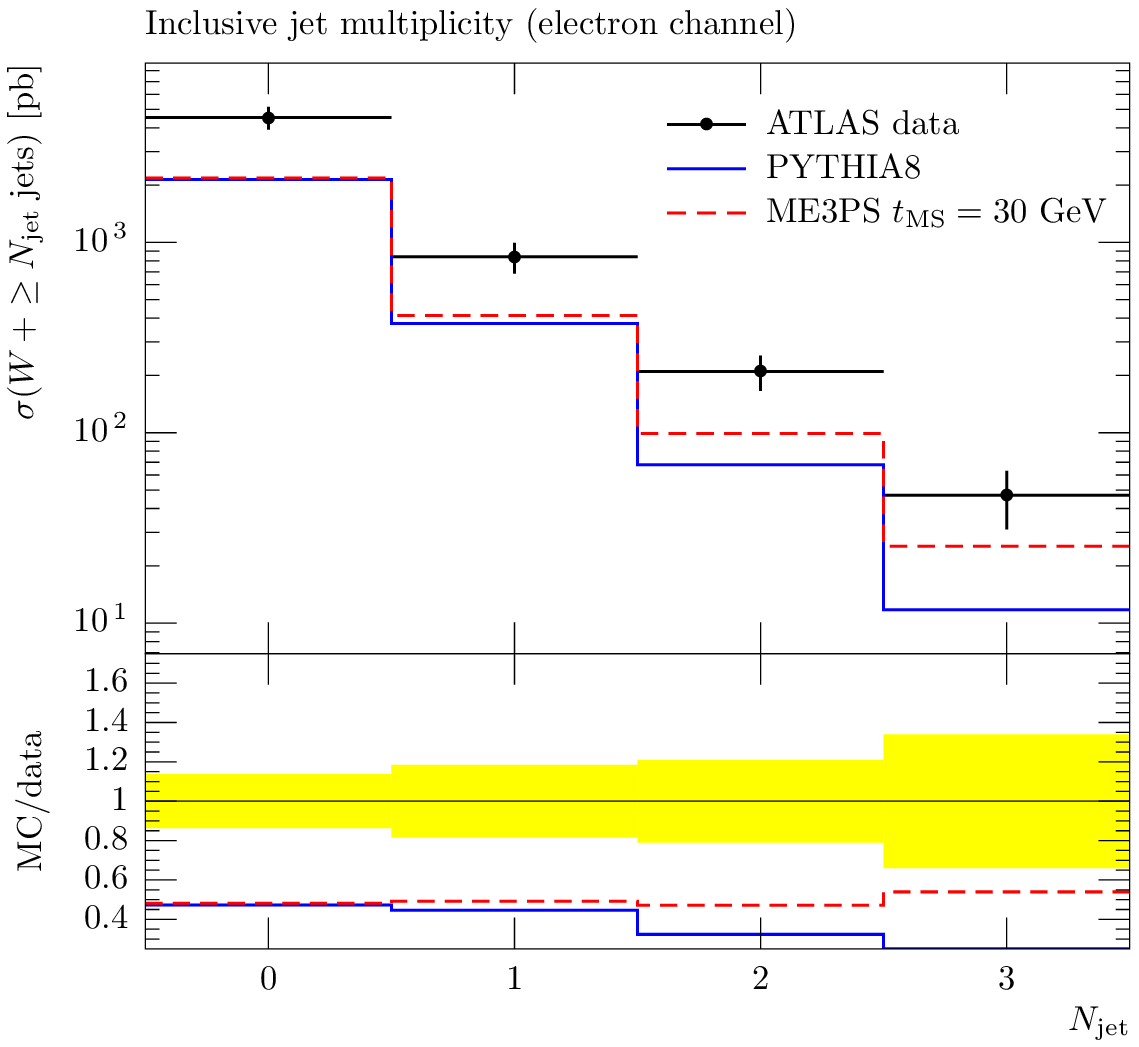}
  \includegraphics[width=0.5\textwidth]
    {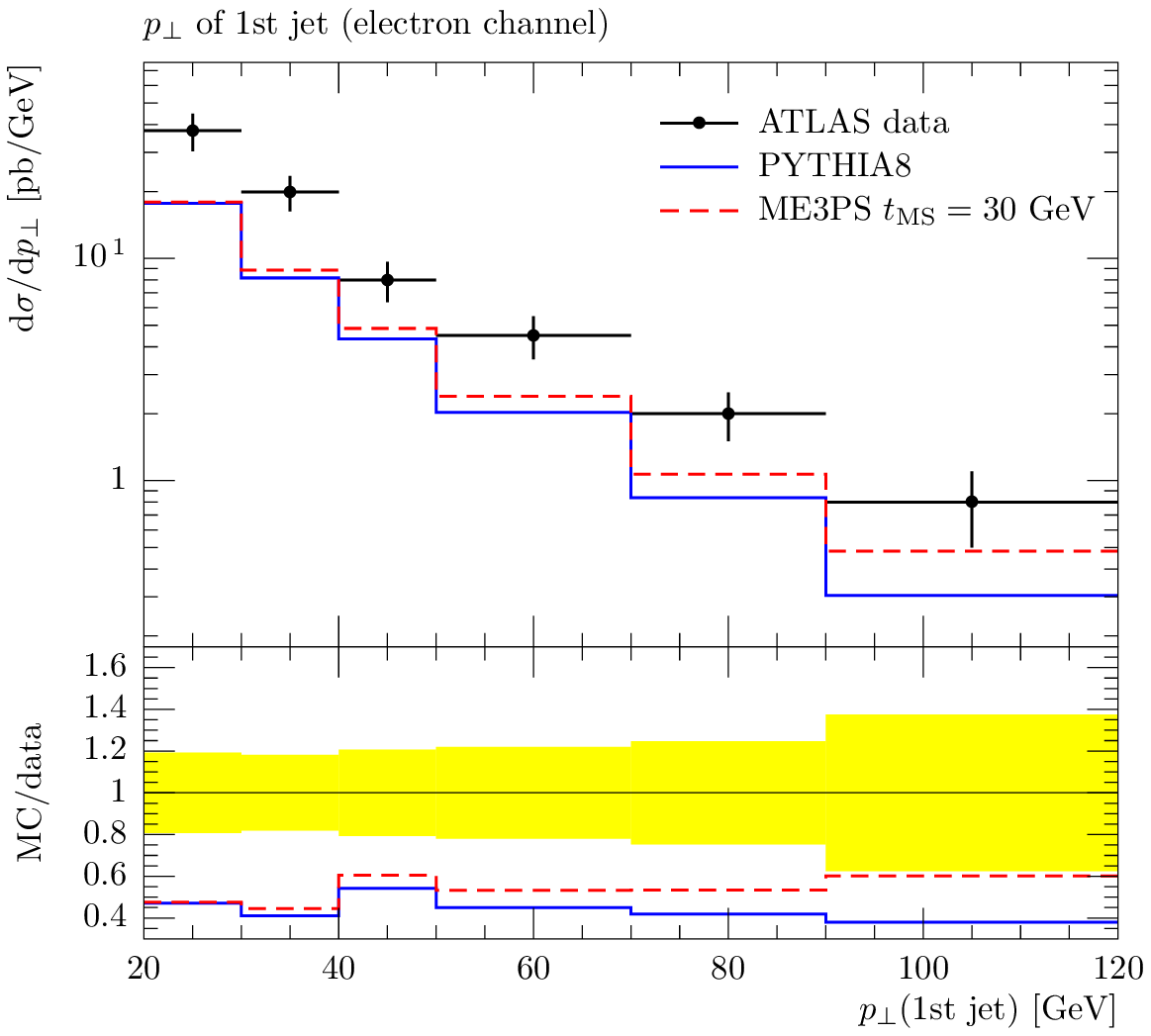}
}
\caption{\label{fig:w-pt}Jet multiplicity and transverse momentum of
  the hardest jet in $\W ~+$ jet events as measured in the electron
  channel by ATLAS \cite{ATLAS:2010pg}. The merging scale is $\tms =
  \min\{k_{\perp i}\} = 15\textnormal{ GeV}$. Effects of multiple
  scatterings and hadronisation are included. The plots were produced
  with \rivet \cite{Buckley:2010ar}.}
}

\FIGURE{
\centerline{
  \includegraphics[width=0.5\textwidth]
    {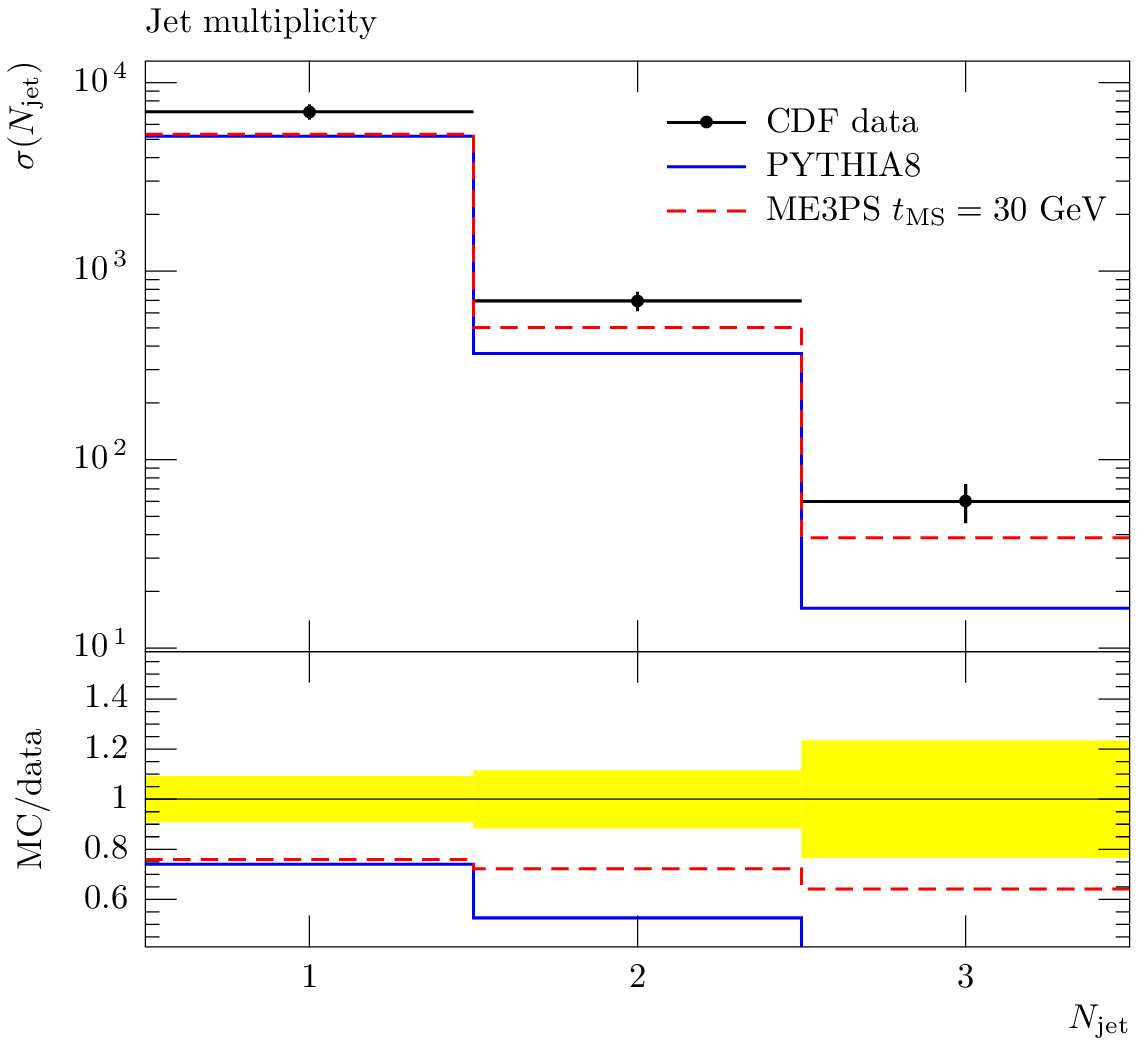}
  \includegraphics[width=0.5\textwidth]
    {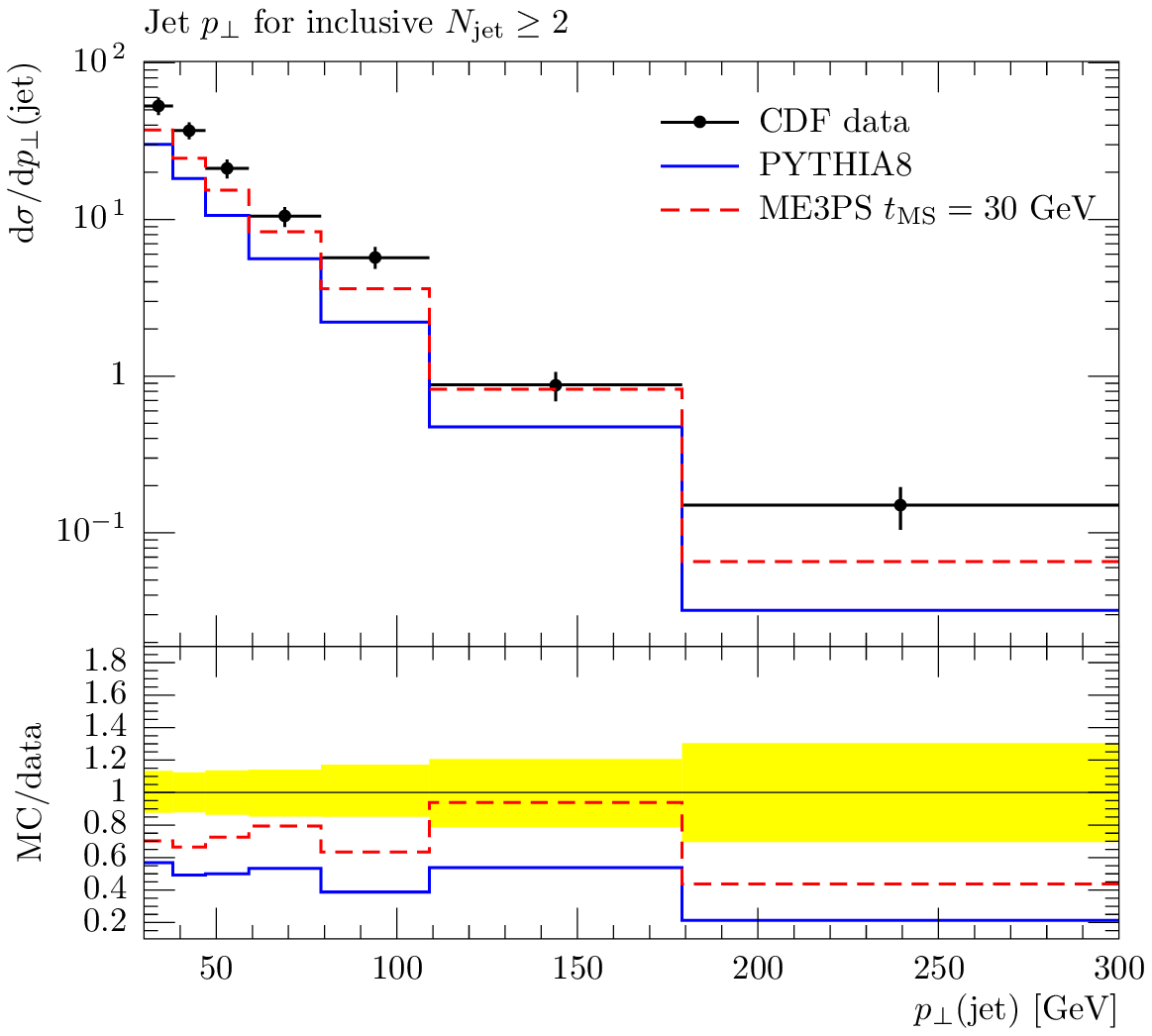}
}
\caption{\label{fig:z-pt}Jet multiplicity and inclusive jet
  transverse momentum in Drell-Yan events, as measured by CDF
  \cite{:2007cp}. The merging scale is $\tms = 30\textnormal{ GeV}$. 
  Effects of multiple scatterings and
  hadronisation are included, and Tune 4C was chosen. The plots were 
  produced with \rivet \cite{Buckley:2010ar}. }
}

In hadron collisions, we can assess the extent of change when
including additional jets by looking at vector boson production with two or more
additional jets. In Figures~\ref{fig:w-pt} and \ref{fig:z-pt}, we compare jet 
$k_\perp$ spectra and jet multiplicities for $\W$ production and in Drell-Yan 
events to data, respectively. In general, we find more jets with high $k_\perp$ 
and better agreement with jet multiplicity data.

It is particularly instructive to investigate the change of $k_\perp$
distributions when increasing the numbers of jets in the matrix
element generation. Figure~\ref{fig:wp:pt-jets-diff-njet} again shows that
the $k_\perp$ spectra develop harder tails when including higher multiplicity 
matrix element configurations. 

\FIGURE{
\centerline{
  \includegraphics[width=0.5\textwidth]
    {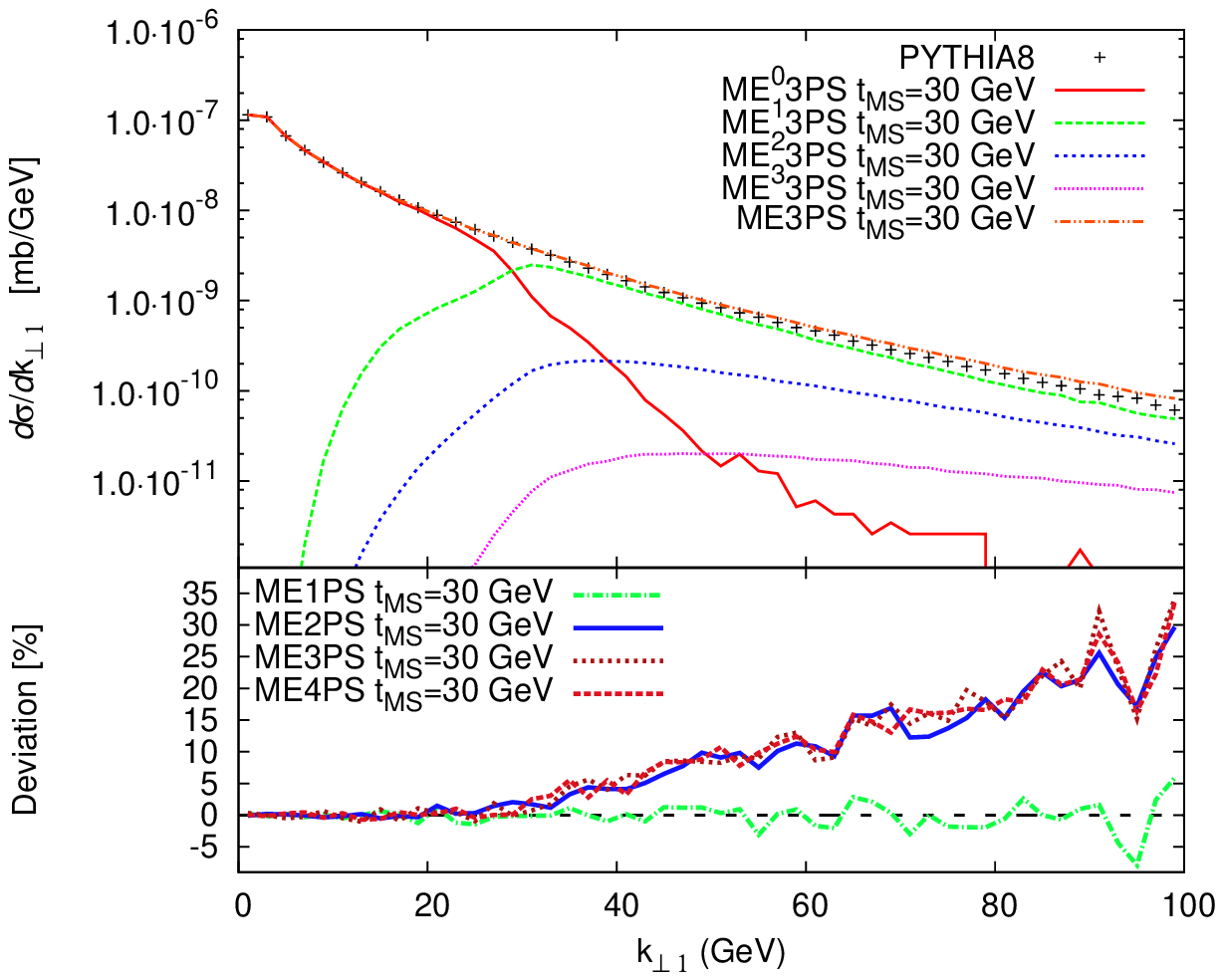}
  \includegraphics[width=0.5\textwidth]
    {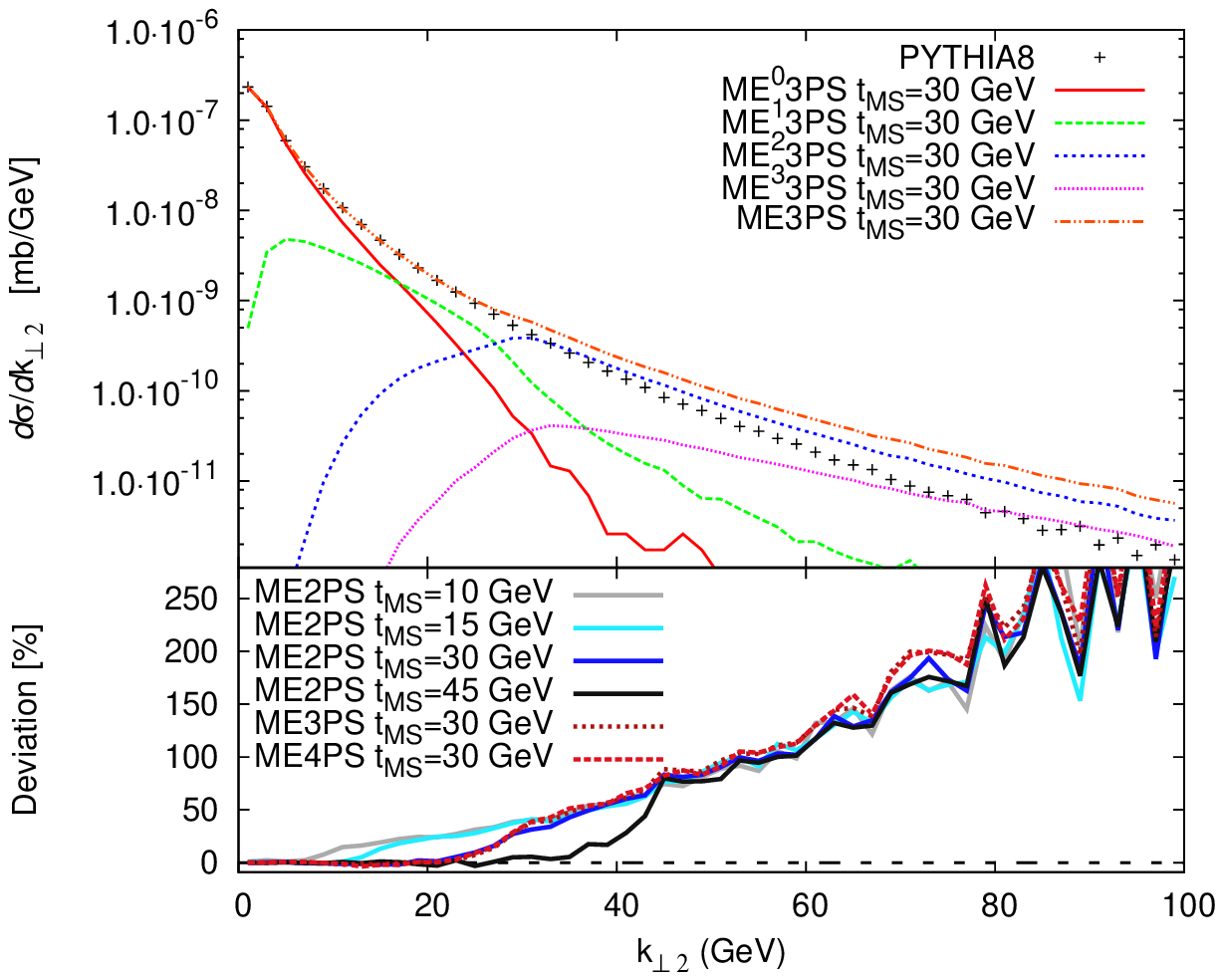}
}
\caption{\label{fig:wp:pt-jets-diff-njet}$k_\perp$ of the hardest and second 
hardest jet, for different number of merged jets, in $\W+$ jets events at 
$\ECM = 1960$ GeV in $\p\pbar$ collisions. The merging scale is defined in
$\tms = \min\{k_{\perp i}\}$. Jets were defined with the 
$k_\perp$-algorithm with $D = 0.4$. Multiple interactions and hadronisation were
switched off. Plots produced with \pytppp Tune 4C. The bottom in-sets show the
deviation of the merged samples from default, first-order corrected \pytppp.}
}

Analysing the $k_{\perp 2}$ separation when two jets are clustered
into a single jet in the right panel, it is interesting to see how
this increase arises. For small merging scales (\eg\ $10,15$
GeV), $k_{\perp 2}$ in two-jet merging quickly grows at the merging
scale and remains flat until a more gradual ascend sets in at
$k_{\perp 2}\approx 60$ GeV.  There, the ME2PS distributions for
$\tms=10,15$ GeV also join the curves for larger merging scales ($30,45$ GeV). 
This behaviour of ME2PS for low $\tms$ can more clearly be seen in the left
panel of Figure~\ref{fig:wp:pt-jets-diff-njet-over-me4ps}.  When
inspecting the ME3PS curves for $\tms=10,15$ GeV, we again see a
hardening of the spectrum, which is to some extent stable when going
to ME4PS.  Such a stabilisation inspires the conclusion that the
$k_{\perp n\leq N}$ separation between the $n$'th and $(n-1)$'th
hardest jets is stable once the maximal number of merged jets is
increased above $n$, as was found in \cite{Hoeche:2009rj}.

One possible argument for this effect is that when looking at the
$k_{\perp 2}$ separation at which a jet $a_1$ and a jet $a_2$ are
clustered into a single jet in ME3PS, the parent jets $b_1,b_2,b_3$
which produced $a_1$ and $a_2$ were harder than in ME2PS, thus again
favouring harder jets $a_{1,2}$, \ie\ larger separations, $k_{\perp
  2}$. The stabilisation could then be explained by assuming that the
parent jets producing $b_{1,2,3}$ in ME4PS will not greatly increase
the hardness of $b_{1,2,3}$ because in ME4PS, most jets will be just
above the merging scale due to a steeply falling $k_{\perp 4}$
spectrum.

\FIGURE{
\centerline{
  \includegraphics[width=0.5\textwidth]
    {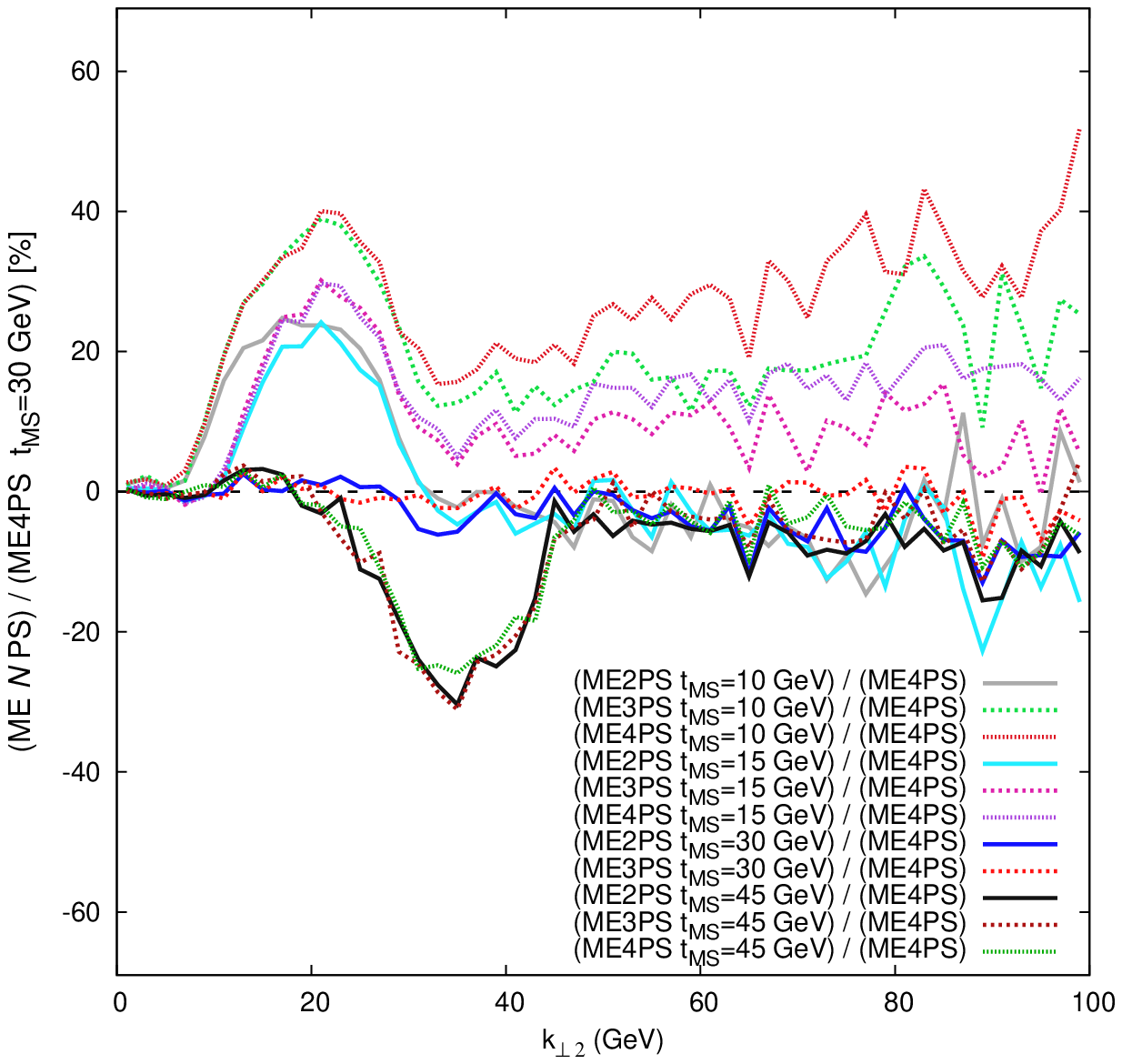}
  \includegraphics[width=0.5\textwidth]
    {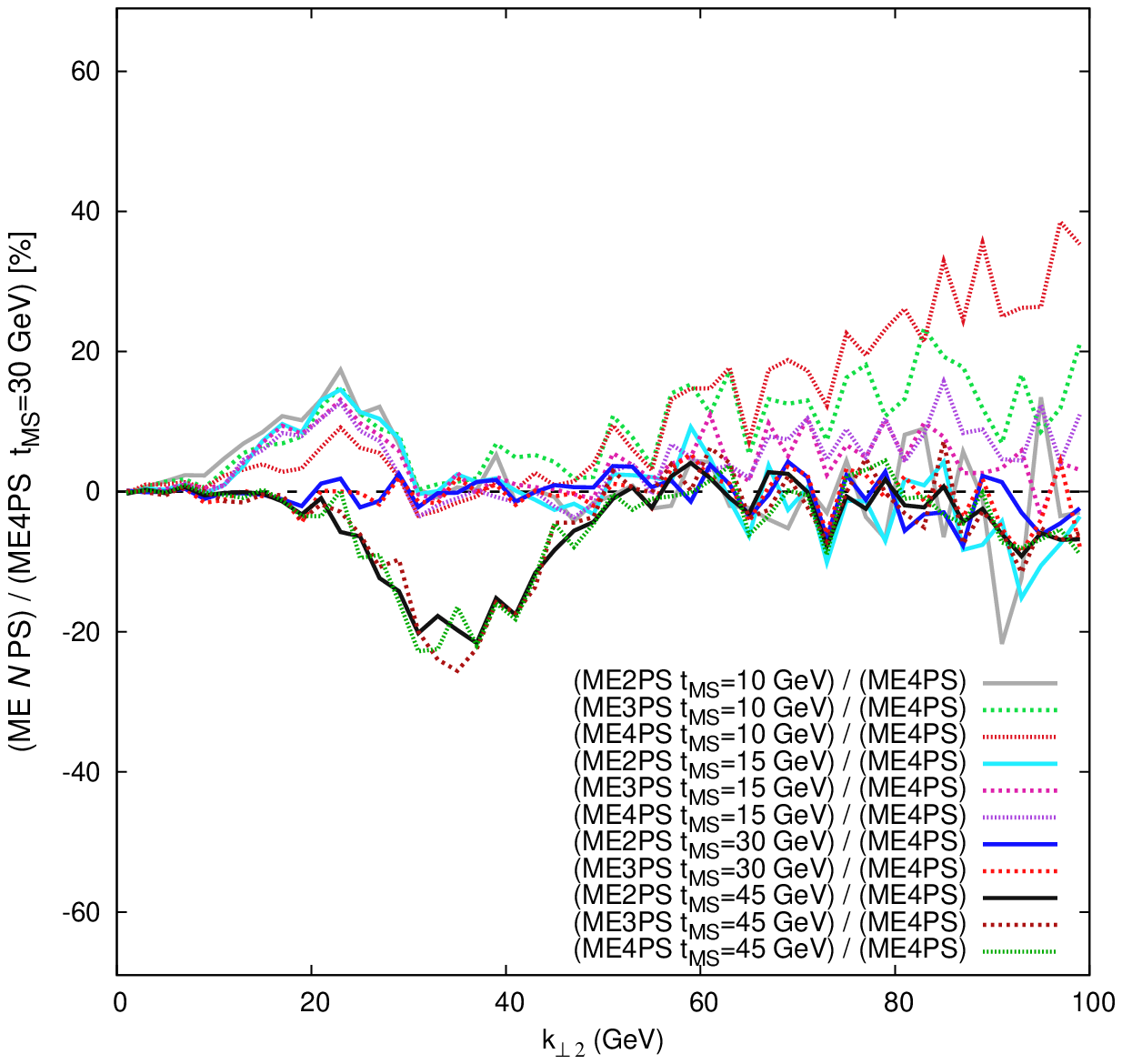}
}
\caption{\label{fig:wp:pt-jets-diff-njet-over-me4ps}$k_{\perp 2}$ separation of
the second jet in $\W+$ jets events at 
$\ECM = 1960$ GeV in $\p\pbar$ collisions. The curves are normalised to the 
$k_{\perp 2}$ distribution in ME4PS at $\tms = 30$ GeV. Jets were defined with 
the $k_\perp$-algorithm with $D = 0.4$. Multiple interactions and hadronisation
were switched off. Left panel: Plots produced with \pytppp Tune 4C. Right panel:
Plots produced with \pytppp Tune 4C, with enforced rapidity ordering switched 
off.}
}

However, in our implementation, the question arises if a stable
$k_{\perp 2}$ distribution will also be stable to changing the value
of the merging scale.  First, notice that there is no shape change in
the $\tms=45$ GeV curves when going from ME2PS to ME3PS (or ME4PS),
even though by the above reasoning, further distortions should be more
pronounced at high merging scales. It is critical to notice (see
Figure~\ref{fig:wp:pt-jets-diff-njet-over-me4ps}) that for low merging
scales, the spectrum in ME4PS is significantly harder than the ME4PS
reference at $\tms=30$ GeV, whereas once their initial ascend is over,
the curves for $\tms=45$ GeV nicely join the $\tms=30$ GeV ones.
These observations can again be explained by unitarity violation for
$\tms=10$ GeV and $\tms=15$ GeV, which stabilise when merging more
jets, but do not decrease. Since the changed cross section is stable
while the sample composition changes between ME2PS and ME3PS, the
shape of $k_{\perp 2}$ has to change.  In support of this rationale,
the right panel of Figure~\ref{fig:wp:pt-jets-diff-njet-over-me4ps}
shows that when reducing unitarity violations by not enforcing
rapidity ordering in the shower, the effects are significantly reduced
as well. These considerations can be applied to jet separations
$k_{\perp n\geq 2}$ as well.

\FIGURE{
\centerline{
  \includegraphics[width=0.6\textwidth]
    {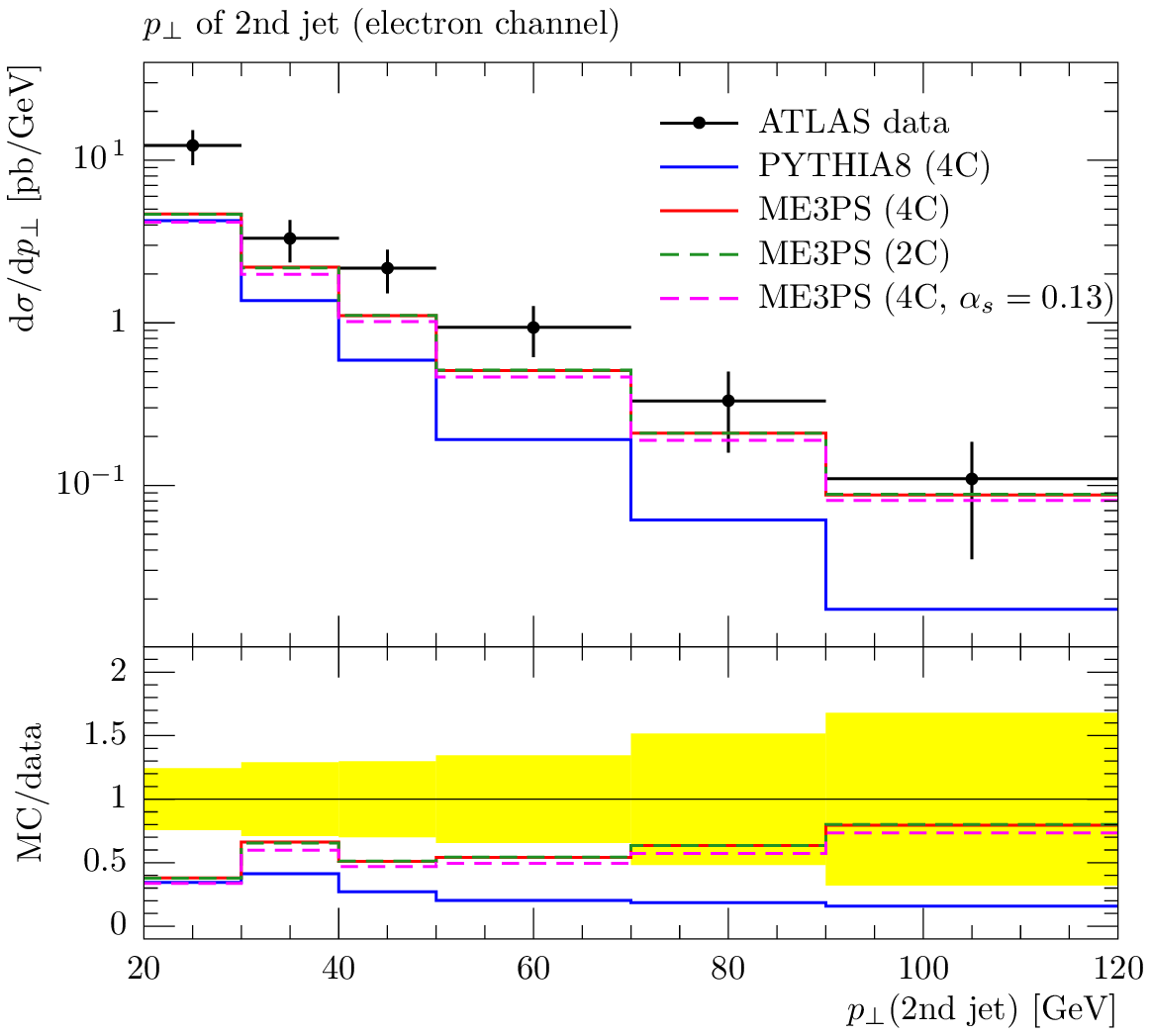}
}
  \caption{\label{fig:wp:diff-in-tunes}Transverse momentum of the second 
  hardest jet in $\W ~+$ jet events as measured in the electron channel by 
  ATLAS \cite{ATLAS:2010pg}, for different tunes. ``2C" indicates that
  Tune 2C was used, while ``4C" uses Tune 4C, the current default tune in 
  \pytppp.157. The label ``$\as = 0.13$" stands for fixing $\as(\mz) = 0.129783$
  in Tune 4C, as discussed in the text. The merging scale is 
$\tms = \min\{k_{\perp i}\} =  30\textnormal{ GeV}$. Effects of multiple
  interactions and hadronisation are included. The plot was produced
  with \rivet \cite{Buckley:2010ar}.}
}

Every parton shower relies on phenomenological models to confine
partons into hadrons, thus making systematic tuning is a critical step
in the development of an event generator. Tuning however should not
hide the shortcomings due to approximations made. If residual tuning
effects because of correlations between tuning parameters remain in phase space
regions with well-separated jets, we expect such changes
to be stabilised when correcting with higher multiplicity matrix
elements. The impact of changing between different tunes in \pytppp is
shown in Figure~\ref{fig:wp:diff-in-tunes}, where we show the results
of using Tune 2C, Tune 4C and forcing $\as(\mz) = 0.129783$ (the
CTEQ6L1 fit value) in all components of \pytppp, in comparison with
ATLAS data \cite{ATLAS:2010pg}.  We find that the $\p\p \to \W$+jets
predictions are fairly stable with respect to changing tunes. As
expected, we observe that the ME3PS sample is harder than default
\pytppp.

\FIGURE{
\centerline{
  \includegraphics[width=0.5\textwidth]
    {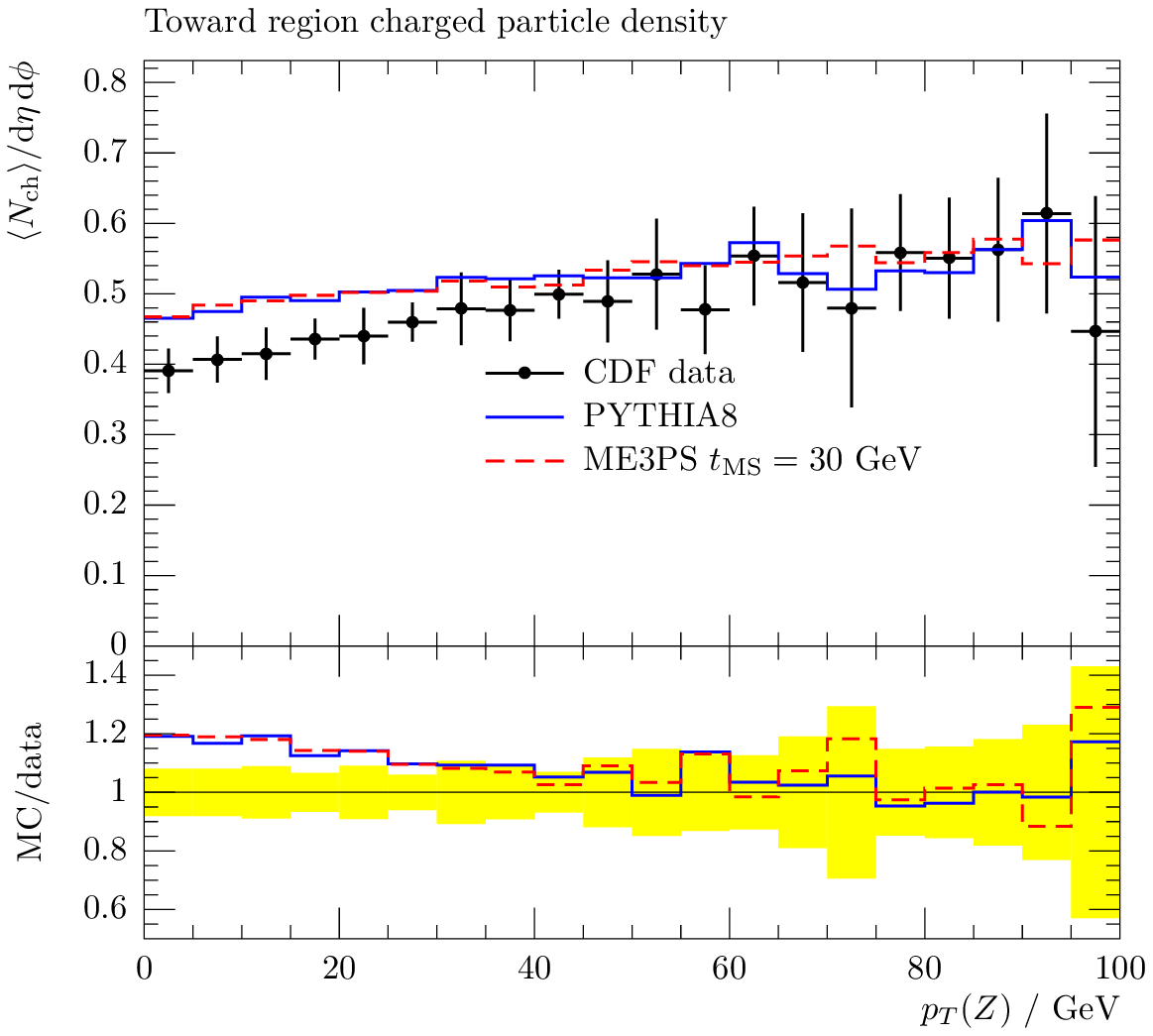}
  \includegraphics[width=0.5\textwidth]
    {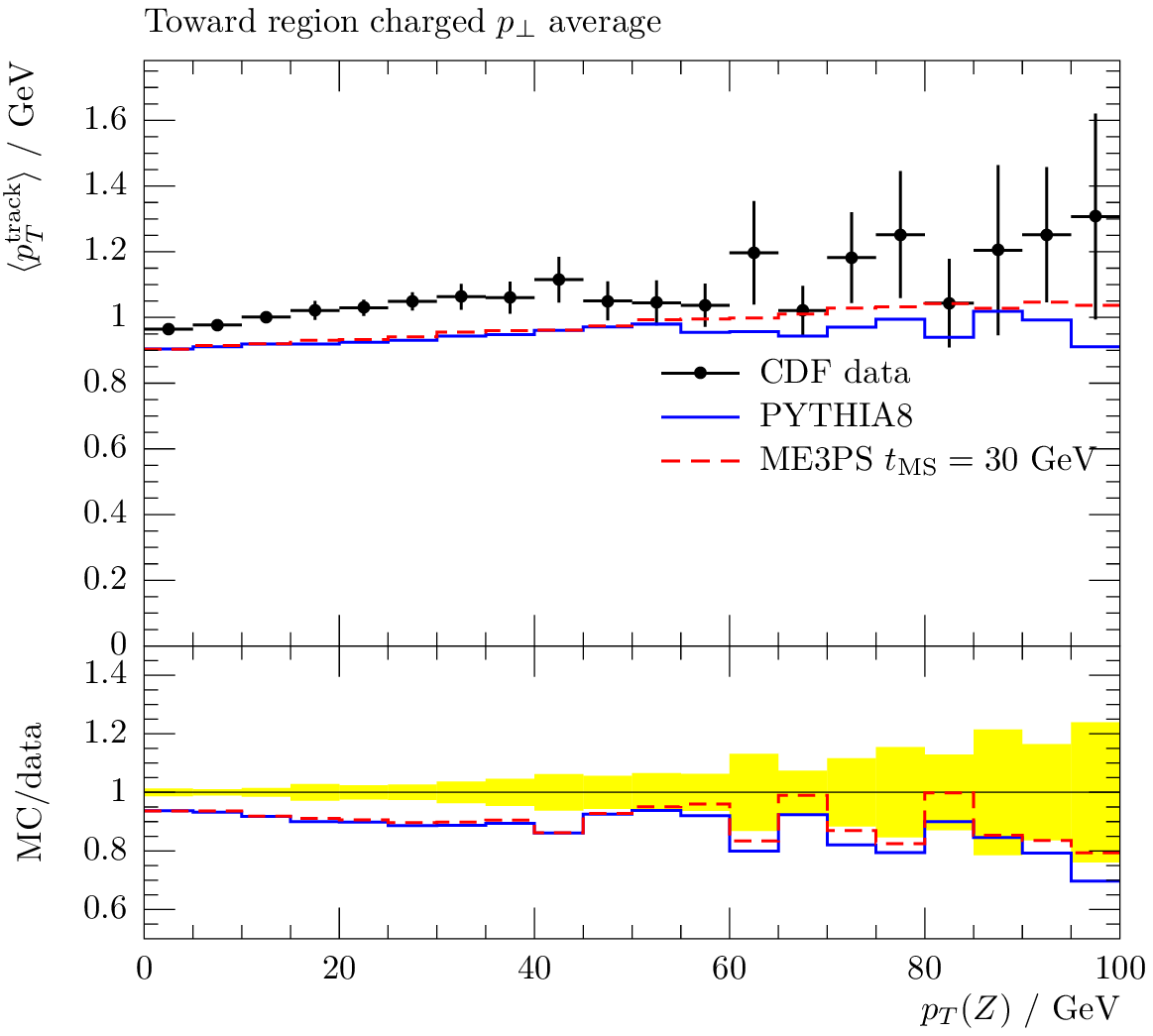}
}
\caption{\label{fig:z-ue}Toward region charged particle density and average
  $p_{\perp}$ in Drell-Yan events, as measured by CDF
  \cite{Aaltonen:2010rm}. The merging scale is $\tms = \min\{k_{\perp i}\} =
  30\textnormal{ GeV}$. Effects of multiple scatterings and
  hadronisation are included, and Tune 4C was chosen. The plots were 
  produced with \rivet \cite{Buckley:2010ar}. }
}

Finally in Figure~\ref{fig:z-ue} we show the effect of our treatment
of multiple interactions. The associated hadronic activity in $\Z$
production events, especially in the azimuthal direction direction of
the $\Z$, is very sensitive to underlying event effects, and hence
also to multiple interactions \cite{Aaltonen:2010rm}. In our merging
scheme we have been very careful to make sure that multiple
interactions are treated exactly the same way as in
standard \pytppp without inclusion of matrix element configurations. And, as 
seen in Figure~\ref{fig:z-ue}, the differences between the merged sample and
default \pytppp are indeed very small.

\subsection{Di-boson and QCD jet production}
\label{sec:diboson-qcdjets}

Our implementation is in principle general enough to be applied to any
process that can be handled by \pytppp. However, in this publication,
we restrict ourselves to two further examples. First, let us examine
di-boson production, with one of the bosons decaying
hadronically. Allowing hadronic decays of weak bosons in the
underlying Born process provides another complication, and for
technical reasons we here restrict the matrix element to only produce
extra jets in from the incoming partons, while additional jets in the
hadronic boson decay are only produced by the shower. As the first
emission in the boson decay is anyway ME-corrected in standard
\pytppp, this is not a severe restriction. Note however that this
means that we have to treat emissions from the boson decay on the same
footing as multiple interactions (and QED radiation). This means that
they are included in the Sudakov form factors generated from
reclustered states, but when showering from a $n<N$ state, if the
first emission is from the boson decay, the event is never
vetoed. The partons from the boson decay are also not involved in the
reclustering of matrix element states.

The performance of our implementation concerning these issues can be
tested when merging $\p\p \to \Wp\Z \to \eplus\nu_{\el}~jj$+jets
matrix elements. The left panel of Figure~\ref{fig:wpz:mjj} shows that
also in the case of di-boson production, the $k_{\perp 3}$ spectrum
becomes harder on inclusion of additional jets. There are no visible
differences in the default \pytppp results when changing between only
ordering emissions in evolution $\ord$ and ordering both in $\ord$ and
rapidity, since $k_{\perp 3}$ is dominated by the hardest shower
emission, which is not affected by the additional rapidity
ordering. 
We observe only small differences between merged samples with and without
enforced rapidity ordering in the shower.
Relative changes in
$k_{\perp 3}$ are, as expected, comparable to the effects on $k_{\perp
  1}$ when including additional jets in $\p\p \to \Wp \to
\eplus\nu_{\el}$ (see \eg\ Figure~\ref{fig:wp:pt-jets-diff-njet}). We
have checked that different jet definitions do not change this trend.

A consequence of harder jets can be seen in the right panel of
Figure~\ref{fig:wpz:mjj}, where we show the di-jet invariant mass
distribution with cuts and jet definition from CDF
\cite{Aaltonen:2011mk}. The spectrum develops a harder tail compared
to default \pytppp. Particularly in the region $140 < m_{jj} < 200$
GeV we find an increase around $10\%$. Also, the distribution is
sensitive to the unitarity violations due to enforced rapidity
ordering, so that care has to taken when comparing MEPS distributions
to experimental data. In \cite{Aaltonen:2011mk}, the shape of the
di-boson backgrounds was modelled by \pythia\!\!6.216, which should
behave similar to default \pytppp. Merging additional jets in $\p\p
\to \Wp\Z \to \eplus\nu_{\el}~jj$ can affect the shape of the di-jet
invariant mass spectrum in a way which will reduce the significance of
the effect found by CDF. We plan to further investigate these issues
in a future publication.

\FIGURE{
\centerline{
  \includegraphics[width=0.5\textwidth]
    {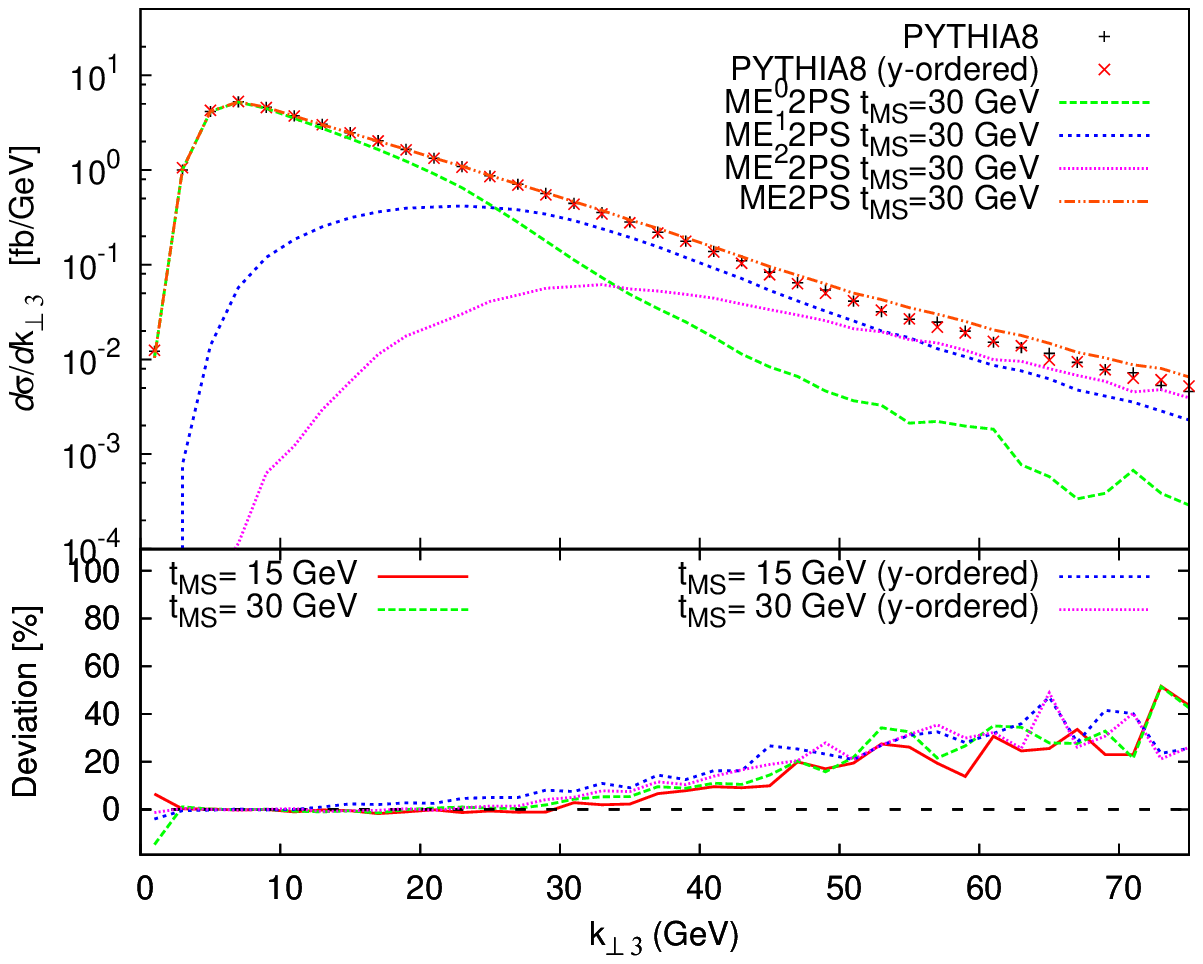}
  \includegraphics[width=0.5\textwidth]
    {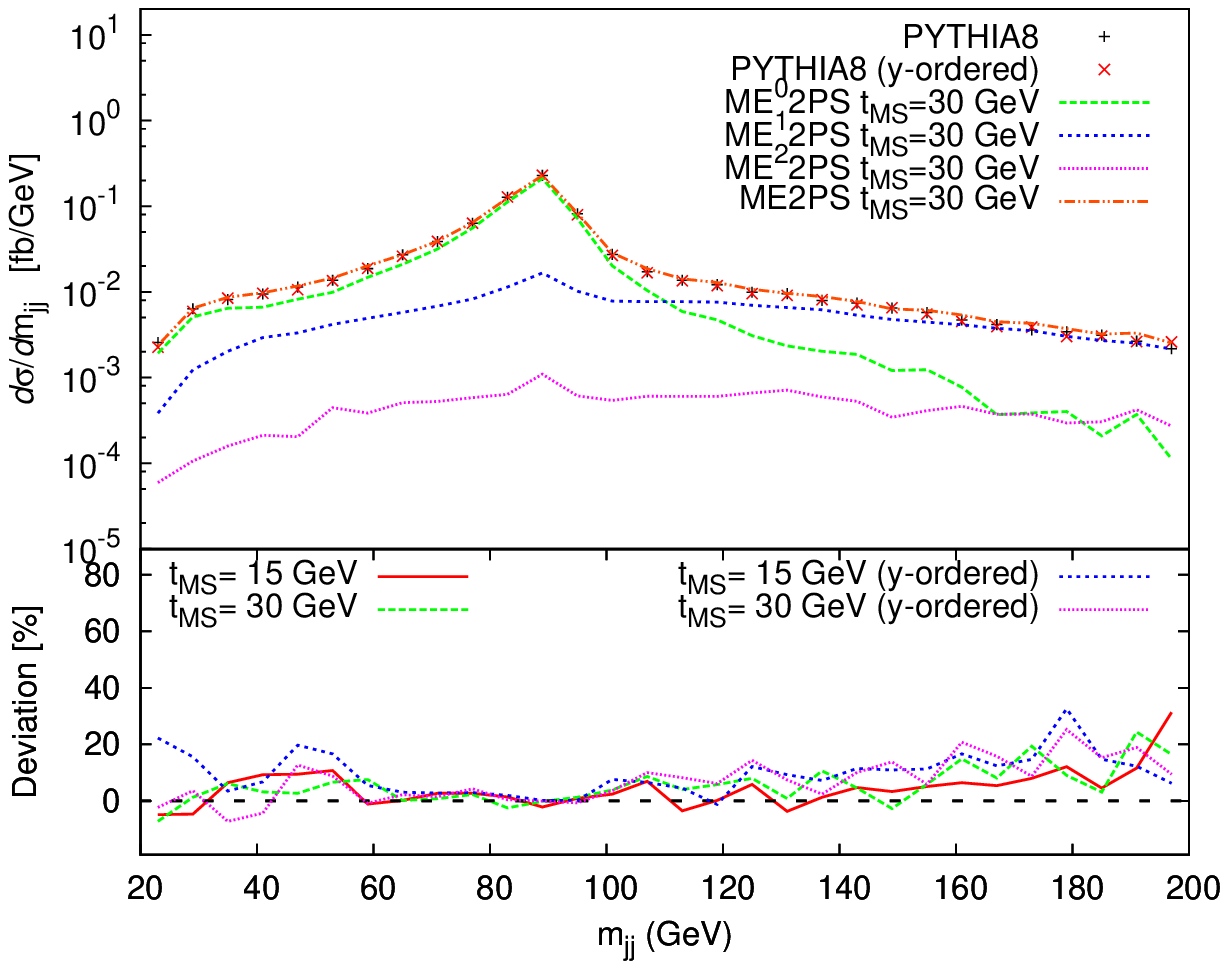}
}
\caption{\label{fig:wpz:mjj}Sample results of including matrix
  elements with additional jets in $\p\p \to \Wp\Z \to
  \eplus\nu_{\el}~jj$ events, at $\ECM = 1960$ GeV in $\p\pbar$
  collisions. The merging scale is defined in $\tms = \min\{k_{\perp
    i}\}$. Multiple interactions are included. Curves with enforced rapidity 
  ordering in the shower carry
  an additional label ``y-ordered''. The bottom in-set shows the
  deviation of the merged samples to \pytppp.  Left panel: $k_{\perp
    3}$ of the third and second hardest jet at hadron level.  Jets
  were defined with the $k_\perp$-algorithm with $D = 0.4$.  Right
  panel: Di-jet invariant mass in at hadron level. Cuts are taken from
  the recent CDF publication \cite{Aaltonen:2011mk}. Jets were defined
  with the CDF JETCLU algorithm \cite{Blazey:2000qt} as implemented in
  {\texttt{fastjet}}.  }
}

Finally, we examine QCD jet production. For such events, we set the shower 
starting scale for the $2\to 2$ process to the transverse momentum of the 
outgoing partons. The maximal scale for secondary scatterings is set to the 
same value. In \pytppp, users are generally allowed to choose a different 
prescription of setting a maximal scale of multiple interactions, \eg\ the 
energy of the colliding hadrons. Adopting this example, we risk double-counting 
configurations, since interactions identical to the hard process can be 
generated.

To remove this double counting, an additional veto on the 
transverse momentum of multiple interactions in the trial shower has to be 
applied. We have checked that when allowing secondary scatterings up to the 
kinematical limit and applying a veto, distributions are not changed with 
respect to setting $\ord_{\textnormal{\tiny{MI}},max} = p_{\perp, 2 \to 2}$. 
The results presented here have been produced with fixing the starting scales 
for the hard process to the transverse momentum, as is the default in \pytppp.

\FIGURE{
  \includegraphics[width=0.75\textwidth]
    {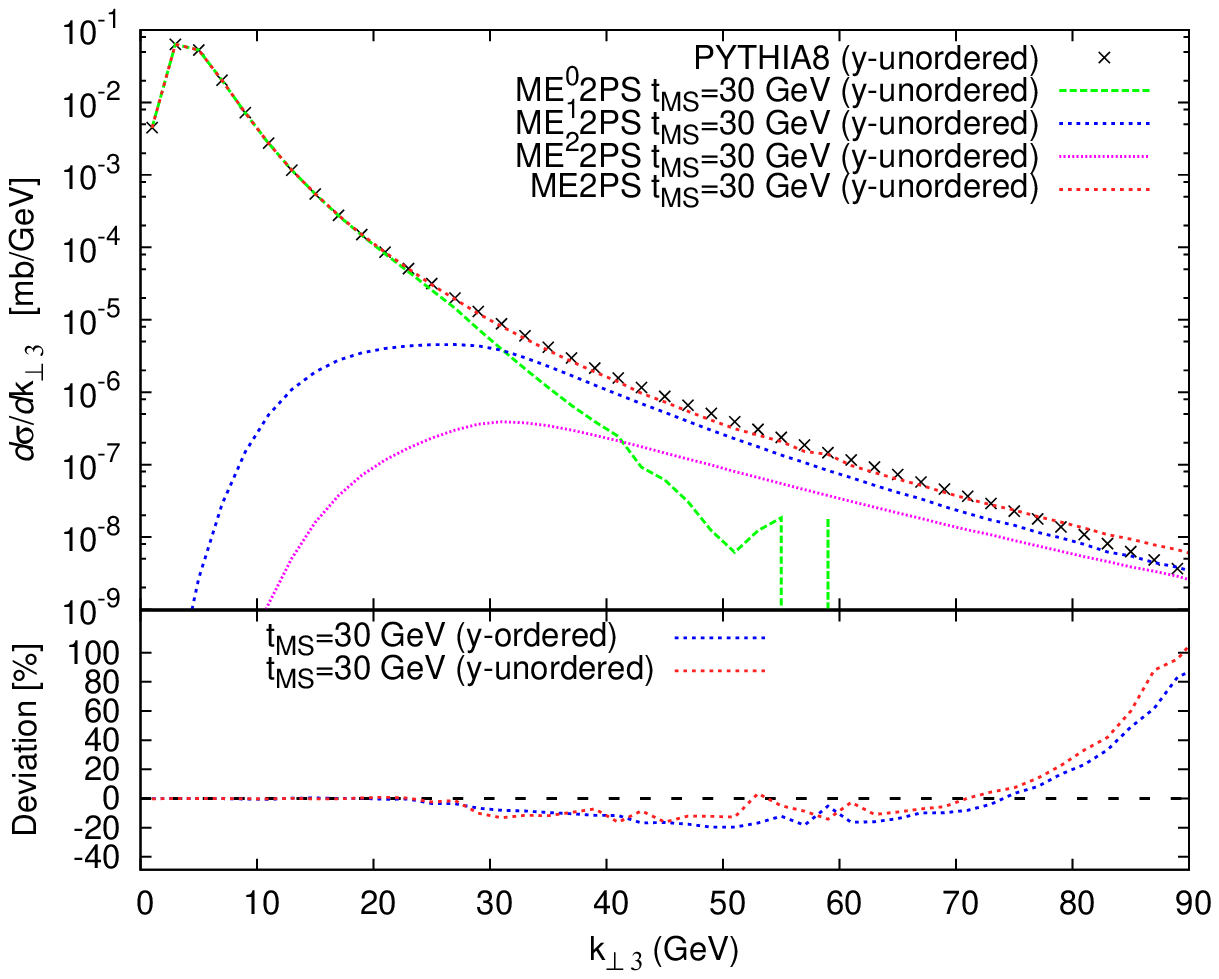}
\caption{\label{fig:jj:pt4}$k_{\perp 3}$ separation of
the third jet in pure QCD jet events at $\ECM = 1960$ GeV in $\p\pbar$ 
collisions. The merging scale is $\tms = 30\textnormal{ GeV}$. Jets were defined
with the $k_\perp$-algorithm with $D = 0.4$. Multiple interactions are included
and hadronisation was switched off. Curves with enforced rapidity ordering in 
the shower carry the label ``y-ordered'', while results without explicit
rapidity ordering are labelled ``y-unordered''. The bottom in-set shows the 
deviation of the merged sample default \pytppp.
}
}

Figure~\ref{fig:jj:pt4} shows that for QCD jets as well, inclusion of
additional jets increases the hardness $k_{\perp 3}$ of the third
jet. Compared to the changes in $k_{\perp 1}$ for $\W + $jets, the
effect is, however, moderate. This is in accord with the findings in
\cite{Corke:2010yf}, which showed good agreement in the $p_{\perp}$ of
the softest of three partons (there called $p_{\perp 5}$), when
comparing $2\to3$ matrix elements to the default shower after the
first emission from a $2\to2$ core process. There, the shower was
slightly harder than the matrix element until $p_{\perp 5}\approx 80$
GeV. A similar effect can be seen in the $k_{\perp 3}$ separation of
jets, which is related to the $p_{\perp 5}$ of partons.

\FIGURE{
  \includegraphics[width=0.75\textwidth]
    {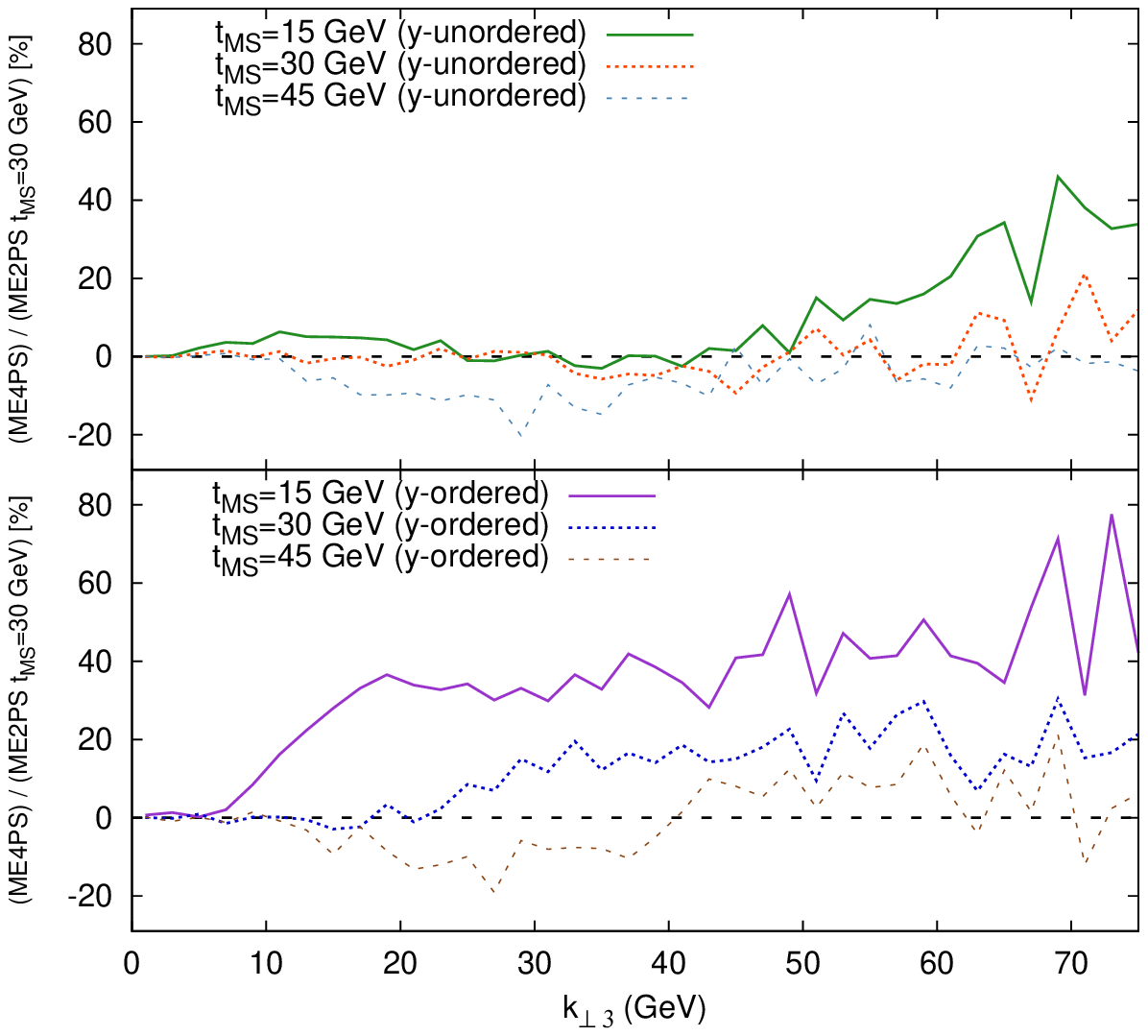}
    \caption{\label{fig:wp:pt-jets-diff-njet-over-me2ps}$k_{\perp 3}$
      separation of the second and third jet in $\W+$jets events at
      $\ECM = 1960$ GeV in $\p\pbar$ collisions. The curve shows the
      deviation in $k_{\perp 3}$ of ME4PS for three different merging
      scales, with respect to ME2PS for $\tms = 30$ GeV. Jets were defined
      with the $k_\perp$-algorithm with $D = 0.4$. Hadronisation and multiple
      interactions were switched off. Curves with enforced rapidity ordering in 
      the shower carry the label ``y-ordered'', while results without explicit
      rapidity ordering are labelled ``y-unordered''.
    }
}

The inclusion of a sample with two additional jets does not change the
situation dramatically, leading us to the conclusion that once the
first few hard jets are generated according to the tree-level matrix
elements, the parton shower does a fairly good job in describing the
hardness of additional jets. This is supported by the upper panel of
Figure~\ref{fig:wp:pt-jets-diff-njet-over-me2ps}, showing the
$k_{\perp 3}$ separation between the third and second hardest jets in
$\W+$jets events. Clearly, there are only little changes in the
hardness of the third jet when going from ME2PS to ME4PS, \ie\ the
merging has less impact once a couple of jets are included from the
matrix element states\footnote{As can be seen in the 
lower panel of Figure~\ref{fig:wp:pt-jets-diff-njet-over-me2ps}, this statement
does not hold if there are major unitarity violations -- which should be 
avoided anyway.}.

\FIGURE{
  \includegraphics[width=\textwidth]
    {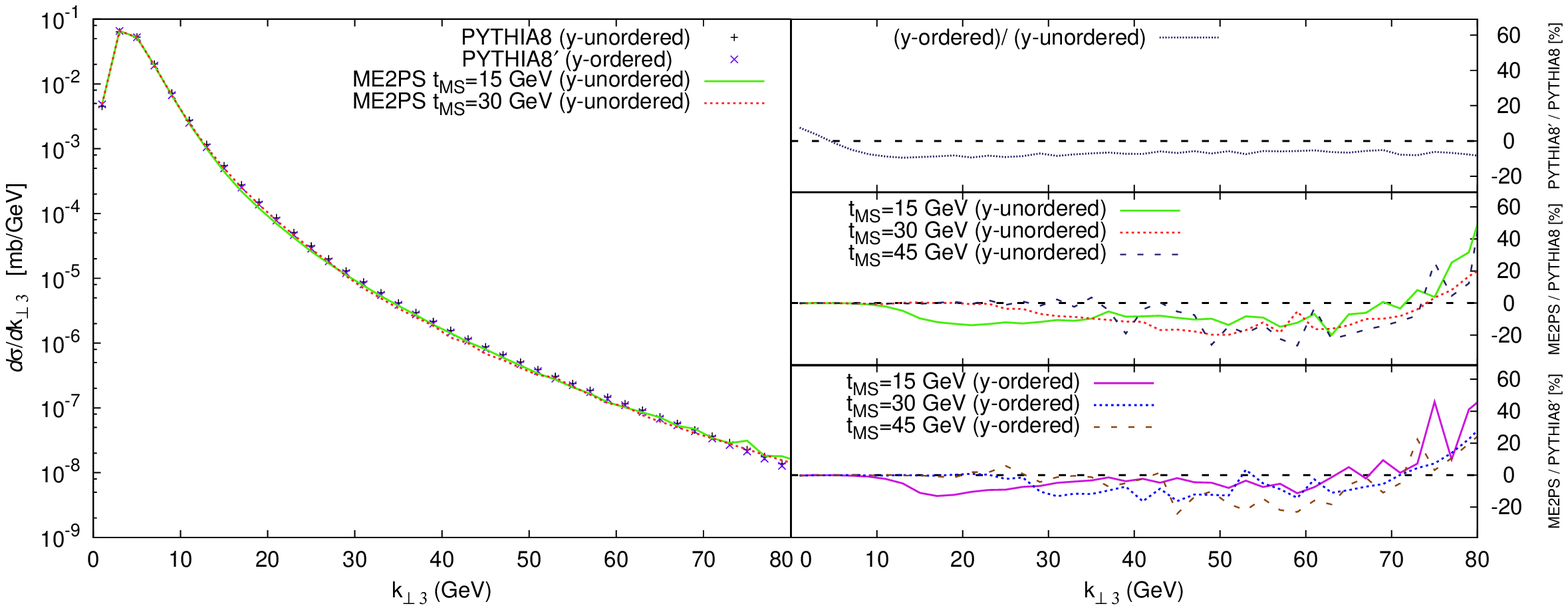}
\caption{\label{fig:jj:pt3-unitarity}$k_{\perp 3}$ separation of
the third jet in pure QCD jet events at $\ECM = 1960$ GeV in $\p\pbar$ 
collisions, for three different merging scales. Jets were defined
with the $k_\perp$-algorithm with $D = 0.4$. Multiple interactions and
hadronisation were switched off. Curves with enforced rapidity ordering in 
the shower carry the label ``y-ordered'', while results without explicit
rapidity ordering are labelled ``y-unordered''. Left panel: $k_{\perp 3}$ 
separation of the third jet. Upper right panel: Deviation in $k_{\perp 3}$
between default \pytppp (Tune 4C) and \pytppp (Tune 4C) with 
no enforced rapidity ordering. Centre right panel: Deviation in $k_{\perp 3}$
between ME2PS sample and \pytppp (Tune 4C) with no enforced rapidity 
ordering. Lower right panel: Deviation in $k_{\perp 3}$ between ME2PS sample 
and default \pytppp (Tune 4C).
}
}

\FIGURE{
  \includegraphics[width=0.75\textwidth]
    {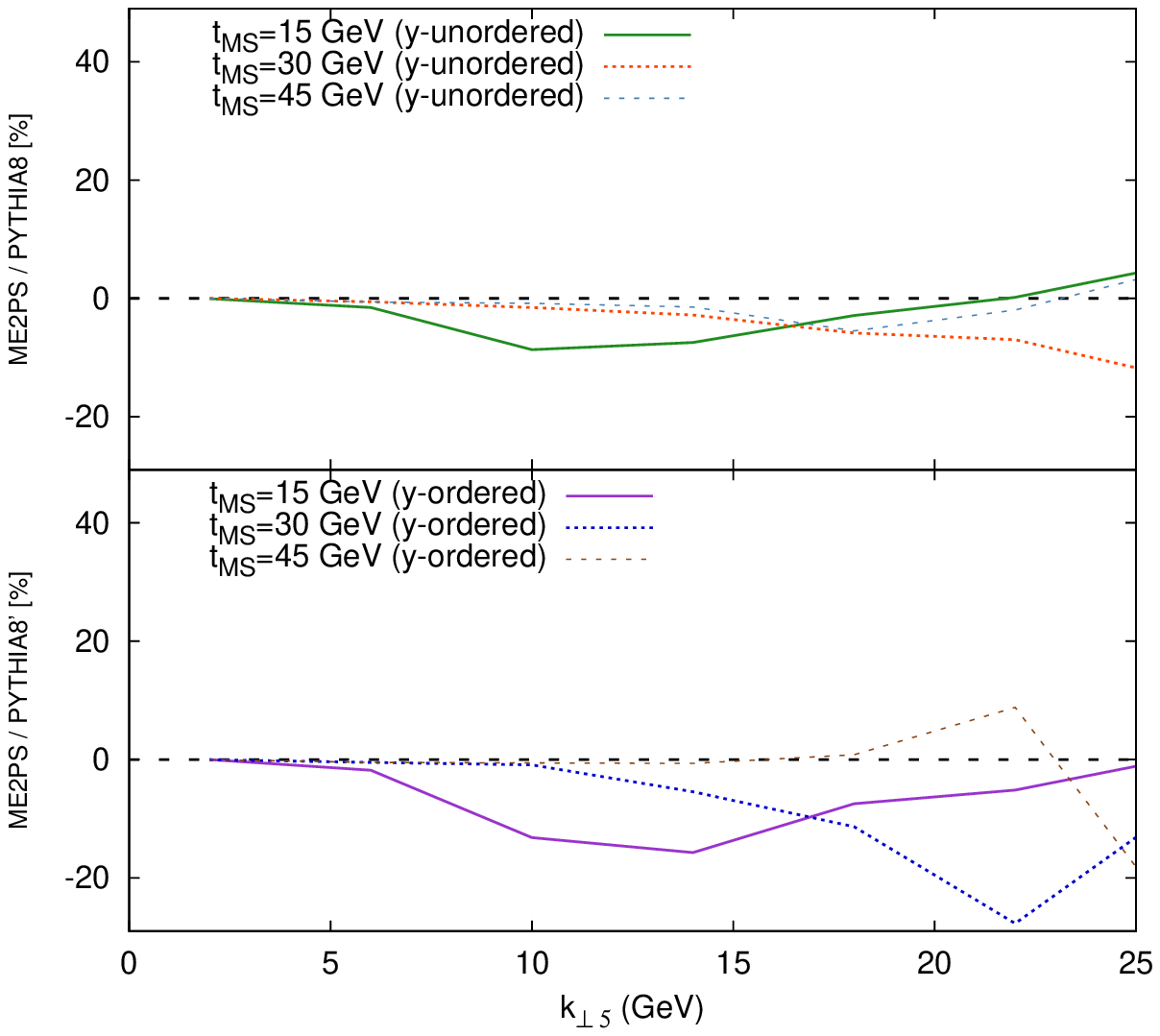}
    \caption{\label{fig:jj:pt5}$k_{\perp 5}$
      separation between the fourth and fifth jet in di-jet events at
      $\ECM = 1960$ GeV in $\p\pbar$ collisions. The curves show the
      deviation of $k_{\perp 5}$--distributions in ME2PS for three different 
      merging scales, with respect to \pytppp. Jets were defined
      with the $k_\perp$-algorithm with $D = 0.4$. Hadronisation and multiple
      interactions were switched off. Curves with enforced rapidity ordering in 
      the shower carry the label ``y-ordered'', while results without explicit
      rapidity ordering are labelled ``y-unordered''.
    }
}

Coming back to pure QCD, we show in Figure \ref{fig:jj:pt3-unitarity}
that also in this case, results for $k_{\perp 3}$ are fairly stable
when changing the merging scale. We register only small unitarity violations
of $\mathcal{O}(10\%)$, which matches the 
changes in $k_{\perp 3}$ in $\W+$jets events when going from ME2PS to ME4PS 
without requiring rapidity ordering, as illustrated in the upper part of
Figure~\ref{fig:wp:pt-jets-diff-njet-over-me2ps}. As $\W+n$ jets contains 
colour configurations similar to di-jet$+(n-2)$ jets, this is another 
indication for the consistency of the implementation.
However, in Figure~\ref{fig:jj:pt3-unitarity}, we find only 
minor changes between different treatments of rapidity ordering for di-jet 
events, whereas for $\W+$jets events, we observe dramatic effects
(see lower plot in Figure~\ref{fig:wp:pt-jets-diff-njet-over-me2ps}). This can
be explained by the fact that when requiring rapidity ordering, \pytppp orders
all emissions \emph{after the first shower emission} in rapidity, meaning 
that for di-jet events, $k_{\perp 3}$ is virtually unaffected by 
the constraint, while in $\W+$jets events, major restrictions on the phase 
space of the second and third jet lead to large unitarity violations. This 
argument is substantiated by Figure~\ref{fig:jj:pt5}, which shows that once 
rapidity ordering becomes relevant, the additional ordering results
in larger deviations for low merging scales. 

\FIGURE{
\centerline{
  \includegraphics[width=0.5\textwidth]
    {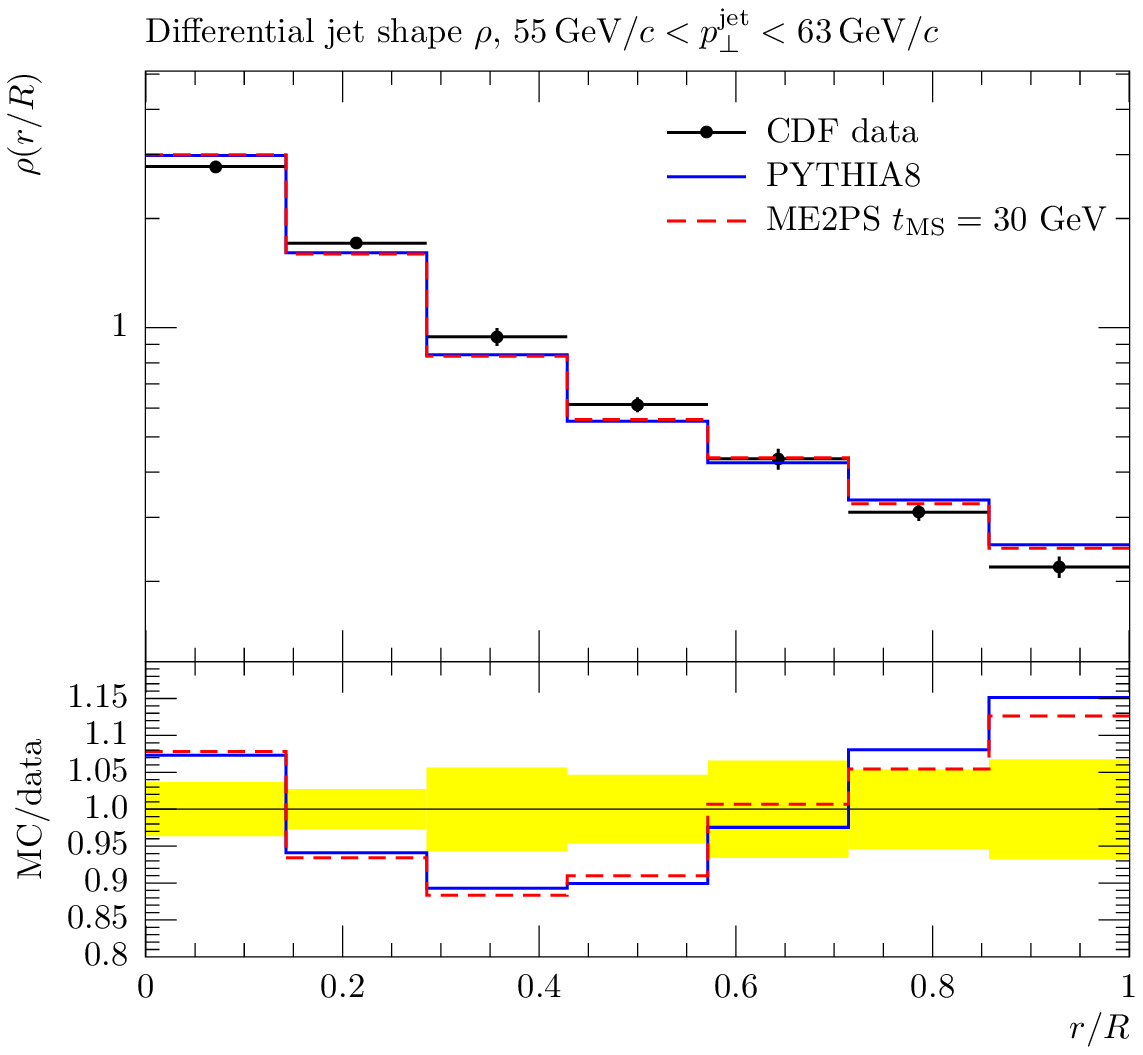}
  \includegraphics[width=0.5\textwidth]
    {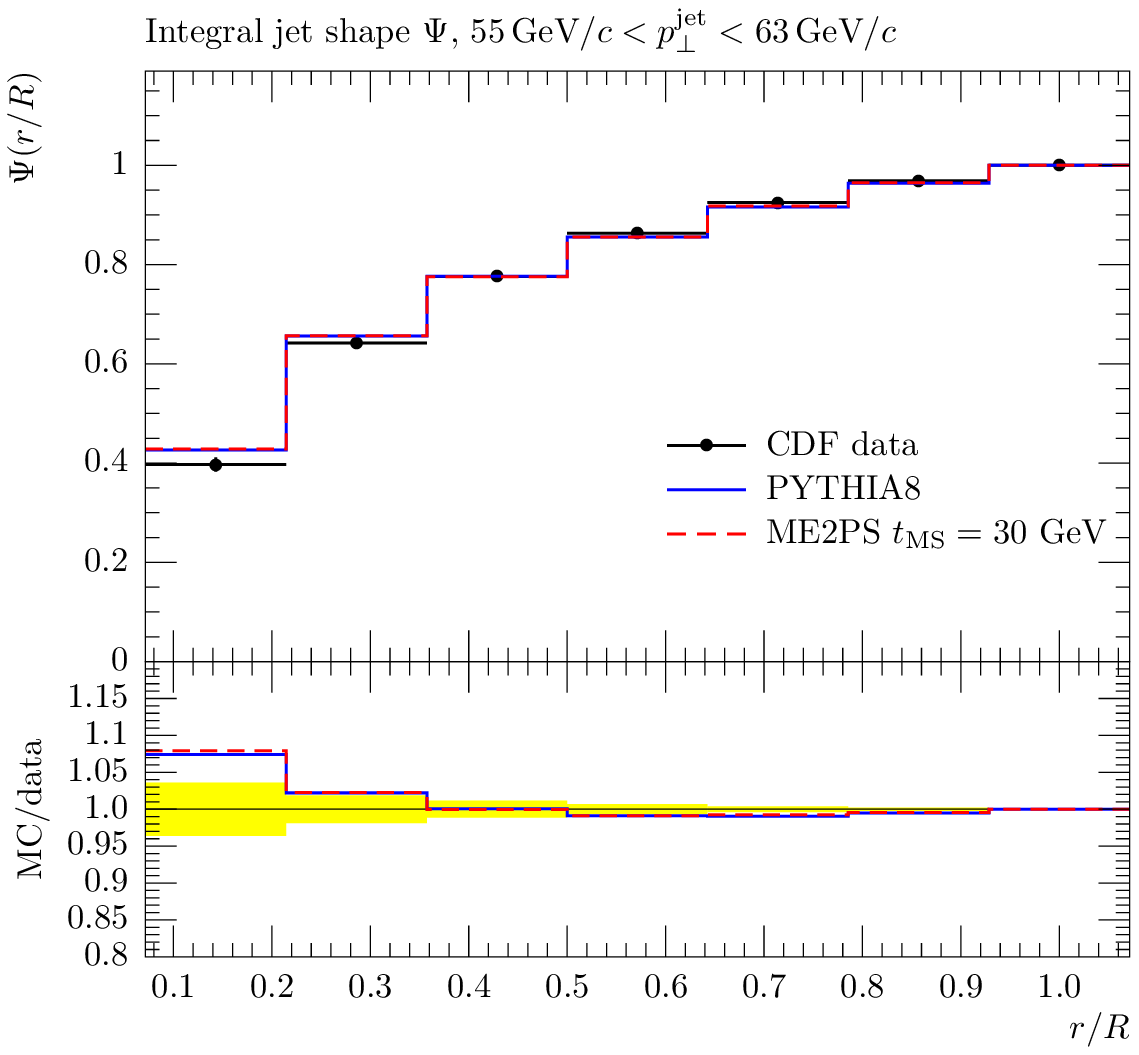}
}
\centerline{
  \includegraphics[width=0.5\textwidth]
    {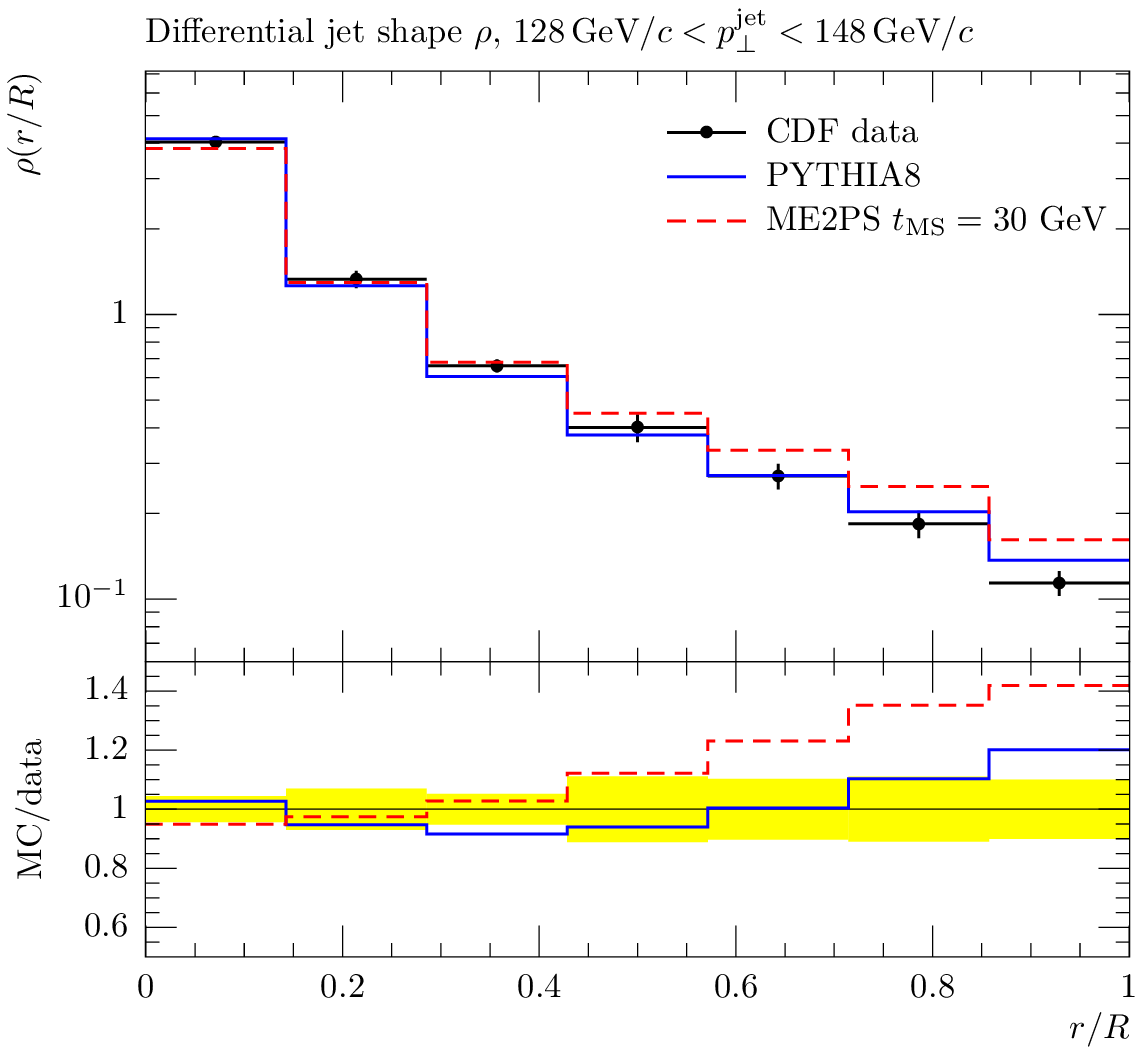}
  \includegraphics[width=0.5\textwidth]
    {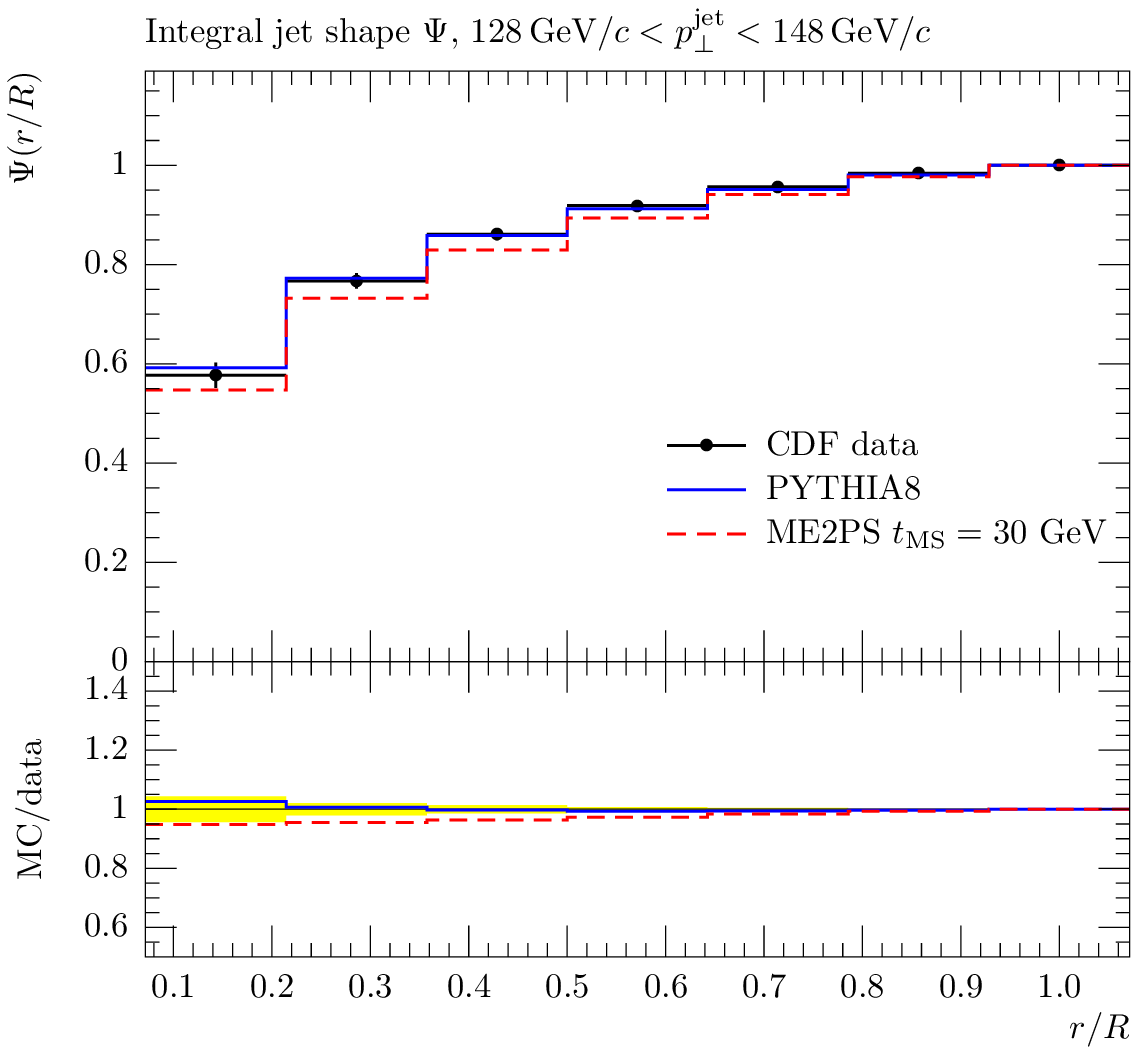}
}
\caption{\label{fig:jj:rho} Jet shapes in QCD events as measured by CDF
  \cite{Acosta:2005ix}. The merging scale is $\tms =
  \min\{k_{\perp i}\} = 30\textnormal{ GeV}$. Effects of multiple
  scatterings and hadronisation are included. The plots were produced
  with \rivet \cite{Buckley:2010ar}.}
}

It is worth noting that since jet spectra are not changed dramatically
when including additional jets, only small differences are expected
when comparing to experimental data. In Figure~\ref{fig:jj:rho}, we
examine the description of CDF jet shapes \cite{Acosta:2005ix} for two
exemplary $p_{\perp}^{jet}$ bins. For a $p_{\perp}^{\textnormal{jet}}$
in $55 \textnormal{ GeV} < p_{\perp}^{\textnormal{jet}} < 63
\textnormal{ GeV}$, we find only very minor changes. However, for
higher $p_{\perp}^{\textnormal{jet}}$ in the region $128 \textnormal{
  GeV} < p_{\perp}^{\textnormal{jet}} < 148 \textnormal{ GeV}$ the
differences between default \pytppp and the merged sample ME2PS with
two additional jets are more pronounced, and we see that the latter
gives a slightly broader shape. This is expected as at high transverse
momentum the effect of the harder third jet in Figure~\ref{fig:jj:pt4}
should come into play, resulting in more jets containing two partons
from the matrix element. Such jets are of course broader.

When checking differential jet shapes for other
$p_{\perp}^{\textnormal{jet}}$ bins, we find that ME2PS does as good
or slightly better than \pytppp for
$p_{\perp}^{\textnormal{jet}}\lesssim 120\textnormal{ GeV}$, while
decreasing too slowly for $p_{\perp}^{\textnormal{jet}} \gtrsim
120\textnormal{ GeV}$.  This indicates that at least some revisions
need to be made when tuning matrix-element-merged \pytppp to pure QCD jet 
data. Since the influence of multiple interactions on jets with 
$p_{\perp}^{\textnormal{jet}} \gtrsim 120\textnormal{ GeV}$ is likely to be 
small, a possible new tune would potentially feature changes in 
$\as\left(\mz\right)$ and other parameters to prescribe the physics of 
hard jets.

\section{Conclusions and Outlook}
\label{sec:outlook}

We have implemented \ckkwl merging inside the \pytppp framework, and
have shown that it works well for several sample processes:
$\ee\to$~jets, (di-) boson and pure QCD jet production in hadronic
collisions. The implementation is, however, quite general and could be
used for any process which \pytppp is able to handle.

The algorithm is true to the \ckkwl spirit, in that if matrix element
samples are provided for up to $N$ extra partons, every event where
the $n\le N$ hardest (in the parton shower sense) partons can be
produced by the matrix element, it will be evolved from the
corresponding matrix element state.

By construction, the dependence on the merging scale vanishes to the
logarithmic precision of the \pytppp parton shower. Nevertheless, we
find visible sub-leading effects due to different choices that can be
made in the procedure. In particular we have investigated
  \begin{itemize}\itemsep 0cm
  \item different ways of choosing parton shower histories,
  \item different strategies for handling unordered histories,
  \item different starting scales for incomplete histories,
  \item different options for including multiple scatterings.
  \end{itemize}
In all these cases we found the effects to be small.

However, we found that in some cases there are large merging scale
dependences from unitarity violations. These problems have been noted
before in other \ckkw-based algorithms
\cite{Hamilton:2010wh,Hoche:2010kg}, and arise from the fact that what
is exponentiated in the Sudakov form factors is only the parton shower
approximation to the matrix elements, rather than the matrix elements
themselves. In addition, the phase space integrated over in the
Sudakov may differ from the full phase space available to the matrix
element.

For our implementation, one would expect the unitarity violations to
be diminished in the cases where \pytppp already include a
matrix-element reweighting of the first parton shower emission
(similar to what is done in \powheg \cite{Nason:2004rx,
  Frixione:2007vw}). However, we found that the effects on the
contrary are very large, and traced the reason for this to the fact
that the default tune of \pytppp uses a rapidity ordering for the
initial-state shower in addition to the ordering in the transverse
momentum evolution scale.
This results in a severe restriction of the
phase space over which Sudakov form factors are integrated, giving 
increased merging scale dependences.
When removing the
rapidity ordering, the unitarity violations are reduced to an almost
negligible effect.

An important result of our investigations is that the \pytppp shower
(without enforced rapidity ordering) actually is quite good at
describing the hardness of multi-jet events, as long as the hardest
few jets are generated according to the exact matrix elements. Of
course, there may be special observables related to details in the
correlations between jets, where merging with high-multiplicity matrix
elements is still necessary to get a correct description, but for the
main features of multi-jet event it seems to be enough to merge with a
limited number of extra jets. 

Before our \ckkwl implementation can be used for reliable predictions
and comparisons with experimental data, the parameters of \pytppp need
to be retuned. We have shown that for $\ee\to$~jets, the merging with
multi-jet matrix elements barely changes the description of data, and
we can assume that the parameters for the hadronisation and final
state showers will not need to be substantially changed. Furthermore,
for pure QCD processes in hadronic collisions, the effects of
multi-jet merging are again very modest, except for very high
transverse momentum jets, so also for minimum bias and underlying
event observables the tuning needed can be assumed to be minor. On the
other hand, for electro-weak processes and for very hard jets in pure
QCD processes in hadronic collisions the merging gives quite
substantial effects, which means retuning is necessary. To get
stable results, this new tune should be done without the rapidity
ordering discussed above.

\section*{Acknowledgements}

We are grateful to Richard Corke and Torbjörn Sjöstrand for useful
discussions and explanations of the technical details of the \pytppp
machinery.

\begin{appendix}

\section{Comments on the logarithmic accuracy of \ckkwl}
\label{sec:comm-logar-accur}

\FIGURE{
\centerline{
  \includegraphics[width=1.0\textwidth]
    {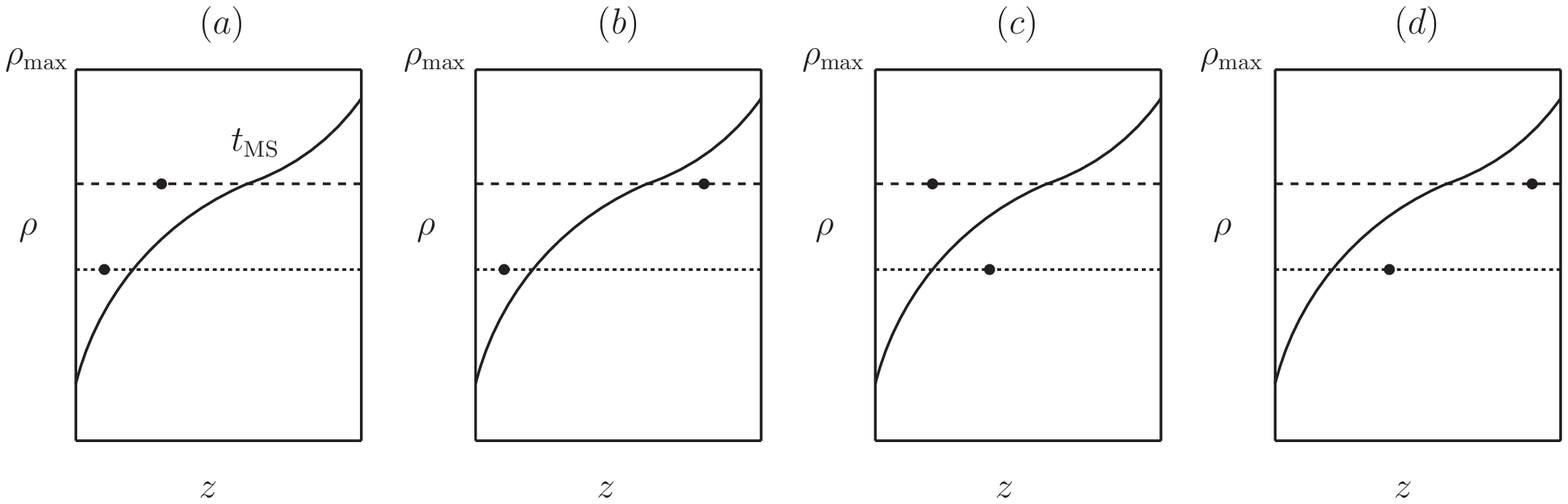}
}
\caption{\label{fig:phasespace-ckkwl}A schematic view of how the
  two-parton phase space is filled in \ckkwl. The figure illustrates
  how two partons at evolution scales $\ord_1$ and $\ord_2$ can be
  classified in terms of a merging scale $\tms$ defined in the
  variable $t$. The vertical axis is the shower ordering scale (which
  is different from the $t$-scale) and the horizontal axis is an
  auxiliary splitting variable. The different states are evolved from
  different matrix-element samples: (a) starts from the 2-jet ME, (b)
  the 0-jet ME, (c) the 1-jet ME, and finally (d) is evolved starting
  from the 0-jet ME. }
}

Here, we would like to elaborate on how the logarithmic accuracy of
the shower is preserved in \ckkwl. This discussion is independent of
the functional form of the merging scale variable. For the sake of
illustration, let us analyse how the phase space for two additional
partons is filled by the merging.  As shown in 
Figure~\ref{fig:phasespace-ckkwl}, there are four different ways in which two
shower emissions at scales $\ord_1$ and $\ord_2$ can be classified in
the variable $t$. Let the merging scale $\tms$ separate the regions of
low and high $t$, as sketched in the Figure~\ref{fig:phasespace-ckkwl}.

In panel (a), the two partons are inside the matrix element region and
as such should be generated by the matrix element generator. Sudakov
form factors are added in the way discussed in
\ref{sec:algorithm-step-step}. Notice that this is done by performing
trial showers, \ie\ that we discard the event if $\ord_1>\ord_{trial}
> \ord_2$, and we never change reclustered states or the matrix
element configuration. In this way, for non-zero weighted events,
every ME configuration is treated exactly as in the parton shower.  It
is also clear that panel (d), a configuration with two jets which are
soft in $t$ should come from the parton shower. Panel (c) provides the
next complicated region. Since region (a) is already correctly
accounted for, we veto shower emissions that would evolve the state
into one in (a).  This means that all states which evolved from a
matrix element state with one jet above $\tms$, and which have
non-zero weight, \ie\ with the first emission below $\tms$ or no
emission at all, are produced by PS evolution. As outlined in section
\ref{sec:algorithm-step-step}, Sudakov form factors below the
reconstructed evolution scale of the ME emission are added by using
the full shower when producing trial emissions. We ensure in this way
that form factors are added to the ME one-jet configuration in the
correct parton shower manner.

To our understanding, the truncated shower approach and \ckkwl only
differ in their treatment of panel (b). Let us clarify this
statement. In \ckkwl, we define the ME region to contain the $n\leq
N$ hardest jets in the evolution variable $\ord$, which are also above
the cut in $t$. Once we are inside the PS region, we believe the
shower is performing well, so that all further emissions will be taken
from the parton shower. The first emission in panel (b) is already in
the PS region. Thus, every further splitting is taken from the
shower. For the example this means that this two jet state is
generated from the 0-jet ME sample. Since the shower is the only
ingredient in how the state is produced, the accuracy of the shower is
preserved.

Truncated showers differ, in that they allow what we call a pure PS
state to evolve into something which could have evolved from a ME
1-jet state. Here, the emission is kept, the reclustered state
changed, and evolved further until the scale $\ord_2$ is
reached. Then, an emission with the reconstructed ME splitting
variables is forced. In this way, the path how the state was reached
is correctly described, and the accuracy of the shower is
retained. Truncated showering is allowed if the emissions that were
inserted before the hard emission are soft and did not change the
flavour of the line that will emit the hard jet.

This example reveals the different philosophies behind the \ckkwl and
Truncated Shower approaches. In \ckkwl, a compromise is made in that
only the $n\leq N$ emissions hardest in the evolution variable, and
above the merging scale cut $\tms$, are corrected with matrix element
configurations. Thus, in comparison to using truncated showers, a
smaller region of phase space have a matrix element
structure. However, we are allowed to use the full shower to generate
no-emission probabilities.

When using truncated showers, the flavour of the splitting lines has to be 
conserved in order to be able to attach the ME emission, \ie\ truncated showers
only allow gluon emissions. Also, splittings in the truncated showers cannot be
allowed to remove too much momentum from the line, since otherwise, the ME 
emission could not be forced. These restrictions make the Sudakov form factors 
differ slightly between the full shower and the truncated shower, though 
differences are sub-leading and might be tiny in an actual implementation.
  
Summarising, we believe that both \ckkwl and the Truncated Shower
approach have to compromise in regions with $\tms$-unordered
splittings. In \ckkwl, only the hardest partons in the evolution
variable will be corrected with tree-level matrix elements, as long as
they are above $\tms$ as well. This effectively means that the shower
evolution variable should be some measure of hardness, since
otherwise, only small regions of the relevant phase space will be
endowed with corrections. Choosing \eg\ a shower with an ordering
variable defined by angles would not be suitable. Truncated Shower
prescriptions allow correcting larger parts of the phase space with ME
configurations, though at the expense of compromising in the
generation of Sudakov form factors. This approach is particularly
suited if the evolution variable does not provide a hardness measure,
since then, the differences in the Sudakov form factors are vanishing,
while large fractions of the phase space can be described by ME
emissions. Since the evolution in transverse momentum provides a good
hardness measure, \ckkwl provides a natural merging scheme in \pytppp.

\section{Reconstructing shower splitting probabilities and intermediate
states}
\label{app:prob-and-kin}

In a numerical fixed order calculation, different Feynman graphs can contribute
to a particular phase space point. The analogue in a parton shower is that 
a multitude of different sequences of shower splittings can fill the same
phase space point. We describe in this appendix the construction and choice
of parton shower histories. The prescriptions below are implemented in \pytppp, 
with the code being publicly available from version $8.157$ onwards. Given
a matrix element state $\state{n}$, all possible intermediate states, 
splitting probabilities and splitting scales are reconstructed.
We first detail how splitting probabilities are calculated and
used to choose a particular path of shower splittings. We will after this 
outline how intermediate states are constructed.

\subsection{Calculation of splitting probabilities}

When assigning a parton shower history to a matrix element 
state, we have to decide on how to choose amongst all possible splitting 
sequences. Our choice of a suitable discriminant between these ``paths" is 
guided by the collinear factorisation of $n$-particle matrix elements:
\begin{eqnarray}
  \label{eq:me-factorize}
d\sigma_{n} &=& \mathcal{L}_n(x_n,t_n) F_{n}
 \left|{\cal M}_{n}\right|^2d\Phi_{n}\approx\nonumber\\
&&
  \frac{\as}{2\pi}\frac{1}{Q^2}
  \splitP(z)
  \mathcal{L}_n(x_{n},t_n)
  F_n \left|{\cal M}_{n-1}\right|^2~
  \frac{d\kTi{}^2dz}{z\left(1 -z\right)}
  \frac{d\phi}{2\pi}d\Phi_{n-1}
\end{eqnarray}
where $F_{m}$ is the flux factor and $\mathcal{L}_m$ the parton
luminosity for the $m$-parton final state using the factorisation
scale $t_n$, while $(\kTi{}^2,z,\phi)$ are the splitting variables,
$Q^2$ the virtuality of the splitting parton, and $\splitP(z)$ is the
DGLAP splitting kernel for the splitting. The integration measure is
given by
\begin{equation}
d\Phi_{m} = d\phi_{m}\frac{dx_m^+}{x_m^+}\frac{dx_m^-}{x_m^-},
\end{equation}
where $d\phi_{m}$ is the $m$-particle phase space volume and 
$x_m^\pm$ are the momentum fractions of the incoming partons moving 
in $\pm z$ direction. Using the fact that 
\begin{eqnarray}
  &&\mathcal{L}_{n}(x_{n},t_n) =
  x_{n}^+f_{n}^+(x_{n}^+,t_n)x_{n}^-f_{n}^-(x_{n}^-,t_n)~,
  \textnormal{ with }
  \mathcal{L}_{n}(x_{n},t_n) = \mathcal{L}_{n-1}(x_{n-1},t_{n-1}) 
  \textnormal{ for FSR,}\nonumber\\
  && \textnormal{and}\\
  && F_n = 
\begin{cases}
F_{n-1} &\textnormal{for FSR}\\
zF_{n-1} &\textnormal{for ISR}\\
\end{cases}\nonumber
\end{eqnarray}
as well as the definition of the evolution variable in \eqsref{eq:ordISR} and
\refeq{eq:ordFSR}, we can write the factorised transition cross section
as
\begin{equation}
  \label{eq:me-factorize-2}
d\sigma_{n} \approx
\begin{cases}
\left[\frac{\as}{2\pi}\frac{\splitP(z)}{\ord}d\kTi{}^2dz
     \frac{d\phi}{2\pi}\right]d\sigma_{n-1}, &\textnormal{for FSR;}\\
\left[\frac{\as}{2\pi}
      \frac{x_n^+f_n^+(x_{n}^+,t_n)}{x_{n-1}^+f_{n-1}^+(x_{n-1}^+,t_{n-1})}
      \frac{\splitP(z)}{\ord}d\kTi{}^2dz
     \frac{d\phi}{2\pi}\right]d\sigma_{n-1} &\textnormal{for ISR.}\\
\end{cases}
\end{equation}
To illustrate initial state radiation, we 
have here chosen the parton moving along $+z$ direction to split.  
Iterating this procedure down to the desired Born-level state 
represented by $m=0$, we can construct one path of collinear 
splittings by which we may have arrived at the
$n$-particle state. We can use the sum over all different
possible paths $p$,
\begin{equation}
  \label{eq:me-ps-path}
  d\sigma_n\approx
  \left[\sum_{p}\prod_{i=1}^n
  \frac{\as}{2\pi}
  \frac{x_{ip}f_{ip}(x_{ip},\ord_{ip})}{x_{i-1p}f_{i-1p}(x_{i-1p},\ord_{ip})}
  \frac{\splitP_{ip}(z_{ip})}{\ord_{ip}}
  d\kTi{ip}^2dz_{ip}\frac{d\phi_{ip}}{2\pi}\right]~d\sigma_0~,
\end{equation}
where $(\ord_{ip}^2,z_{ip},\phi_{ip})$ and $\splitP_{ip}$ are the
reconstructed splitting variables and splitting function for the
$i$'th splitting in the path $p$, as an approximation of the
$n$-parton cross section. The PDF ratio will equal unity for final
state splitting. We can make this correspondence exact for the very first 
splitting, by adding matrix element corrections to the splitting kernel 
and finding a common integration measure for the joined evolution 
along the possible paths, as will be addressed in the following. The very
first emission can be attributed to either a splitting of dipole end ``1",
or of dipole end ``2". If the momentum of dipole end $i \in \{1,2\}$ after 
the splitting is $p_i$, and the momentum of the emitted parton is $p_3$,
we define
\begin{eqnarray}
&&Q_{1i}^2 = 
  \begin{cases}
    (p_{i} + p_{3})^2 &\textnormal{ for FSR}\\
    (p_{i} - p_{3})^2 &\textnormal{ for ISR}
  \end{cases}
\qquad
z_{1i}^2 = 
  \begin{cases}
    \frac{x_i}{x_i+x_3},\textnormal{with}~
    x_k = \frac{2p_k\sum_{j=1}^3 p_j }{(p_1+p_2+p_3)^2}
      &\textnormal{ for FSR}\\
    \frac{(p_1+p_2-p_3)^2}{(p_1+p_2)^2} &\textnormal{ for ISR}
  \end{cases}\nonumber\\
&&
Q_{1q\neq p}^2 = 
  \begin{cases}
    Q_{11}^2 &\textnormal{if } p=2\\
    Q_{12}^2 &\textnormal{if } p=1
  \end{cases}
\qquad\qquad
m_{Dip}^2 =  (p_{1} + p_{2} + p_{3})^2 \textnormal{ for FSR}
\nonumber
\end{eqnarray}
With this notation, a joined evolution equation is given by
\begin{eqnarray}
&&d\mathcal{P}_{\textnormal{\tiny{FSR}}}
 = \left[\sum_{p=1}^2
    \frac{\as}{2\pi}\frac{\splitP_{1p}(z_{1p})
    \splitP_{p\textnormal{\tiny{MEcorr}}}}{Q_{1p}^2}
   ~\frac{(1-z_{1p})m_{Dip}^2}{Q_{11}^2+Q_{12}^2}\right]
 ~ d\pT^2 dy \label{eq:joined-prob-fsr}\quad\\
&&d\mathcal{P}_{\textnormal{\tiny{ISR}}}
 = \left[\sum_{p=1}^2
    \frac{\as}{2\pi}
~ \frac{x_{1p} f_{1p}(x_{1p},\ord_{1p})}{x_{0p} f_{0p}(x_{0p},\ord_{1p})}
  \right.\nonumber\\
&&\qquad\qquad\left.~\splitP_{1p}(z_{1p})\splitP_{p\textnormal{\tiny{MEcorr}}} 
\frac{(1-z_{1p}) Q_{1q\neq p}^2 + \ord_{reg}}
     {(1-z_{1p})^2 Q_{11}^2 Q_{12}^2 + (1-z_{1p})^2\ord_{reg}\hat s +\ord_{reg}^2}
\right] ~dQ^2 dz~.\quad\label{eq:joined-prob-isr}
\end{eqnarray}
The common integration measures for both paths were defined by introducing the
variables
\begin{eqnarray}
&&p_\perp^2 = \frac{Q_{11}^2Q_{12}^2}{m_{Dip}^2}~, \quad y = \frac{1}{2}\ln
    \frac{Q_{12}^2}{Q_{11}^2} \quad \textnormal{for the FSR case, and}\\
&&dQ^2 = d\left|Q_{11,12}\right| \quad z = z_{11} = z_{12} \quad
    \textnormal{for the ISR case.}
\end{eqnarray}
For initial state splittings, the weight takes a more
complicated form since in \pytppp, infrared singularities are
regularised by the introduction of a small scale $\ord_{reg}$. This is
inspired by the regularisation of multiple interactions using
arguments relating to colour screening
effects\cite{Corke:2010yf}. For vanishing $\ord_{reg}$,
the weight for the first splitting of initial particles again takes
the form given in \eqref{eq:me-factorize-2}. Note that we keep \as\
fixed, as the running of \as\ is corrected for later in the
algorithm. Also, the change in incoming parton content, compensated by
ratios of parton distributions, will be corrected later on. Hence, the
weight for each splitting should not contain \as\ or PDF ratio
factors. The product of these weights of individual splittings in a
path will then be used as the weight when choosing a path with the
normalised probability
\begin{eqnarray}
\label{eq:path-prob-1}
  w_p &=& \frac{ w_{1p}(z_{1p}) \prod_{i=2}^n
          \frac{\splitP_{ip}(z_{ip})}{\ord_{ip}}}%
  {\sum_{r} w_{1r}(z_{1r})
 \prod_{i=2}^n\frac{\splitP_{ir}(z_{ir})}{\ord_{ir}}}
 \quad\textnormal{where}\quad\\
\label{eq:path-prob-2}
w_{1p}(z_{1p}) &=& 
\begin{cases}
\frac{\splitP_{1p}(z_{1p})\splitP_{p\textnormal{\tiny{MEcorr}}}}{Q_{1p}^2} 
  ~\frac{(1-z_{1p})m_{Dip}^2}{Q_{11}^2+Q_{12}^2} & \textnormal{for FSR}\\
\splitP_{1p}(z_{1p})\splitP_{p\textnormal{\tiny{MEcorr}}}
\frac{(1-z_{1p}) Q_{1q\neq p}^2 + \ord_{reg} }
{(1-z_{1p})^2 Q_{11}^2 Q_{12}^2 + (1-z_{1p})^2 \ord_{reg}  \hat s + \ord_{reg}^2}
 & \textnormal{for ISR}
\end{cases} 
\end{eqnarray}
In section \ref{sec:results}, we compare this probabilistic prescription with a
winner-takes-it-all strategy based on the smallest sum of
transverse momenta, and observe minor, though visible, differences.

\subsection{Reconstruction of intermediate states}

Given an n-parton phase space point $\state{n}$ from a matrix element 
generator, we explicitly construct all possible intermediate states 
$\state{0}\dots,\state{n-1}$ in all paths $p$ by reclustering allowed 
shower emissions. For the construction of the state $\state{i}$, given that
we have the state $\state{i+1}$, this rather complicated step is 
achieved by inverting all the changes the shower would have applied in the 
construction of the emission. This means that we need construct
\begin{enumerate}
\item The underlying momenta
$\tilde{p} = \{\tilde{p}_{0},\dots,\tilde{p}_{k+i}\}$ from the momenta
$p = \{{p}_{0},\dots,{p}_{k+i+1}\}$;
\item The underlying flavour configuration $\flst_{+i}$ from the 
configuration $\fls_{+i+1}$;
\item The underlying colour configuration $\clst_{+i}$ from $\cls_{+i+1}$.
\end{enumerate}
In the following, write ``before" for values before the clustering, and 
``after" for values after the clustering.

\subsubsection*{Reclustering of momenta}
The construction of the reclustered momenta $\tilde{p}$ from the momenta 
$p$ is achieved by exactly reverting all changes \pytppp would have done if the
showers would have constructed an emission resulting in the momenta $p$. 
Formally speaking, this means that we invert the radiative phase space mapping
of the shower. The construction of the momentum of an emission in \pytppp 
differs between initial state and final state splittings, leading to different
prescriptions how underlying kinematics $\tilde{p}$ are constructed.

For a final state emission with a final 
state recoiler, this means that the momenta of the reconstructed radiator and 
recoiler in the rest frame of the dipole are set to
\begin{eqnarray*}
p^{\mu}_{\textnormal{radiator, after}} =
 \left( ~0, ~0, ~~\frac{m_{Dip}}{2},~\frac{m_{Dip}}{2}\right)~,\qquad
p^{\mu}_{\textnormal{recoiler, after}} =
 \left( ~0, ~0, -\frac{m_{Dip}}{2},~\frac{m_{Dip}}{2}\right)
\end{eqnarray*}
and then rotated and boosted from the rest frame of the decaying 
dipole\footnote{Defined by $\vec{p}_{\textnormal{radiator, before}} + 
\vec{p}_{\textnormal{emitted, before}}$ aligned along $+z$-direction, 
$\vec{p}_{\textnormal{recoiler, before}}$ aligned along $-z$-direction} to the 
event centre-of-mass frame\footnote{Defined by the orientation of the momenta 
$\vec{p}_{\textnormal{radiator, before}} + 
\vec{p}_{\textnormal{emitted, before}}$ and 
$\vec{p}_{\textnormal{recoiler, before}}$ taken from the 
unchanged $2\rightarrow n$ state, where these are not anti-parallel.}.

For
final state splittings with an initial state recoiler, the shower would have 
taken the energy (four-momentum) for the emitted particle from the beam. 
This steps is undone after the boost to the event centre-of-mass frame 
by setting the recoiler momentum according to
\begin{eqnarray*}
p^{\mu}_{\textnormal{recoiler, after}} =
   2\cdot p^{\mu}_{\textnormal{recoiler, before}}
 ~-~ p^{\mu,\textnormal{ after Lorentz transformation}}
      _{\textnormal{recoiler, after}}.
\end{eqnarray*}

For initial state splittings, \pytppp distributes
the recoil among all final state particles, making the inversion of 
this momentum mapping more complicated. We will denote all unchanged 
momenta of the original $2\rightarrow n$ process by $p_i$. The momenta 
of the partons involved in the splitting are denoted by 
$p_{\textnormal{mother}}$, 
$p_{\textnormal{sister}}$ and $p_{\textnormal{partner}}$. When referring to
$p_{\textnormal{mother}}$, $p_{\textnormal{sister}}$ or 
$p_{\textnormal{partner}}$ in the following we always think about the momenta
of these particles at the current step in the construction of 
reclustered kinematics. Inverting the construction of kinematics of \pytppp
proceeds as follows:
\begin{enumerate}
\item Undo the rotation with
\begin{eqnarray*}
 \phi = \arctan\left( 
 \frac{[p_{\textnormal{sister}}]_y}{[p_{\textnormal{sister}}]_x}\right)
\end{eqnarray*}
that \pytppp would have done, by rotating all momenta with $-\phi$.
\item Transform all momenta from the event centre-of-mass 
frame\footnote{Defined by the orientation of the momenta 
$p_{\textnormal{daughter}}$ and $p_{\textnormal{recoiler}}$ in the rotated,
but otherwise unchanged $2\rightarrow n$ process.} to the centre-of-mass 
frame\footnote{Defined by $\vec{p}_{\textnormal{daughter}}$ being aligned to
the +$z$-direction and $p_{\textnormal{recoiler}}$ aligned to -$z$-direction.} 
of the momenta $p_b^\mu = p_{\textnormal{daughter}}^\mu 
         = p_{\textnormal{mother}}^\mu - p_{\textnormal{sister}}^\mu$
and $p_{\textnormal{recoiler}}^\mu = p_{\textnormal{partner}}^\mu$. Notice 
that we transform to the centre-of-mass frame of the off-shell momentum
$p_{\textnormal{daughter}}$.
\item Undo the
\begin{eqnarray*}
 -\theta = -\arctan\left( \frac{\sqrt{[p_{\textnormal{mother}}]_x^2
         + [p_{\textnormal{mother}}]_y^2 } }{[p_{\textnormal{mother}}]_z}
           \right)
\end{eqnarray*}
rotation that \pytppp would have done by rotating all momenta with $\theta$.
\item Construct the on-shell momenta $p_{\textnormal{daughter}}$
and $p_{\textnormal{recoiler}}$ by resetting
\begin{eqnarray*}
&p^{\mu}_{\textnormal{daughter}} =
 \left( ~0, ~0, ~~\frac{1}{2}\hat s,~\frac{1}{2}\hat s\right)
\quad
&p^{\mu}_{\textnormal{recoiler}} =
 \left( ~0, ~0, -\frac{1}{2}\hat s,~\frac{1}{2}\hat s\right)~,\nonumber
\end{eqnarray*}
where
\begin{eqnarray*}
&\hat s = z x_1 x_2\cdot \ECM^2 \quad \textnormal{and} \quad
&x_1 = \frac{2E_1}{\ECM}, ~~x_2 = \frac{2E_2}{\ECM}, ~~ z
     = \frac{\left(p_1^\mu + p_2^\mu - p_3^\mu\right)^2}
            {\left(p_1^\mu + p_2^\mu\right)^2}\nonumber~.
\end{eqnarray*}
\item Boost along the z-axis to the frame where the energy fraction of 
the newly constructed $p^{\mu}_{\textnormal{recoiler}}$ is the original 
value $x_2$, i.e.\ along the vector
\begin{eqnarray*}
 \vec{p}_{\textnormal{boost}}
  = \left( ~0, ~0, \frac{ z\cdot x_1 - x_2 }{z\cdot x_1 + x_2} \right)~.
\end{eqnarray*}
After this boost, the newly constructed $p^{\mu}_{\textnormal{recoiler}}$ 
should be identical to $p^{\mu}_2$
\item Undo the initial $-\phi$ rotation by a rotation with $\phi$.
\end{enumerate}
These changes allow us to reconstruct the state from which \pytppp would have
constructed the matrix element momenta if the shower would have produced
a splitting at the reconstructed splitting scale. We tested that this method
of inverting the shower splitting kinematics exactly reproduces lower 
multiplicity states from states with additional shower emissions. We found
complete agreement, since this construction explicitly inverts the momentum
mapping of the shower.

\subsubsection*{Reconstruction of the underlying flavour structure}

To reconstruct the intermediate state $\state{i}$, we further have to assign
the correct flavour structure $\flst_{+i}$. With the notation $f(k)$
and $\bar{f}(k)$ for the flavour of particle $k$ and the antiparticle to $k$,
the flavour mapping can be accomplished by following the rules:
\begin{enumerate}
\item If the emitted parton is a gluon, then
\begin{eqnarray*}
f(\textnormal{radiator, after}) = f(\textnormal{radiator, before})
\end{eqnarray*}
\item If the emitted parton is a quark, and the radiating parton is a gluon,
then
\begin{eqnarray*}
f(\textnormal{radiator, after}) = 
\begin{cases}
 f(\textnormal{emitted, before}) & \textnormal{ in FSR,}\\
  \bar{f}(\textnormal{emitted, before}) & \textnormal{ in ISR,}
\end{cases}
\end{eqnarray*}
\item If the emitted parton is a quark, and the radiating parton is the 
corresponding antiquark, then
\begin{eqnarray*}
f(\textnormal{radiator, after}) = 
\begin{cases}
 \g & \textnormal{ in FSR,}\\
 \textnormal{not possible} & \textnormal{ in ISR,}
\end{cases}
\end{eqnarray*}
\item  If the emitted parton is a quark, and the radiating parton is a quark
of the same flavour, then
\begin{eqnarray*}
f(\textnormal{radiator, after}) = 
\begin{cases}
 \textnormal{not possible} & \textnormal{ in FSR,}\\
 \g & \textnormal{ in ISR,}
\end{cases}
\end{eqnarray*}
\end{enumerate}
This exhausts the list of QCD flavour mappings in \pytppp, so that following
these rules, the flavour configuration $\flst_{+i}$ of the state $\state{i}$
can be reconstructed.

\subsubsection*{Reconstruction of the underlying colour structure}

Finally, we need to construct the colour configuration $\clst_{+i}$.
Let us write $c$ ($\bar{c}$) for the colour (anticolour) of partons, and 
indicate the parton to be reconstructed by a subscript $r$. After flavours have
been assigned, the colour of the parton $\p_r$ can be found by following the 
rules
\begin{enumerate}
\item For final state splittings with a gluon involved as either emitted or
radiating parton, i.e.\
\begin{eqnarray*}
&& \q_r\to\q\g_{emt}~,\quad
   \q_r\to\g\q_{emt}~,\quad
 \bar{\q}_r\to\bar{\q}\g_{emt}~,\quad
   \bar{\q}_r\to\g\bar{\q}_{emt}~,\quad
   \g_r\to\g\g_{emt}\quad\textnormal{(FSR)}
\end{eqnarray*}
remove the index appearing both as colour and anticolour
in the (emitted, radiating)--parton pair. Set the leftover colour and anticolour
as the colour and anticolour of $\p_r$, i.e.
\begin{eqnarray*}
&& c_{\q_r} = c_{g_{emt}}~,~ \bar{c}_{\q_r} = 0\\
&&   c_{\bar{\q}_r} = 0~,~ \bar{c}_{\bar{\q}_r} = \bar{c}_{g_{emt}}\\
&& c_{\g_r} = c_{g_{emt}}~,~ \bar{c}_{\g_r} = \bar{c}_{\g_{rad}}\qquad
\textnormal{or}\qquad
   c_{\g_r} = c_{g_{rad}}~,~ \bar{c}_{\g_r} = \bar{c}_{g_{emt}}
\end{eqnarray*}
The second possibility for $\g_r\to\g\g_{emt}$ can occur if the matrix element
generator produced a non-planar colour flow.
\item For final state splittings with quark and antiquarks as emitted and
radiating partons, i.e.\
\begin{eqnarray*}
&& \g_r\to\q\bar{\q}_{emt}~,\quad
   \g_r\to\bar{\q}\q_{emt}\qquad\textnormal{(FSR)}
\end{eqnarray*}
set the colour of $\g_r$ to the quark
colour, the anticolour to the antiquark anticolour. This means
\begin{eqnarray*}
&& c_{\g_r} = c_{\q}~,~ \bar{c}_{\g_r} = \bar{c}_{\bar{q}}
\end{eqnarray*}
irrespectively of whether the quark or the antiquark is considered the 
emitted parton.
\item For initial state splittings with an emitted gluon, i.e.\
\begin{eqnarray*}
&& \g\to\g_r\g_{emt}~,\quad
   \q\to\q_r\g_{emt}\qquad\textnormal{(ISR)}
\end{eqnarray*}
remove the index appearing as colour (or anticolour) both in the emitted 
and radiating parton. If a colour (anticolour) index remains in the initial 
state, set the colour (anticolour) of $\p_r$ to the remaining initial 
state colour (anticolour), and set the $\p_r$ anticolour (colour) to the 
remaining final state colour (anticolour) index.
\item For initial state splittings with an emitted quark (antiquark) and a 
gluon radiator, i.e.
\begin{eqnarray*}
&& \g\to\q_r\bar{\q}_{emt}~,\quad
   \g\to\bar{\q}_r\q_{emt}\qquad\textnormal{(ISR)}
\end{eqnarray*}
set the $\p_r$ colour (anticolour) to the colour (anticolour) of
the radiating gluon.
\item For initial state splittings with an emitted quark (antiquark) and a 
quark (antiquark) radiator, i.e.\
\begin{eqnarray*}
&& \q\to\g_r\q_{emt}~,\quad
   \bar{q}\to\g_r\bar{\q}_{emt}\qquad\textnormal{(ISR)}
\end{eqnarray*}
set the $\g_r$ colour (anticolour) to the anticolour (colour) of
the emitted parton. Set the reconstructed gluon anticolour (colour) to the 
anticolour (colour) of the radiating parton.
\end{enumerate}
Once a pair of radiating and emitted partons is chosen, these rules can be 
applied to deduce the colour configuration $\clst_{+i}$ of the state 
$\state{i}$.

Combining the inversion of the parton shower kinematics, the construction of the
underlying flavour configuration and reclustering of colours, the complete
state $\state{i}$ can be generated. We have extensively tested that, given a
state $\state{n}$, our implementation exactly reproduces all states
$\state{(m<n)}$, if the states $\state{m+1},\dots \state{n}$ were generated by
shower splittings, verifying
that we have used the exact inversion of the radiative mappings of \pytppp.

\end{appendix}

\bibliographystyle{utcaps}  
\bibliography{references} 

\end{document}

%% file: mydefs.tex
\usepackage{cite}
\usepackage{epsfig}
\usepackage{xspace}
\usepackage{graphics}
\usepackage{relsize}

\def\hide#1{}

\newcommand{\asme}{\alpha_{\textnormal{s\tiny{ME}}}}

\newcommand{\tms}{t_{\textnormal{\tiny{MS}}}}

\newcommand{\mz}{\textnormal{M}_{\textnormal{\tiny{Z}}}}
\newcommand{\mw}{\textnormal{M}_{\textnormal{\tiny{W}}}}
\newcommand{\ckkwl}{\textnormal{\textsc{Ckkw-l}}\xspace}
\newcommand{\ckkw}{\textnormal{\textsc{Ckkw}}\xspace}

\newcommand{\fls}{\mathcal{F}}
\newcommand{\cls}{\mathcal{C}}
\newcommand{\flst}{\widetilde{\mathcal{F}}}
\newcommand{\clst}{\widetilde{\mathcal{C}}}

\newcommand{\alpgen}{A{\smaller LP}G{\smaller EN}\xspace}

\newcommand{\madgraph}{M{\smaller AD}G{\smaller RAPH}\xspace}
\newcommand{\madevent}{M{\smaller AD}E{\smaller VENT}\xspace}

\newcommand{\rivet}{R{\smaller IVET}\xspace}

\newcommand{\powheg}{P{\smaller OWHEG}\xspace}
\newcommand{\sherpa}{S{\smaller HERPA}\xspace}

\newcommand{\pythia}{P{\smaller YTHIA}\xspace}

\newcommand{\pytppp}{P{\smaller YTHIA}8\xspace}

\newcommand{\herpp}{{\smaller HERWIG++}\xspace}

\newcommand{\ariadne}{A{\smaller RIADNE}\xspace}

\newcommand{\as}{\ensuremath{\alpha_{\mathrm{s}}}}

\newcommand{\pT}{\ensuremath{p_{\perp}}}

\newcommand{\kTi}[1]{\ensuremath{k_{\perp #1}}}

\newcommand{\ECM}{\ensuremath{E_{\mathrm{CM}}}}

\newcommand{\particle}[1]{\ensuremath{\mathrm{#1}}}
\newcommand{\antiparticle}[1]{\ensuremath{\bar{\mathrm{#1}}}}
\newcommand{\el}{\particle{e}}
\newcommand{\eplus}{\ensuremath{\el^+}}

\newcommand{\g}{\particle{g}}
\newcommand{\p}{\particle{p}}
\newcommand{\q}{\particle{q}}

\newcommand{\W}{\particle{W}}
\newcommand{\Wp}{\ensuremath{\W^+}}
\newcommand{\Wm}{\ensuremath{\W^-}}
\newcommand{\Z}{\particle{Z}}

\newcommand{\qu}{\particle{u}}
\newcommand{\qd}{\particle{d}}

\newcommand{\qc}{\particle{c}}

\newcommand{\pbar}{\antiparticle{p}}
\newcommand{\qbar}{\antiparticle{q}}

\newcommand{\ubar}{\antiparticle{u}}
\newcommand{\dbar}{\antiparticle{d}}

\newcommand{\cbar}{\antiparticle{c}}

\newcommand{\ee}{\ensuremath{\el^+\el^-}}
\newcommand{\tee}{\ensuremath{\el^+\el^-}\xspace}

\newcommand{\ord}{\ensuremath{\rho}}
\newcommand{\aux}{\ensuremath{{\boldmath z}}}
\newcommand{\state}[1]{\ensuremath{S_{+#1}}}
\newcommand{\splitP}{\ensuremath{P}}

\newcommand{\done}[1]{}

\def\mrm#1{\mathrm{#1}}
\newcommand{\ie}{i.e.\xspace}
\newcommand{\eg}{e.g.\xspace}

\def\sub#1{\ensuremath{_{\mrm{#1}}}}
\def\sup#1{\ensuremath{^{\mrm{#1}}}}

\def\sud#1{\ensuremath{\Delta_{\state{#1}}}}
\def\Pnoem{\ensuremath{\Pi}}
\def\noem#1{\ensuremath{\Pnoem_{\state{#1}}}}

\def\f2d3{\ensuremath{F_2^{\mrm{D}3}}}

\providecommand{\eqref}[1]{eq.~(\ref{#1})\xspace}
\renewcommand{\eqref}[1]{eq.~(\ref{#1})\xspace}
\newcommand{\eqsref}[1]{eqs.~(\ref{#1})\xspace}
\newcommand{\refeq}[1]{(\ref{#1})\xspace}

\newcounter{aenumct}

\newcounter{ienumct}

{\end{list}}

\newcounter{enumct}
\renewenvironment{enumerate}{\begin{list}{\arabic{enumct}.}%
{\usecounter{enumct}\setlength{\topsep}{1mm}%
\setlength{\partopsep}{1mm}\setlength{\itemsep}{0mm}%
\setlength{\parsep}{1mm}}}{\end{list}}